\documentclass[pre,superscriptaddress,longbibliography,footinbib,twocolumn,preprintnumbers,amsmath,amssymb]{revtex4-2}

\usepackage{siunitx}

\makeatletter
\newcommand*{\addFileDependency}[1]{
	\typeout{(#1)}
	\@addtofilelist{#1}
	\IfFileExists{#1}{}{\typeout{No file #1.}}
}
\makeatother

\usepackage{amssymb}
\usepackage{amsmath}
\usepackage{float} 

\usepackage{color}
\usepackage[pdftex]{graphicx}

\usepackage{subcaption}

\usepackage{ragged2e} 
\DeclareCaptionJustification{justified}{\justifying}

\usepackage{empheq}

\usepackage{mathtools}

\usepackage{makecell}

\usepackage{enumerate}
\usepackage{mathtools}
\usepackage{stmaryrd}

\usepackage{enumitem}

\usepackage{textcomp}
\usepackage{gensymb}

\newcommand{\dd}[0]{\mathrm{d}}

\newcommand{\rr}[0]{\boldsymbol{r}}

\newcommand{\kB}[0]{k_{\mathrm{B}}}

\usepackage{lipsum}  
\definecolor{darkblue}{rgb}{0,0,0.6}
\definecolor{darkred}{rgb}{0.6,0,0}
\usepackage[colorlinks=true,urlcolor=darkblue,citecolor=darkblue,linkcolor=darkred]{hyperref}

\begin{document}

\title{A Brownian dynamics study of liquid-liquid phase separation\\ in multi-scale chromatin networks}

\author{L\'ea Beaul\`es}
\affiliation{Sorbonne Universit\'e, CNRS, Physico-Chimie des \'Electrolytes et Nanosyst\`emes Interfaciaux (PHENIX), 4 Place Jussieu, 75005 Paris, France}

\author{Judith Miné-Hattab}
\affiliation{Sorbonne Universit\'e, Institut de Biologie Paris-Seine, CNRS, Laboratory of Computational, Quantitative and Synthetic Biology (CQSB), 4 Place Jussieu, 75005 Paris, France}

\author{Pierre Illien}
\affiliation{Sorbonne Universit\'e, CNRS, Physico-Chimie des \'Electrolytes et Nanosyst\`emes Interfaciaux (PHENIX), 4 Place Jussieu, 75005 Paris, France}

\author{Vincent Dahirel}
\thanks{vincent.dahirel@sorbonne-universite.fr}
\affiliation{Sorbonne Universit\'e, CNRS, Physico-Chimie des \'Electrolytes et Nanosyst\`emes Interfaciaux (PHENIX), 4 Place Jussieu, 75005 Paris, France}

\begin{abstract}

In living cells, proteins involved in specialized biochemical functions are often spatially organized within biomolecular condensates. Increasing evidence suggests that some of these condensates, including DNA repair condensates, emerge through liquid–liquid phase separation (LLPS). In the nucleus, however, condensates form within a highly heterogeneous environment composed of chromatin fibers, RNA, and additional protein scaffolds such as PAR chains, all of which may interact with phase-separating proteins. Moreover, condensate formation is frequently associated with specific chromatin conformations; for instance, loop extrusion has been proposed as a mechanism promoting DNA repair condensates.
Here, we investigate how the surrounding fibrous environment controls the morphology and spatial organization of phase-separated condensates. Using Brownian dynamics simulations of minimal models combining Lennard-Jones particles with fixed fibrous substrates, we examine the respective roles of local fiber geometry and large-scale network organization, reflecting the multiscale architecture of chromatin. We show that protein–fiber interactions strongly influence droplet positioning relative to the substrate, in a manner analogous to wetting transitions in soft condensed matter systems. Both local geometric constraints and global network organization markedly affect droplet size, morphology, and multiplicity. In addition, large-scale asymmetries in fiber organization can induce robust spatial localization of the dense phase. Our results thus highlight how multiscale structural heterogeneity of the nuclear environment can regulate the emergence and organization of biomolecular condensates.

\end{abstract}

\date{\today}

\maketitle

\section{Introduction}  

Biomolecular condensates, also referred to as membraneless organelles, are mesoscale structures with radii ranging from about 10 nm to several micrometers~\cite{Brangwynne_2009, Shin_2017, Shin_2018, Gibson_2019, Weber_2019, Sabari_2020, Gouveia_2022, Rouches_2024}. These assemblies are found both in the cytoplasm (e.g., stress granules or P-bodies)~\cite{Brangwynne_2009, Shin_2017} and in the nucleoplasm of eukaryotic cells, where they include nucleoli, repair focii, and transcription factories~\cite{Hnisz_2017, Sabari_2020, Rippe_2022, Wu_2022, Mann_2023}. Many of these condensates display liquid-like properties and are thought to form via liquid–liquid phase separation (LLPS), a process in which a homogeneous mixture demixes into two coexisting liquid phases~\cite{Weber_2019, Brangwynne_2015, Banani_2017}. Hallmark features of LLPS include condensate fusion~\cite{Kilic_2019}, internal rearrangement, and dynamic exchange of components between dilute and dense phase~\cite{Hyman_2014, Alberti_2019, Heltberg_2021}.

In the nucleus, LLPS of RNA and protein factors underlies the formation of condensates involved in DNA replication and repair~\cite{Mine-hattab_2021, Heltberg_2022, GarciaFernandez_2023}, transcriptional regulation~\cite{Mann_2023}, and RNA processing~\cite{Sabari_2020, Shin_2018}. These processes are tightly coupled to the structural organization of chromatin, a heteropolymer of DNA and histone proteins arranged into epigenetic domains~\cite{Filion_2010, Boettiger_2016, Cattoni_2017, Szabo_2018}, characterized by preferential intra-domain contacts~\cite{Sexton_2012}. The interplay between chromatin folding and condensate formation is therefore increasingly recognized as a key regulator of nuclear organization and gene expression~\cite{Arnould_2020,Arnould_2021,Arnould_2023, Sabari_2020, Shin_2019}.

Recent advances in imaging enable quantitative characterization of condensate geometry and dynamics~\cite{Keber_2024, Broedersz_2014, Brangwynne_2009, Heltberg_2021}. A striking observation is the coexistence of many mesoscopic condensates within single cells, pointing to robust mechanisms to control growth mechanisms. While non-equilibrium biochemical reactions have been proposed to arrest Ostwald ripening~\cite{Weber_2019,Zwicker_2022}, alternative mechanisms may arise from the nuclear environment itself. In particular, chromatin fibers provide a heterogeneous scaffold that can influence nucleation, wetting, and droplet stability~\cite{Strom_2024}. In this context, LLPS and wetting are intrinsically coupled, as both are governed by the balance of interfacial free energies.

Phase separation is primarily characterized by the number, size, and dispersity of droplets. In the presence of fibers or surfaces, the spatial positioning of the dense phase with respect to these interfaces becomes an additional key property. This is particularly relevant for nuclear condensates: for instance, DNA repair condensates are observed to form specifically at damaged sites~\cite{Mine-hattab_2021}. We refer here to this phenomenon as localization of the dense phase. Several mechanisms may account for such localization, including local nucleation triggered by a perturbation~\cite{Mine-hattab_2021}, colocalisation of a pre-existing droplet and a specific site~\cite{Du_2024}, or redistribution of material via coarsening dynamics~\cite{Heltberg_2022}. Disentangling these contributions requires understanding how geometry alone can bias condensate positioning.

The physics of wetting on fibers has been extensively studied at macroscopic scales~\cite{Rayleigh_1878, DeGennes_2004, Eggers_2008}, while at smaller scales thermal fluctuations and finite-size effects become important~\cite{Zhang_2021, Gopan_2014, Zhang_2020}. In biological systems, additional complexity arises from the multiscale organization of chromatin, including its geometry, topology, and biochemical heterogeneity. On the LLPS side, minimal theoretical models such as Flory–Huggins descriptions have provided key insights into condensate formation~\cite{Flory_1942, Huggins_1942, Zwicker_2022, Tiani_2025, Ronceray_liquid_2022, Rosowski_2020}. These approaches capture the essential role of weak multivalent interactions between biomolecules, often modeled as effective isotropic interactions, and highlight the sensitivity of phase behavior to small parameter variations. 

While protein–protein interactions have been extensively characterized, the role of substrate-binding interactions and of the underlying geometry remains less understood~\cite{Strom_2024, Style_2018}. In particular, although the effects of network elasticity and concentration on embedded droplets have been studied~\cite{Style_2018, Lee_2021, Zhang_2021, Qi_2021, Ronceray_liquid_2022}, it is still unclear how the geometry of a fibrous scaffold controls condensate size, number, and spatial organization. However, some studies indirectly suggest that geometric changes of chromatin are critical to condensate formation. For instance, some experiments correlate loop extrusion, chromatin domain formation and condensate growth~\cite{Arnould_2021}.

In this article, we address this question using minimal coarse-grained models. We perform Brownian Dynamics simulations of phase-separating fluids interacting with fixed fibers, focusing on how both local fiber geometry and large-scale network organization affect the macroscopic structure of the dense phase. By neglecting chromatin flexibility, we isolate the role of geometry and interactions, and provide a baseline description of LLPS in the presence of an immobile scaffold.

As a reference, we first consider a Lennard-Jones fluid in bulk, where phase separation leads to the formation of a single droplet at equilibrium. We then progressively introduce chromatin-like fibers with increasing geometric complexity. We analyze single fibers (straight, zig-zag, and looped) to probe the role of curvature and intersections, and extend the study to multi-fiber networks with controlled spatial organization, ranging from regular to disordered configurations.

We show that fiber geometry provides a robust control of condensate organization. First, protein--chromatin interactions induce wetting transitions that modify droplet morphology and shift the effective phase behavior. Second, local geometric features such as loops or intersections act as preferential binding sites, enabling the coexistence of multiple droplets in regimes where bulk systems would coarsen into a single domain. Third, the large-scale organization of the fiber network controls the spatial localization of the dense phase, with structural inhomogeneities driving condensates toward chromatin-specific regions. Together, these results demonstrate that chromatin geometry alone can regulate condensate size, multiplicity, and positioning, providing an equilibrium physical mechanism that complements biochemical and non-equilibrium models of condensate organization.

\section{Model}
\label{sec:model}

\subsection{Model for the proteins condensate in presence of chromatin}

\begin{figure}
    \centering
    \begin{subfigure}{1\linewidth}
        \subcaption{}
        \includegraphics[width=1\linewidth]{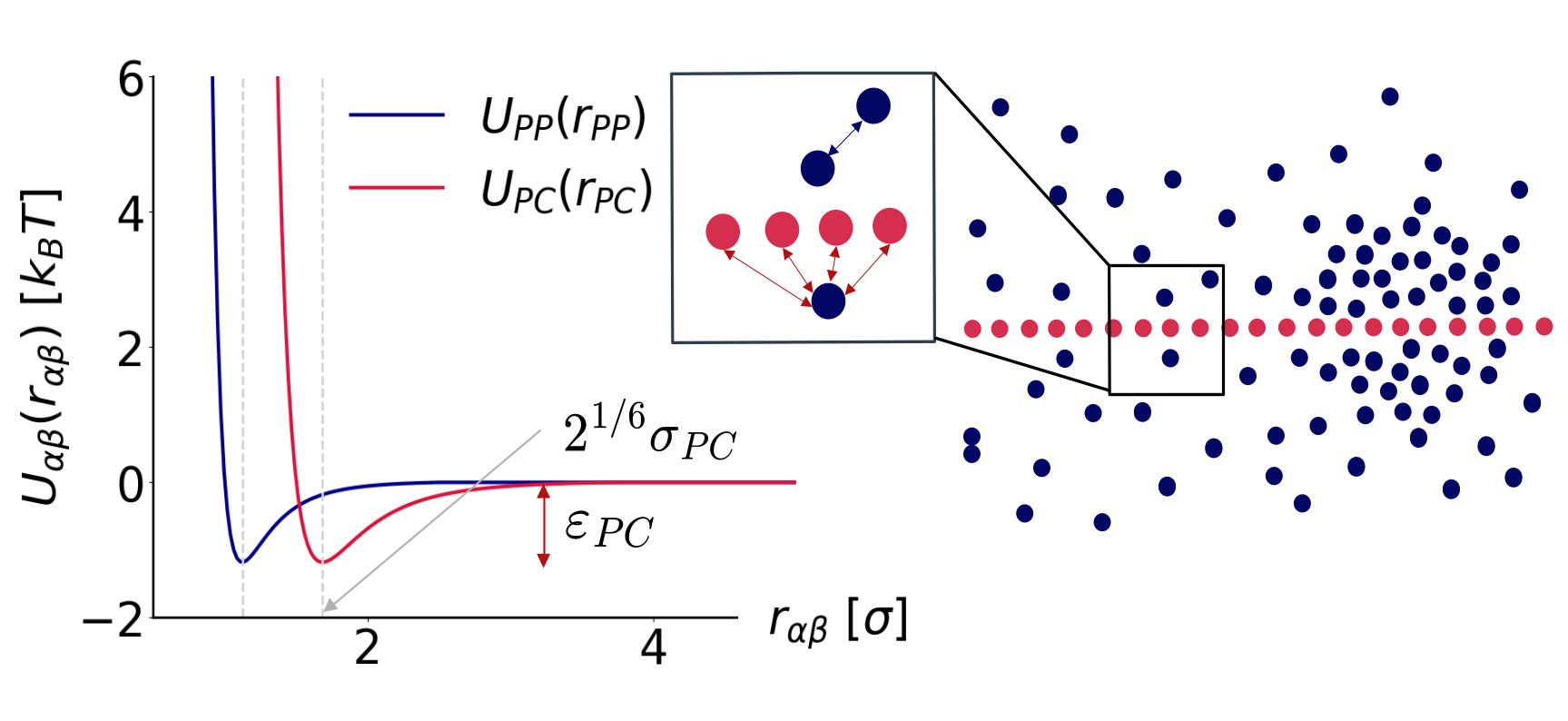}
        \label{fig:chromatin_geometries_LJ}
    \end{subfigure}
    
    \begin{subfigure}{1\linewidth}
        \subcaption{}
        \includegraphics[width=1\linewidth]{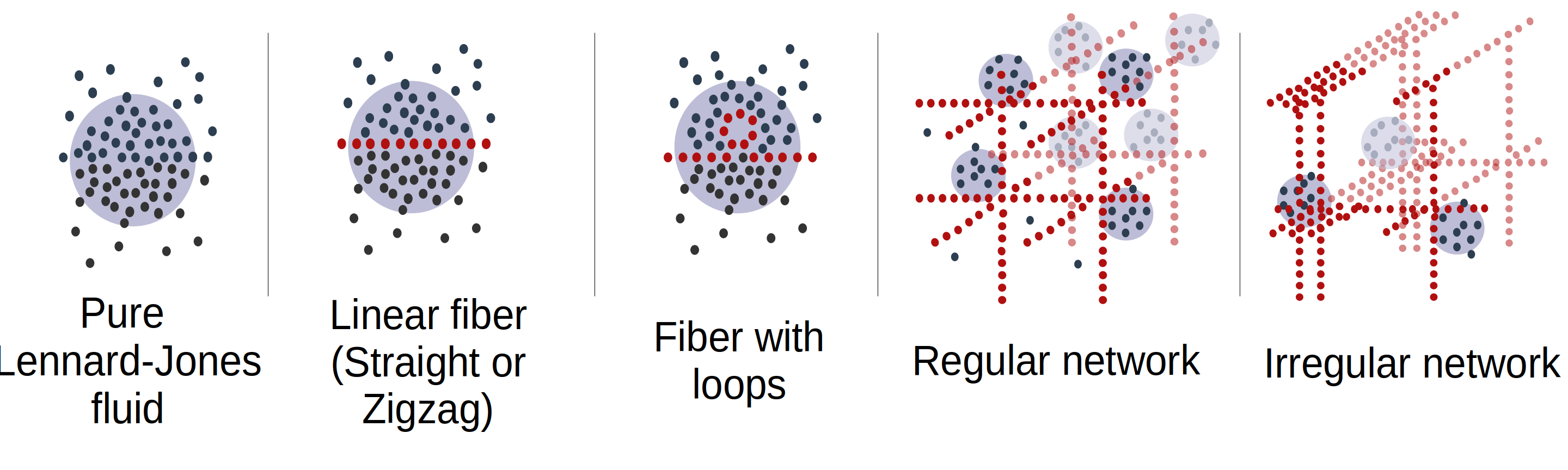}
        \label{fig:chromatin_geometries_GEOM}
    \end{subfigure}
    
    \begin{subfigure}{0.42\linewidth}
        \subcaption{}
        \includegraphics[width=1\linewidth]{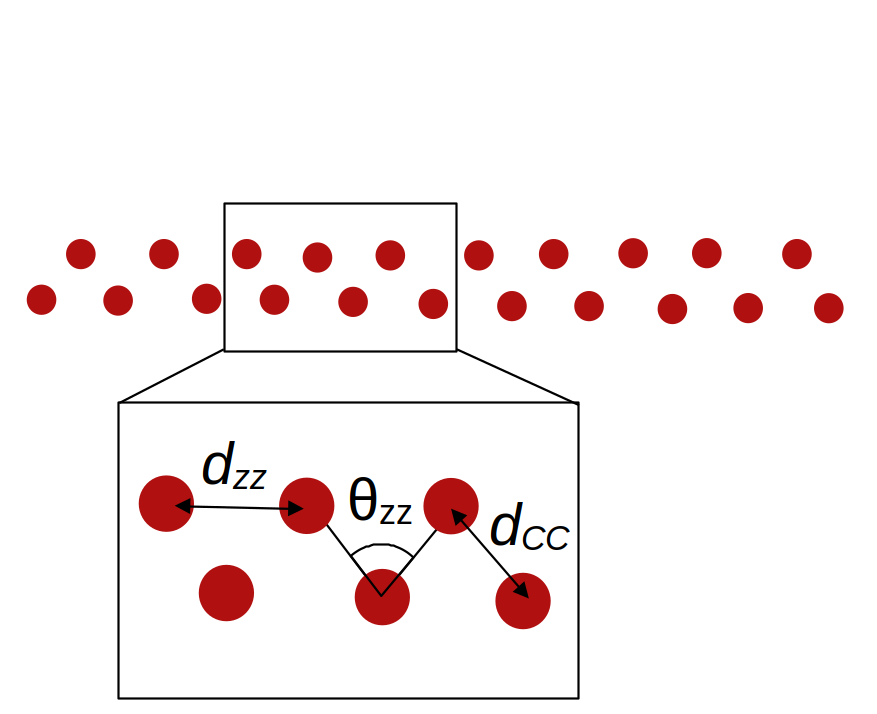}
        \label{fig:chromatin_geometries_ZZ}
    \end{subfigure}
    \hspace{0.0\linewidth}
    \begin{subfigure}{0.55\linewidth}
        \subcaption{}
        \includegraphics[width=1\linewidth]{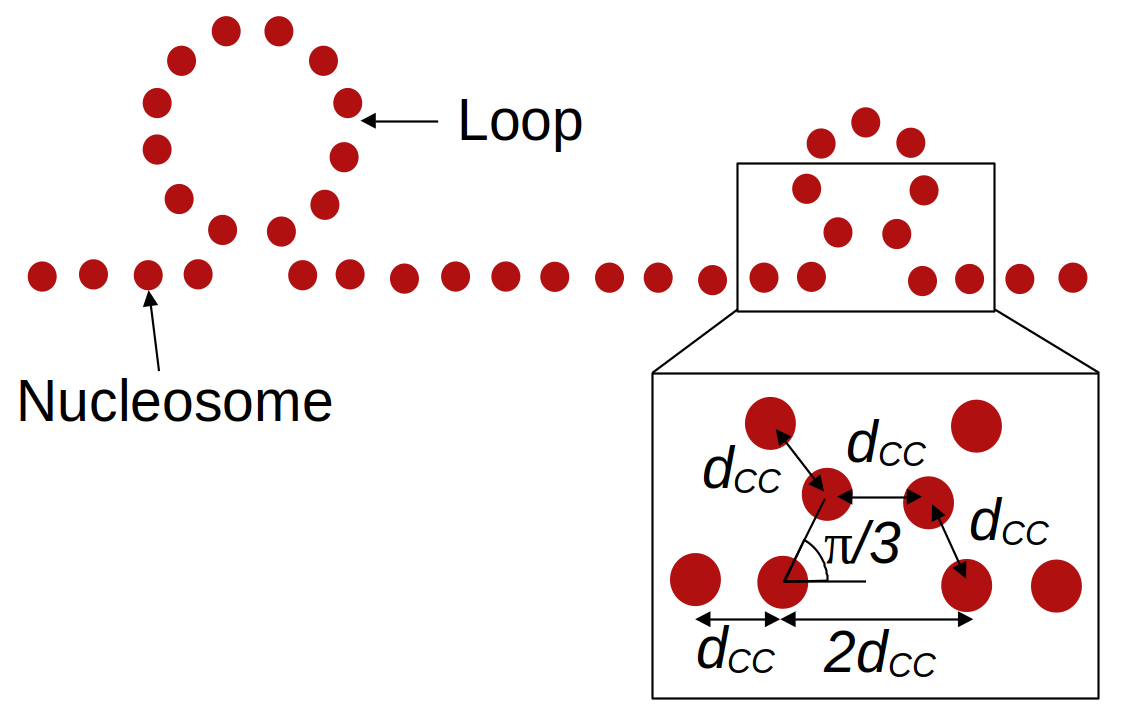}
        \label{fig:chromatin_geometries_LOOP}
    \end{subfigure}
    
    \caption{\subref{fig:chromatin_geometries_LJ}. Illustrative sketch of the inter-particle potentials $U_{\alpha\beta}$, for $\varepsilon_{PP}=\varepsilon_{PC}=1.2$, and $\sigma_{PC} = 1.5 \, \sigma_{PP}$ \subref{fig:chromatin_geometries_GEOM}. Geometries of the chromatin model. Nucleosomes are represented by red spheres, and free proteins by blue spheres. 
    \subref{fig:chromatin_geometries_ZZ}. Details of the zigzag geometry. \subref{fig:chromatin_geometries_LOOP}. Details of the fiber with loops. Note that neighbouring loops lies in parallel planes distant of $d_{CC}$ to ensure a minimum distance between nucleosomes of $d_{CC}$.}
    \label{fig:chromatin_geometries}
\end{figure}

As a minimal model of a phase separating fluid around fibers, we use an effective two-species system, including explicit proteins $P$ and fixed chromatin $C$ into an implicit solvent.
As chromatin beads are fixed, there are only two kinds of pair interaction potentials, denoted $U_{PP}$ for the protein-protein interactions and $U_{PC}$ for protein-chromatin interactions. These interaction potentials account for the mean influence of solvent particles. The interaction between two particles $m$ and $n$ depends on their species $P$ or $C$ (hereafter labelled $\alpha$ and $\beta$) and on their relative distance $r_{mn}=|\rr_m-\rr_n|$. 
The evolution equations of the protein positions obey Brownian dynamics (i.e., overdamped Langevin) \cite{Allen_2017}, which are integrated via the LAMMPS computational package \cite{Thompson_2022}:
\begin{equation}
    \frac{\dd \rr_n}{\dd t} = \sqrt{2D} \boldsymbol{\eta}_n(t) -\frac{D}{\kB T} \sum_{m\neq n}\nabla U_{\alpha\beta}  (r_{mn}), 
    \label{overdampedLangevin}
\end{equation}

where we assume that all the proteins have the same bare diffusion coefficient $D$, and where $\boldsymbol{\eta}_n(t)$ is a Gaussian white noise of zero mean and unit variance $\langle \eta_{n,i}(t)\eta_{m,j}(t') \rangle = \delta_{ij}\delta_{nm}\delta(t-t')$, with $\eta_{n,i}$ being the $i$-th component of $\boldsymbol{\eta}_n$. Since the dynamics is overdamped, the velocities of the proteins at a given time are irrelevant, and the state of the system is completely described by the positions of the proteins. 
 The particles interact with each other through a Lennard-Jones (LJ) potential, which is truncated at a distance $r_c=2.5\sigma_{\alpha\beta}$ ($\sigma_{\alpha\beta}$ being the distance at which $U_{\alpha\beta}=0$ which is equivalent to the particle diameter for $\alpha=\beta$), and shifted in order to ensure continuity of the potential at $r=r_c$. It reads $U_{\alpha\beta}(r)=[U^{\text{LJ}}_{\alpha\beta}(r)-U^{\text{LJ}}_{\alpha\beta}(r_c)]\theta(r_c-r)$, where $U^{\text{LJ}}_{\alpha\beta}(r)=4\varepsilon_{\alpha\beta}\left[\left(\frac{\sigma_{\alpha\beta}}{r}\right)^{12} -   \left(\frac{\sigma}{r}\right)^6 \right]$ is the standard LJ potential and $\theta(r)$ denotes the Heaviside function ($U_{PP}$ and $U_{PC}$ shown in Figure \ref{fig:chromatin_geometries_LJ}). 
 
 The energy parameters of the interaction potentials are $\varepsilon_{PP}$ and $\varepsilon_{PC}$. Throughout the paper, the distances will be measured in units of the protein diameter $\sigma = \sigma_{PP}$,  the energies in units of $\kB T$ and time in units of $\sigma^2/D$.

\subsection{Biological length scales}

Since thermal fluctuations of chromatin are ignored in our model, the bead size does not need to be directly related to the persistence length of chromatin. In line with minimal models of protein phase separation\cite{Brackley_2016, Zhang_mechanical_2021, Tortora_2023}, we represent one chromatin bead at the same scale as a typical DNA-binding protein, which naturally leads us to choose the nucleosome as the basic unit of chromatin fiber.  The nucleosome is the fundamental building block of chromatin. It consists of $200 \pm 40$ DNA base pairs wrapped around a histone octamer~\cite{Mcghee_1980, Olins_2003}. Its diameter is approximately $11$ nm~\cite{Mcghee_1980}, which we set as $\sigma_{PC} = 1.5 \, \sigma_{PP}$. 
In human interphase cells, the genome spans $\sim 3 \times 10^9$ base pairs, corresponding to roughly $1.5 \times 10^7$ nucleosomes per nucleus. In our systems, the number of chromatin beads is referred to as the number of nucleosomes $N_{\text{nucle}}$, and is varied to mimic different conditions.

We do not aim to infer the spatial distribution of nucleosomes from real imaging data of a living cell, nor to choose the potentials between nucleosomes and nuclear proteins as those of a specific DNA-binding protein. Instead, our goal is to systematically vary:  
(1) chromatin geometry, represented by nucleosome positions, using either biologically realistic concentrations or lower concentrations for isolated fibers as detailed hereafter; and  
(2) protein–chromatin interactions. 

\subsection{Coarse-grained chromatin}
\subsubsection{Linear fiber}

There is a large variety of chromatin geometries within the length scales under study here (nucleosomes to condensates, $10$~nm to $1$~$\mu$m). In order to understand the possible interactions of chromatin with condensate-forming proteins, we choose to concentrate on generic variations of the surface seen by condensate proteins near chromatin. We use both models with single fibers in an elongated simulation box, and models with several fibers in a larger cubic simulation box (Table \ref{tab:System_parameters}). The reference geometry is an ideal case where proteins only feel chromatin as a single straight alignment of nucleosomes, each placed at a distance $d_{CC}=1.67 \sigma \approx$ \qty{8.35}{\nm} of each other (straight fiber in Figure \ref{fig:chromatin_geometries_GEOM}). In this geometry, the condensate only experiences interactions with one fiber and its periodic images. A first variation to this ideal case consists of adding the typical bending of the chromatin fiber, through a zig-zag arrangement~\cite{Widom_1992, Schalch_2005} (shown in Figure \ref{fig:chromatin_geometries_ZZ}). This adds some rugosity to the fiber at a small length scale compared to that of a typical simulated condensate with a radius of gyration on the order of $80\sigma$ (in other words, in our simulations, a droplet wetting the fiber includes several periods of the zig-zag pattern). This zig-zag geometry allows systematic variation of the total number of nucleosomes $N_{\text{nucle}}$ while keeping an effective linear geometry. 

\subsubsection{Fibers with loops and networks}

We then then progressively increase the geometrical complexity by adding two related elements: (1) intersections, defined as regions of space where two nucleosomes, which are not neighbors in the chain, come close to each other, and (2) the presence of several fibers in the simulation box. There can be intersections within the same fiber if there is a loop~\cite{Lieberman-aiden_2009, Fudenberg_2016}, or between different fibers. Intersections generically mimic chromatin contacts of various origins, which can be detected by Hi-C experiments~\cite{Lieberman-aiden_2009}.
Among models with intersections, we first consider the case of a single fiber with loops of variable number and size. The loops are constructed as a circular arrangement of nucleosomes, with the same nucleosome-nucleosome distance $d_{CC}$ as the rest of the fiber, as shown in Figure \ref{fig:chromatin_geometries_LOOP}. The size of the loops are defined by the number of nucleosomes composing them and vary from 4 to 21 nucleosomes. This geometry allows us to study how the presence of intersections can stabilize droplets and localize the droplet material at specific locations  along a 1D fiber. 
We then consider systems with intersections in 3D, belonging to a network of intercrossing fibers in larger simulation boxes ($V=(200\sigma)^3$, around 54 times the box volume used for a single fiber, see details in Table \ref{tab:System_parameters}). We design systems with multiple fibers arranged either (i) on a regular cubic grid, where intersections contain three orthogonal fibers (regular network in Figure \ref{fig:chromatin_geometries_GEOM}), or (ii) in a randomized configuration, where the same number of fibers per direction is preserved (the fibers are still parallel to one of the box axes), but their relative positions are randomized, removing systematic intersections (Irregular network in Figure \ref{fig:chromatin_geometries_GEOM}). In the randomized case, the fiber positions are either all uniformly distributed across the box, or a subset of the fibers ($30\%$ or $50\%$) is confined to a sub-volume ($V_{\text{sub}} = 0.25^3 V_{\text{box}}$), thereby creating a density contrast, while the remaining fibers are still uniformly distributed.  

This comparison allows us to disentangle the effects of regularity versus disorder in fiber organization on condensate formation and stability, as well as the role played by network inhomogeneity. Precise values of the system input parameters are given in Table \ref{tab:System_parameters}.

\begin{table*}[t]
    \centering
    \begin{tabular}{l|lllll}
 &Straight fiber &Zigzag fiber&Fiber with loops& Dilute network &Dense network\\\hline
 Box volume in $\sigma^3$& $4\times (33.44)^3 $&$4\times (33.44)^3 $&$4\times (33.44)^3 $&  $(200.4)^3  $&$(200.4)^3  $\\
 Box volume in \unit{\cubic \um}& \num{1.87e-2}&  \num{1.87e-2}&  \num{1.87e-2}&  \num{1}&\num{1}\\
 Nbr. of proteins& $7560$& $7560$& $7560$&  $48289$ &$48289$\\
 Nbr. of proteins per $\sigma^3$& $0.050$& $0.050$& $0,050$&  0.006 &0.006\\
 Nbr. of proteins per \unit{\cubic \um} & 40.43& 40.43& 40.43&  $48289.00$&$48289.00$\\
Nbr. of fibers &$1$ &$1$ &$1$ & $3\times4^2=48$ &$3\times 8^2=192$\\
Total nbr. of nucleosome&$80$&\numrange{92}{160}&\numrange{83}{160}& $5760$&$23040$\\
 Nucleosome per $\sigma^3$& \num{5.3e-4} & \numrange{6.1e-4}{1.1e-3}& \numrange{5.5e-4}{1.1e-3}&  \num{7.2e-4} &\num{2.9e-3} \\
 Nucleosome per \unit{\cubic \um} & \num{4.28e3}&\numrange{4.92e3}{8.56e3}& \numrange{4.44e3}{8.56e3}& \num{5.76e3}&\num{2.30e4}\\
 Nbr. intersections& 0&0& 1 to 8& 64  &512 \\
 Spacing between fibers& - &-& -& $50\sigma \approx$ \qty{250}{\nm}&$25\sigma \approx$ \qty{125}{\nm}\\\end{tabular}
    \caption{Input parameters for the systems in all geometries. The simulation is set up with periodic boundary conditions to mimick larger volumes. Note that the spacing between fibers is  exact only in the case of a regular network, else it is an average. The values given in \unit{\um} are estimated using $\sigma \approx$ \qty{5}{\nm}. For the larger systems, the nucleosome concentration corresponds to the average number of nucleosomes per \unit{\cubic\um} in a human nucleus of \qty{1}{\um} of radius and \qty{5}{\um} respectively, taking into account the exact length of the genome and the average distance between nucleosomes in base pairs. }
    \label{tab:System_parameters}
\end{table*}

\section{Phase separation in the presence of a linear fiber}

In this section, we restrict ourselves to linear fibers, either straight or with a weak zig-zag modulation, such that no structural features along the fiber axis can act as a preferred interaction site. The protein–protein interaction is fixed at $\varepsilon_{PP}=1.2$, a value that leads to the formation of a single spherical droplet (of roughly 2500 proteins) in bulk conditions without chromatin.

\begin{figure}
    \centering
    \begin{subfigure}{0.6\linewidth}
        \subcaption{}
        \includegraphics[width=\linewidth]{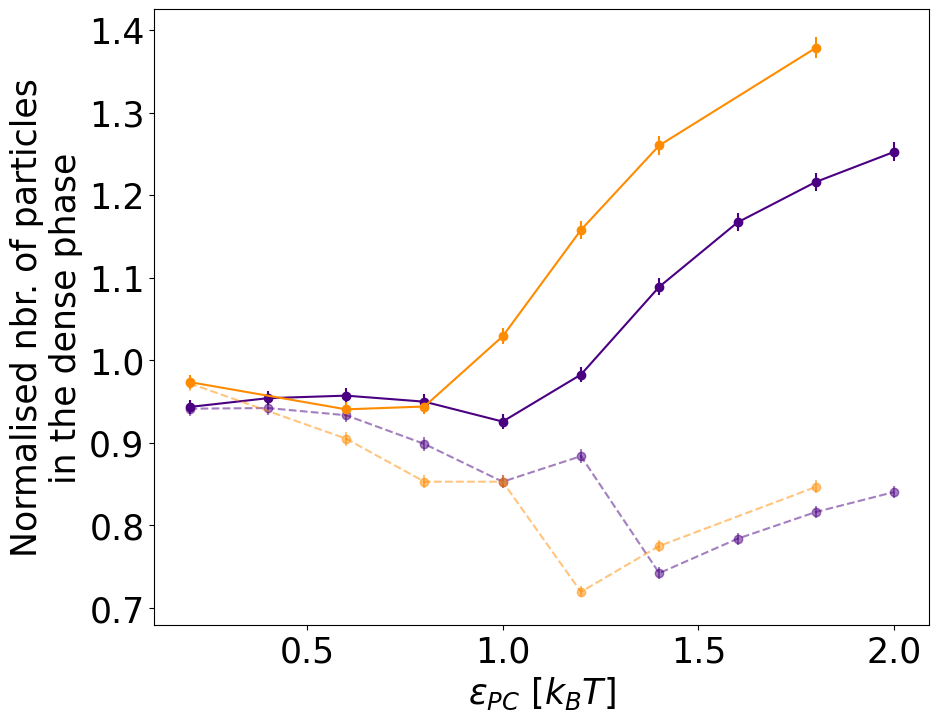}
        \label{fig:Phase_shift_singlefiber_NBR}
    \end{subfigure}

    \begin{subfigure}{0.6\linewidth}
        \subcaption{}
        \includegraphics[width=\linewidth]{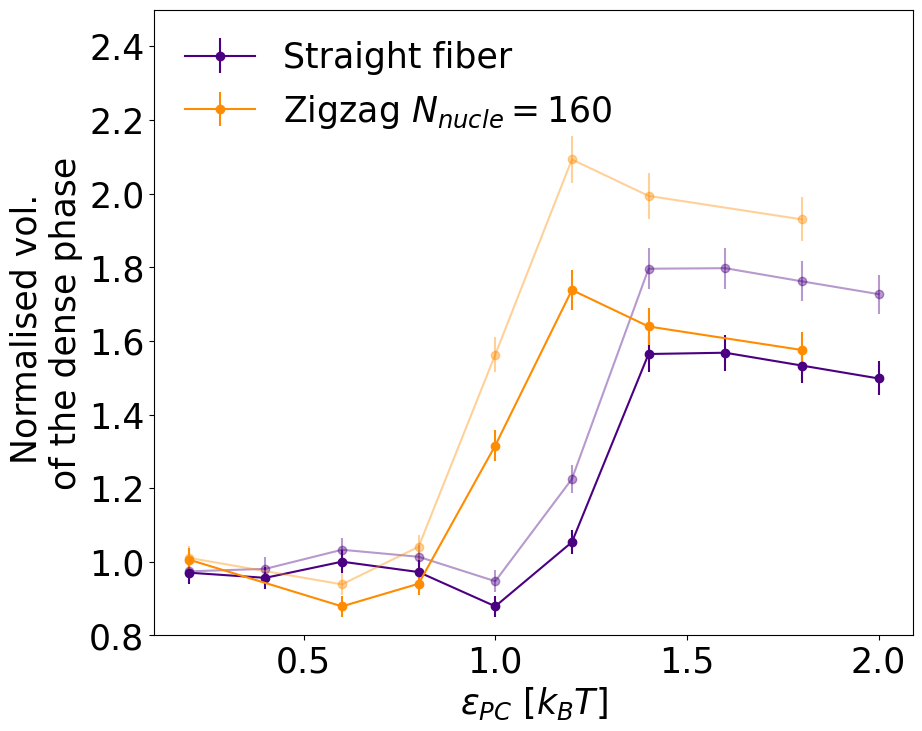}
        \label{fig:Phase_shift_singlefiber_VOL}
    \end{subfigure} 

    \begin{subfigure}{0.6\linewidth}
        \subcaption{}
        \includegraphics[width=\linewidth]{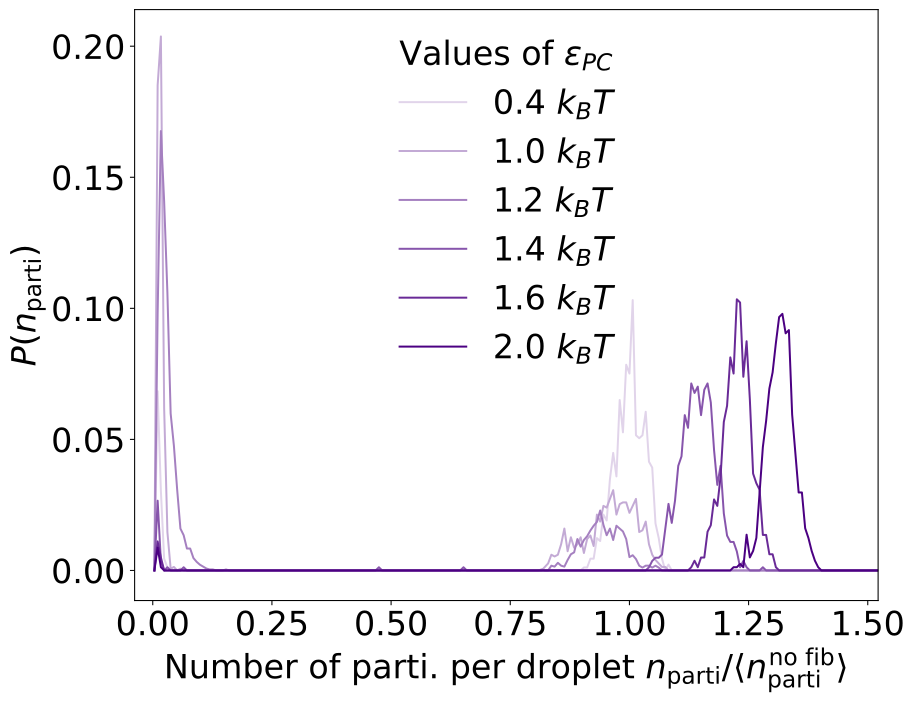}
        \label{fig:Phase_shift_singlefiber_DISTRIB}
    \end{subfigure}
    
    \caption{\subref{fig:Phase_shift_singlefiber_NBR}. Number of particles in the dense phase for various protein-chromatin attraction strength ($\varepsilon_{PC}$) and chromatin linear densities. The counting method is described in Appendix~\ref{sec:methods_track}. This number is normalized by the number of particles in the dense phase in the absence of fiber. The light dashed line corresponds to the number of particles in the dense phase that are not in contact with the fiber. \subref{fig:Phase_shift_singlefiber_VOL}. Volume of the dense phase as a function of $\varepsilon_{PC}$, normalized by the volume of the dense phase in the absence of fiber. The lighter lines correspond to the droplet volume $V$. The darker lines represent the droplet volume excluding the volume occupied by the fiber, i.e., $V-V_{\text{fib}}^{\text{drop}}$. Calculation details of $V$ and $V_{\text{fib}}^{\text{drop}}$ are given Eq.~\eqref{equ:volume_drop} and Eq.~\eqref{equ:volume_fiber_in_drop} respectively, in Appendix~\ref{sec:methods_vol}.  \subref{fig:Phase_shift_singlefiber_DISTRIB}. Distribution of the number of particles per droplet $n_{\text{parti}}$ for the straight fiber system ($N_{\text{nucle}}=80$), where $n_{\text{parti}}$ is normalized by the average number of particles in the droplet in the absence of fibers, $n_{\text{parti}}^{\text{no fib}}$.}
    \label{fig:Phase_shift_singlefiber}
\end{figure}

Figures \ref{fig:Phase_shift_singlefiber_NBR} and \ref{fig:Phase_shift_singlefiber_VOL} show the protein content and volume of the denser phase as a function of chromatin–protein attraction strength $\varepsilon_{PC}$, for straight fibers. As $\varepsilon_{PC}$ grows, the dense phase expands, with a marked change for $\varepsilon_{PC}>1.0$.
As Brownian Dynamics simulations include thermal fluctuations, the fluctuations of the size of dense phase domains can be quantified.
The distributions are shown in Figure \ref{fig:Phase_shift_singlefiber_DISTRIB}. In all cases, the distribution is divided into two peaks, one representing a large stable domain, and the other peak which is typical of small unstable clusters. As expected, the change of regime around $\varepsilon_{PC} = 1.0$  is reflected in the growth of the largest domains. Interestingly, the regime change also results in strong variations in the distribution of small clusters, with more clusters as $\varepsilon_{PC}$ grows to $1.2$, in contrast with a sharp decrease of their probability for larger values of $\varepsilon_{PC}$. 
The evolution of the dense phase volume is similar for zig-zag fibers, but the transition occurs at lower values of $\varepsilon_{PC}$ and with a more pronounced effect (shown in Figure \ref{fig:Phase_shift_singlefiber_NBR} and \ref{fig:Phase_shift_singlefiber_VOL} in light orange). This difference reflects the higher linear density of nucleosomes in the zig-zag configuration, which enhances the effective chromatin–protein attraction.


These results indicate that protein-chromatin interactions shift the local phase boundary and stabilize liquid-like configurations near the surface. Such a substrate-induced modification of the phase behavior is reminiscent of wetting transitions in macroscopic systems, where the onset of complete wetting is described in terms of spreading parameters and interfacial free energies~\cite{Cahn_1977,DeGennes_1985}. At the microscopic level, the wetting regimes emerge from the balance of pairwise interaction parameters. However, establishing a direct connection with macroscopic theories is not appropriate in our case. Classical wetting descriptions rely on thermodynamic interfaces, continuum density profiles, and diverging film thicknesses~\cite{DeGennes_2004}, whereas our simulations operate in a finite-size molecular regime. In such small systems, film thickness is inherently restricted to only a few particle layers, and the usual distinction between partial and complete wetting therefore loses its physical meaning~\cite{Binder_2003}. 

While a quantitative connection between our system and classical wetting descriptions is not relevant, qualitative links exist. In addition to the volume of the liquid phase, wetting affects the shape and position of the liquid around the wettable surface. We therefore quantify how increasing $\varepsilon_{PC}$ alters droplet morphology (sphericity, size distribution) and positioning relative to the chromatin substrate,  for six values of $\varepsilon_{PP}$ ($1.0$, $1.1$, $1.2$, $1.3$, $1.4$, $1.5k_BT$) across a range of ten $\varepsilon_{PC}$ values (from $0.2$ to $2.0k_BT$ in steps of $0.2k_BT$).

\begin{figure}[h!]
    \centering
    \begin{subfigure}{\linewidth}
        \subcaption{}
        \includegraphics[width=\linewidth]{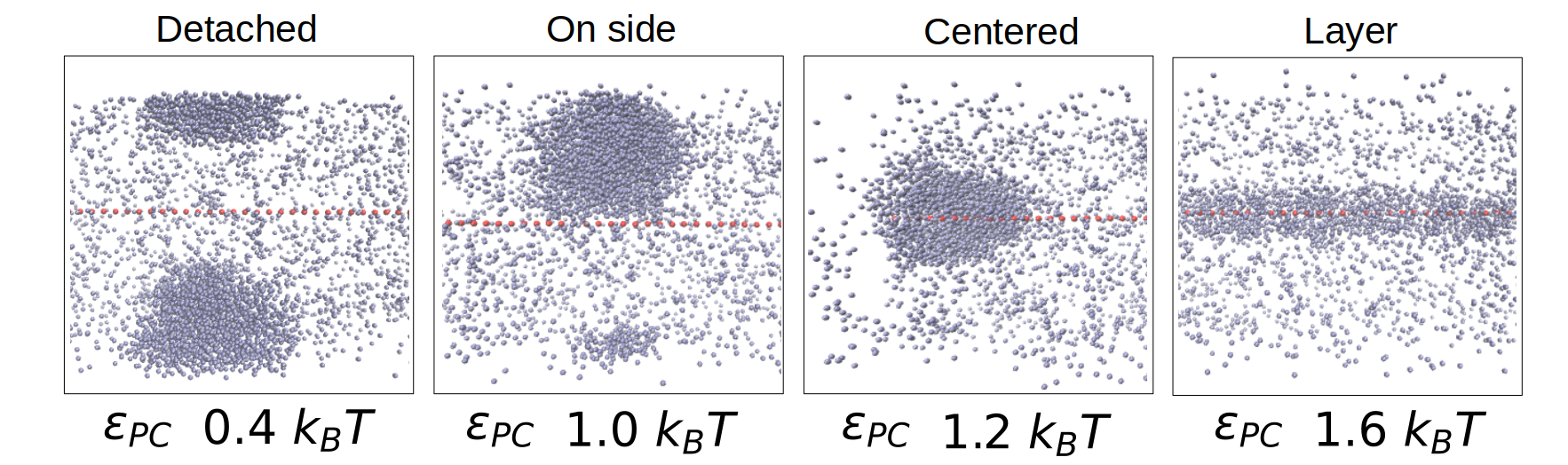}
        \label{fig:linear_localize_SNAP}
    \end{subfigure}
    
    \begin{subfigure}{0.85\linewidth}
        \subcaption{}
        \includegraphics[width=\linewidth]{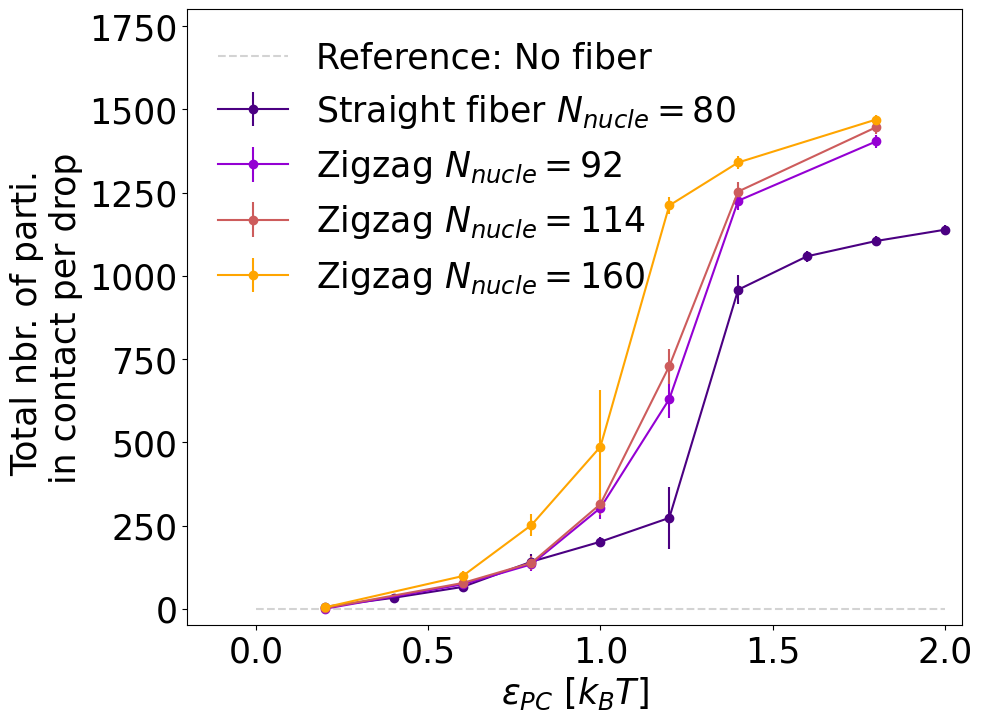}
        \label{fig:linear_localize_NCONT}
    \end{subfigure} 
    
    \begin{subfigure}{0.8\linewidth}
        \subcaption{}
     \includegraphics[width=\linewidth]{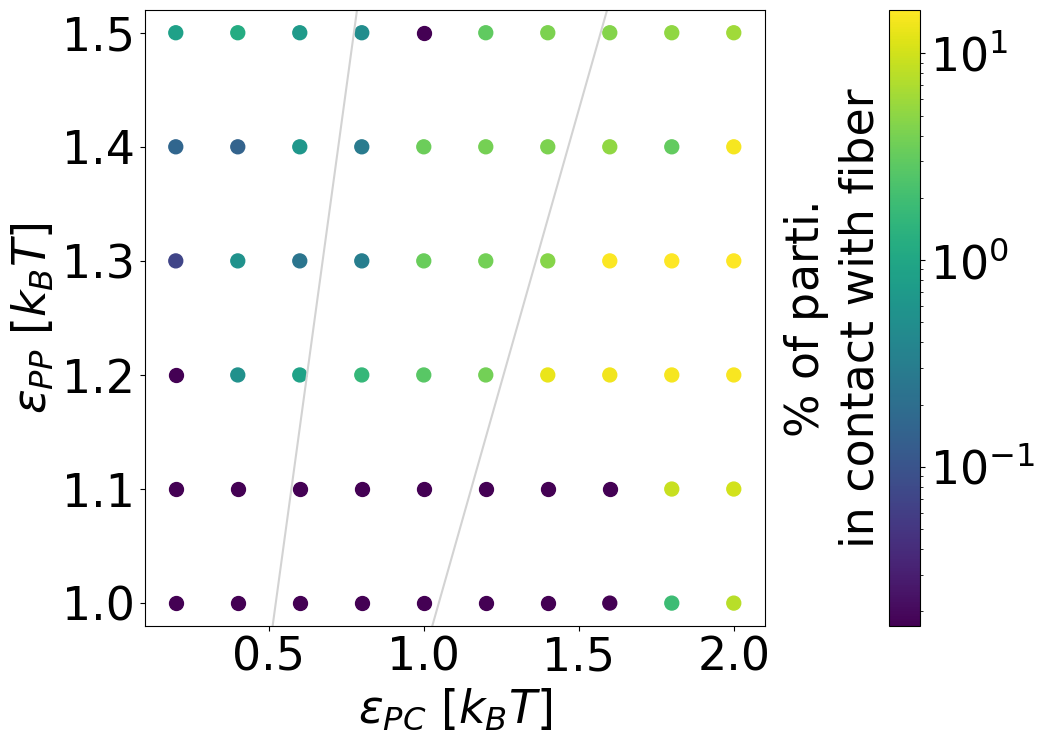}
        \label{fig:linear_localize_C}
     \end{subfigure}
    
     \caption{\subref{fig:linear_localize_SNAP}. Snapshots of typical system configurations with a linear fiber. \subref{fig:linear_localize_NCONT}. Number of particles in contact with the fiber as a function of the protein-chromatin attraction strength $\varepsilon_{PC}$. \subref{fig:linear_localize_C}. Percentage of particles in the droplet being in contact with the fiber as a function of $\varepsilon_{PC}$ for the system with a straight fiber. The grey lines correspond to the transition from detached to side and side to centered found with the analytical model with $n_{\text{parti}}=2600$. This number of particles corresponds to the droplet size in the absence of fiber with $\varepsilon_{PP}=1.2k_BT$ (details in Appendix~\ref{Apx:Analytical_model}). }
    \label{fig:linear_localize}
\end{figure}

In the straight fiber geometry, for a protein-protein interaction strength $\varepsilon_{PP}=1.2$, we observe the following regimes (representative configurations shown in Figure \ref{fig:linear_localize_SNAP}):
\begin{itemize}
    \item For $\varepsilon_{PC}<1.0$, a droplet forms away from the fiber.
    \item For $\varepsilon_{PC}\approx 1.0$, the droplet partially wets the fiber, with the contact area increasing with $\varepsilon_{PC}$.
    \item At sufficiently high $\varepsilon_{PC}$, 
    a protein-rich layer coats the fiber along its entire length.
\end{itemize}
We will hereafter refer to the change of regime as a wetting transition. 

This transition is clearly visible in the number of proteins in contact with the fiber, which exhibits a sharp increase at the wetting transition (Figure \ref{fig:linear_localize_NCONT}). Notably, in the zig-zag geometry, the transition occurs at lower values of $\varepsilon_{PC}$ due to the increased nucleosome density. 

The values of $\varepsilon_{PC}$ at which the change of regime happens increase with $\varepsilon_{PP}$, shown in Figure \ref{fig:linear_localize_C}. 
This behavior can again be interpreted as a mesoscale analogue of macroscopic wetting phenomena: in macroscopic systems, spreading parameters are governed by differences in surface tensions, whereas mesoscopic wetting is expected to depend on differences between interaction strengths.
As the dense phase partially overlaps the fiber, the number of protein–chromatin contacts can be estimated from simple geometrical considerations. Motivated by this observation, we developed a minimal analytical model describing the system energy as a function of the distance between the droplet center of mass and the fiber. The model includes only energetic contributions arising from protein–protein contacts within the droplet and protein–chromatin contacts at the interface
 (see Appendix~\ref{Apx:Analytical_model} for more details). 
 The model reveals energy minima corresponding to the observed geometries of our systems, from a detached droplet to a centered one when $\varepsilon_{PC}$ and $\varepsilon_{PP}$ vary. The grey lines in Figure~\ref{fig:linear_localize_C} show the transition limits found with the analytical model (see Appendix~\ref{Apx:Analytical_model}).  These transition limits qualitatively correlate with percentage of proteins in contact with the fiber in our simulations, unless $\varepsilon_{PP}$ gets lower than $1.2$, in which case the system is no longer in the phase-separating region.

Indeed, for values of $\varepsilon_{PP}\leq1.1k_BT$ (see Figure~\ref{fig:linear_localize_C}), for which no phase separation is observed in the absence of fibers, only a very thin layer forms around the fiber for large values of $\varepsilon_{PC}$. Such surface condensation has been observed in the case of DNA \cite{Morin_2020, Quail_2020} and associated with the prewetting regime. In this regime, a thin layer form near the surface, in the single-phase region of the phase diagram where droplets would dissolve in bulk~\cite{Cahn_1977, DeGennes_1985}.

\section{Phase separation in more complex geometries}

We next investigate phase separation in systems where the spatial density of fiber elements varies along the fiber axis. In the context of chromatin, such heterogeneity arises from organization at multiple length scales. 
Here, a local increase in the effective nucleosome density mimics either higher chromatin compaction or chromatin-associated proteins forming network-like structures (such as PAR chains \cite{Alemasova_polyadp-ribose_2022}). 

We focus on two classes of systems: (1) Single fibers with loops and (2) Networks of fibers, where both loops and intersections offer a larger contact area than a straight or isolated fiber, hence behaving as preferred binding sites. When several loops or intersections are present, phase separation results from a competition between maximizing contacts with these favorable sites and minimizing the interfacial free energy, which favors the formation of a single large droplet. 
Geometric details are given in Section~\ref{sec:model}. These two types of fiber models allow us to systematically study differences in local concentration (by varying the loop size), spatial frequency (number of loops or intersections), and spatial regularity (unequal distances between neighbouring loops or fibers).

\subsection{Single fiber with loops}

We first examine this interplay using fibers containing several equivalent loops, i.e., loops of identical size and regularly spaced along the fiber.
Specifically, we study fibers with four loops of equal size (either 9 or 21 nucleosomes) while varying the protein-chromatin attraction strength $\varepsilon_{PC}$. We then fix the interaction parameters ($\varepsilon_{PP}=\varepsilon_{PC}=1.2$) and investigate how the morphology of the dense phase depends on the loop size, which is varied from 4 to 21 nucleosomes.

\begin{figure}[ht!]
    \centering
    \begin{subfigure}{0.47\linewidth}
        \subcaption{}
        \includegraphics[width=\linewidth]{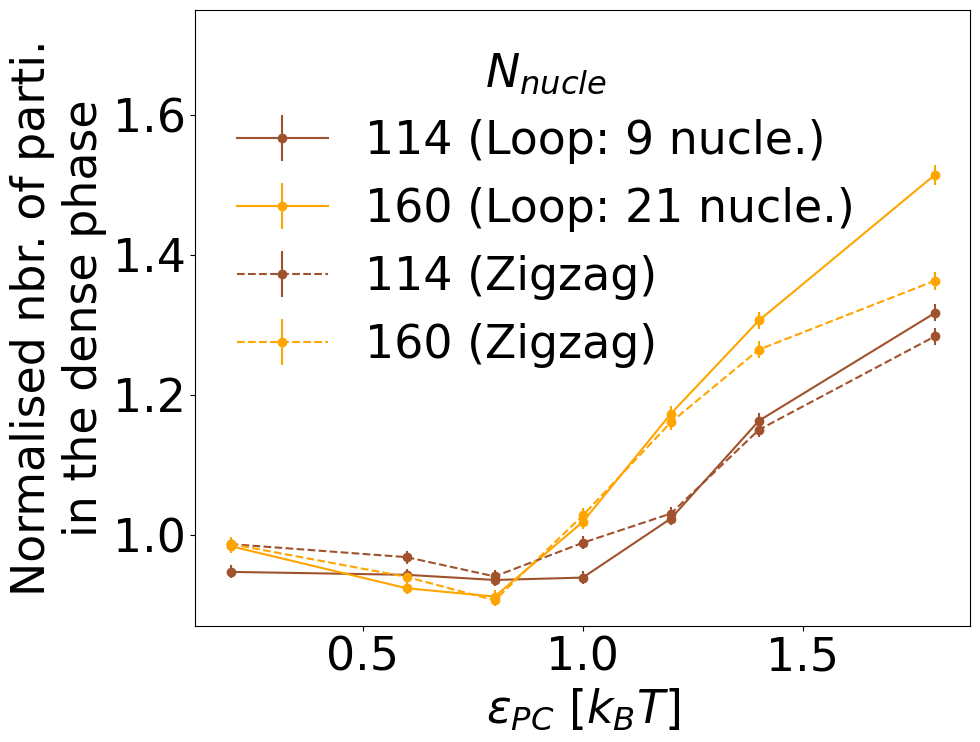}
        \label{fig:Phase_shift_loops_NBR}
    \end{subfigure} 
    \hspace{0.0\linewidth}
    \begin{subfigure}{0.45\linewidth}
        \subcaption{}
        \includegraphics[width=\linewidth]{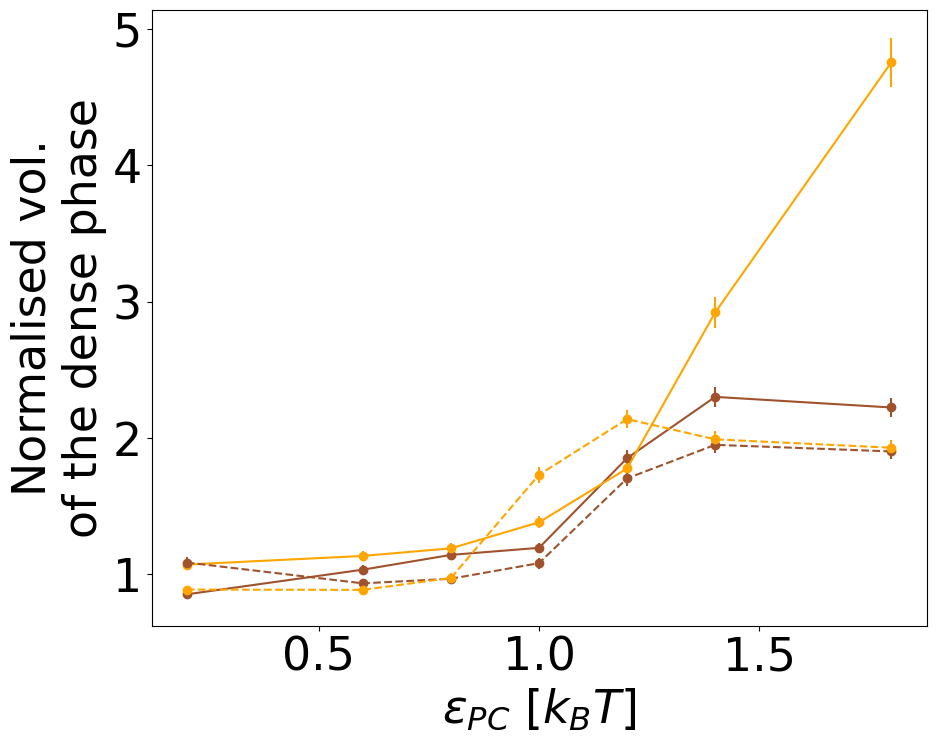}
        \label{fig:Phase_shift_loops_VOL}
    \end{subfigure} 
    \caption{ \subref{fig:Phase_shift_loops_NBR}. Number of particles in the dense phase and \subref{fig:Phase_shift_loops_VOL}. volume of the dense phase, for 4 different chromatin geometries. Both quantities are normalised by the reference values in the absence of fibers. The data are shown for systems containing fibers with four loops of 9 nucleosomes or 21 nucleosomes. These results are compared with zig-zag fibers (without loops) containing the same total number of nucleosomes.}
    \label{fig:Phase_shift_loops}
\end{figure}

\begin{figure}
    \centering
    \begin{subfigure}{\linewidth}
        \subcaption{}
        \includegraphics[width=\linewidth]{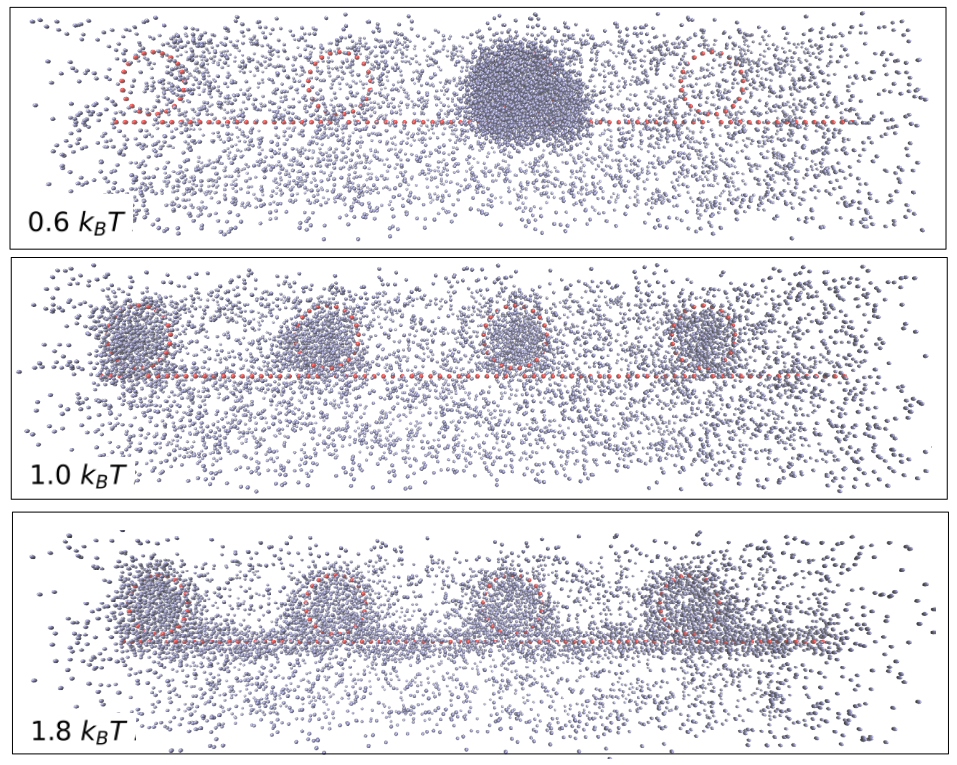}
        \label{fig:size_distribution_loops_SNAP}
    \end{subfigure} 

    \begin{subfigure}{0.45\linewidth}
        \subcaption{}
        \includegraphics[width=\linewidth]{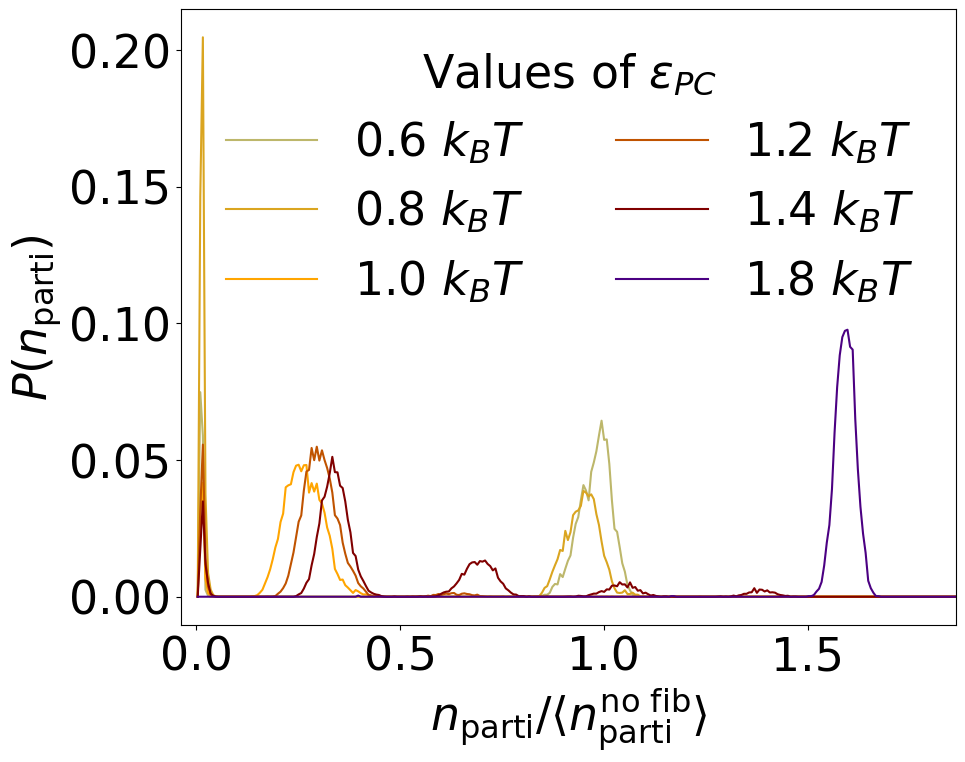}
        \label{fig:size_distribution_loops_EPS}
    \end{subfigure}
    \hspace{0.0\linewidth}
    \begin{subfigure}{0.45\linewidth}
        \subcaption{}
        \includegraphics[width=\linewidth]{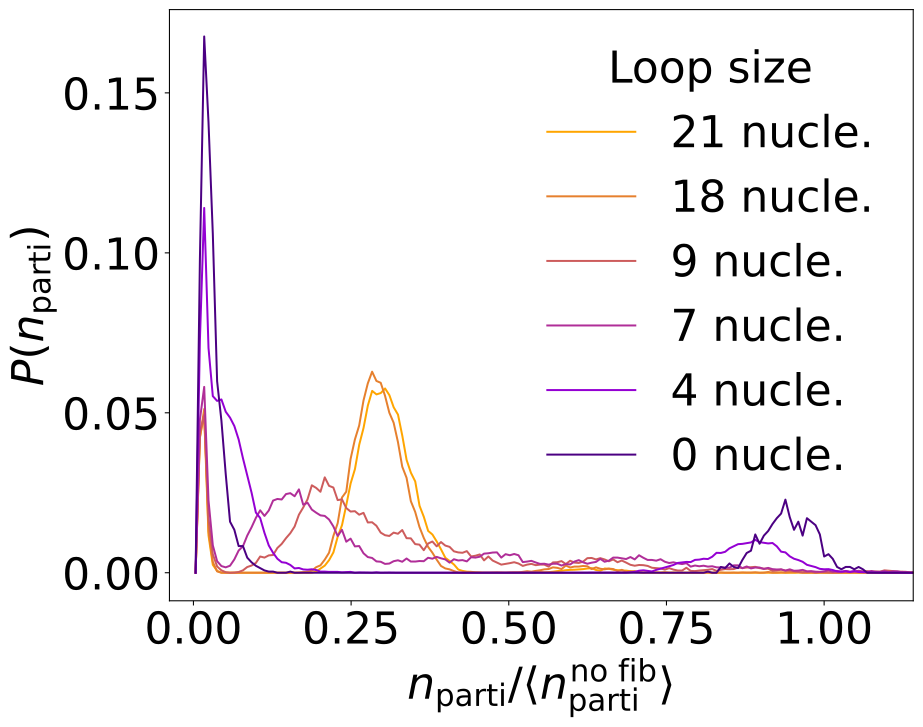}
    \label{fig:size_distribution_loops_SIZE}
    \end{subfigure}

    \begin{subfigure}{0.6\linewidth}
        \subcaption{}
        \includegraphics[width=\linewidth]{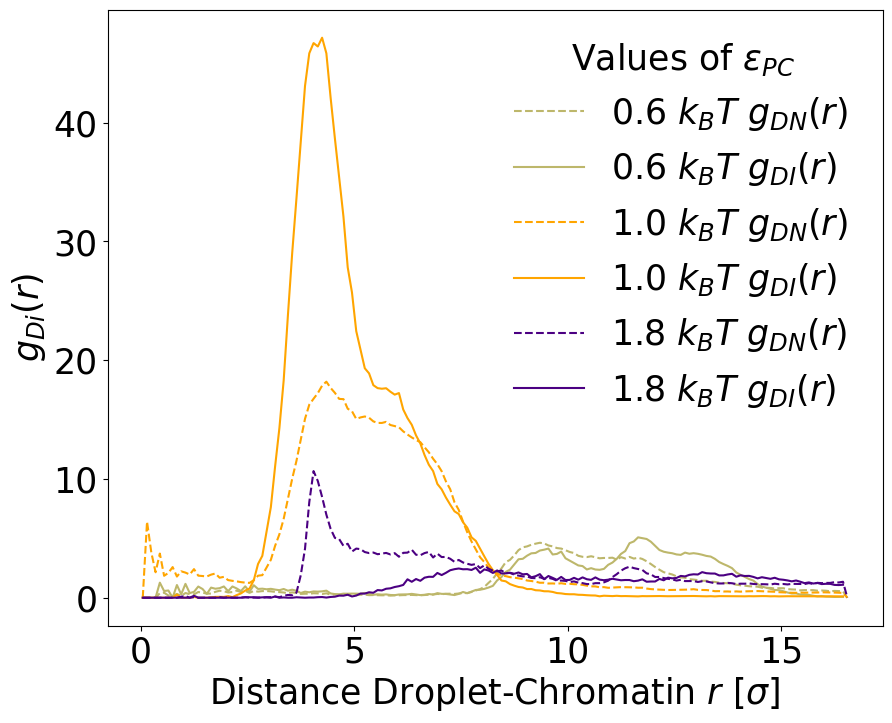}
        \label{fig:localisation_loops_GR_reg}
    \end{subfigure}
    
    \caption{Impact of the attraction strength $\varepsilon_{PC}$, loop size, and loop number on the morphology of the dense phase. 
    \subref{fig:size_distribution_loops_SNAP}. Representative snapshots for fibers with loops for  $\varepsilon_{PC}= 0.6; 0.8$ or $1.8 k_BT$. 
    \subref{fig:size_distribution_loops_EPS}. Effect of $\varepsilon_{PC}$ on the droplet size distribution for a fiber with four loops of 21 nucleosomes. 
    \subref{fig:size_distribution_loops_SIZE}. Effect of loop size on the droplet size distribution for fibers with four loops at $\varepsilon_{PP}=\varepsilon_{PC}=1.2k_BT$.
    \subref{fig:localisation_loops_GR_reg}. Quantification of droplet localization for a system with four regularly spaced loops of 21 nucleosomes. $g_{DN}(r)$ and $g_{DI}(r)$ are the pair correlation functions between the droplet centers of mass ($D$) and nucleosomes, considering either all nucleosomes ($g_{DN}(r)$) or only nucleosomes near fiber intersections ($g_{DI}(r)$). 
    }
    \label{fig:size_distribution_loops}
\end{figure}

Within the explored parameter range, the fiber geometry has little influence on the total number of particles in the dense phase or on its volume; instead, the total number of nucleosomes is the dominant control parameter. As shown in Fig.~\ref{fig:Phase_shift_loops}, systems with linear fibers and fibers with loops exhibit a similar dependence on $\varepsilon_{PC}$ for the same number of nucleosomes $N_{\text{nucle}}$. A notable deviation is observed for fibers with four large loops (21 nucleosomes) at high $\varepsilon_{PC}$. In this regime, the dense phase fully coats the fiber surface (Fig.~\ref{fig:size_distribution_loops_SNAP}), leading to non-ellipsoidal droplet shapes and a consequent overestimation of the dense-phase volume.

Representative configurations are shown in Fig.~\ref{fig:size_distribution_loops_SNAP}. As for straight fibers, the system undergoes a wetting transition upon increasing $\varepsilon_{PC}$: droplets move far from the fiber at low $\varepsilon_{PC}$ and are localized on the fiber at stronger $\varepsilon_{PC}$. For $\varepsilon_{PC} \gg \varepsilon_{PP}$, strong wetting leads to the formation of a continuous dense phase coating the whole fiber. In contrast to the linear case, droplets are also localized along the direction of the fiber axis, preferentially forming around the loops. In addition, stronger protein-chromatin attractions and larger loop sizes both reduce droplet size fluctuations, as reflected by the narrowing of the droplet size distributions (Fig.~\ref{fig:size_distribution_loops_EPS}).

Compared to linear fibers, the key qualitative difference introduced by loops is the emergence of preferred nucleation sites, which can stabilize the coexistence of several droplets. This contrasts with bulk systems or straight fibers (for intermediate $\varepsilon_{PC}$), where Ostwald ripening and coalescence typically lead to a single dominant droplet for the system sizes considered here. This behavior is apparent both in the simulation snapshots and in the droplet size distributions, where droplet sizes fluctuate around a fraction of the size of a whole dense phase (Fig.~\ref{fig:size_distribution_loops_EPS}).

For fibers containing multiple loops, three distinct regimes can be identified:
\begin{itemize}
    \item \textit{Single-droplet regime}: a single loop nucleates and stabilizes a droplet, while the other loops remain uncoated.
    \item \textit{Multi-droplet regime}: droplets are stabilized at several or all loops, leading to droplet coexistence.
    \item \textit{Coating regime}: a single droplet spreads over and coats the entire fiber surface.
\end{itemize}

The stability of loop-associated droplets depends sensitively on both loop size and loop number (Fig.~\ref{fig:size_distribution_loops_SIZE}). Larger loops provide an increased contact area, enhancing droplet stabilization, whereas smaller loops typically support only a small droplet. Increasing the number of loops introduces an additional competition: although each loop can act as a potential nucleation site, the finite amount of dense-phase material limits the number of stable droplets.

Droplets form with a higher probability at loop regions than along other parts of the fiber. This effect is quantified using pair correlation functions $g(r)$ between the droplet center of mass ($D$) and nucleosomes. We compute these correlations both with respect to all nucleosomes, $g_{DN}(r)$, and by restricting the analysis to nucleosomes belonging to loop regions, $g_{DI}(r)$ (see Section~\ref{sec:methods_gr}). The results are shown in Figure~\ref{fig:localisation_loops_GR_reg} for systems with regularly spaced loops. For parameter values at which several droplets are stabilized (illustrated here for $\varepsilon_{PC}=1.0$), droplets are significantly more likely to be located near loop regions than near randomly selected nucleosomes. By contrast, this preferential localization is not observed for detached droplets or for droplets that coat the entire fiber.

When loop sites are no longer equivalent, droplet localization can extend to the scale of the entire system, thereby mimicking a biological situation in which a condensate forms at a specific chromatin location. To illustrate this effect, we consider fibers containing four loops, in which the two central loops are displaced, breaking the regular spacing while keeping the total fiber length and the positions of the outer loops fixed. In this context, localization refers to the preferential formation of droplets at specific positions along the fiber (here, between the two central loops), rather than to the formation of a single droplet at an arbitrary position within the simulation box. These specific positions are defined as those of the nucleosomes with the highest local nucleosome density (see Section~\ref{sec:methods_gr} for details). Hereafter, we refer to this region as the specific site. This localization effect is quantified using the pair correlation function between the droplet center of mass and the positions of nucleosomes within the specific site, $g_{DS}(r)$.

\begin{figure}
    \centering    
    \begin{subfigure}{0.9\linewidth}
        \subcaption{}
        \includegraphics[width=\linewidth]{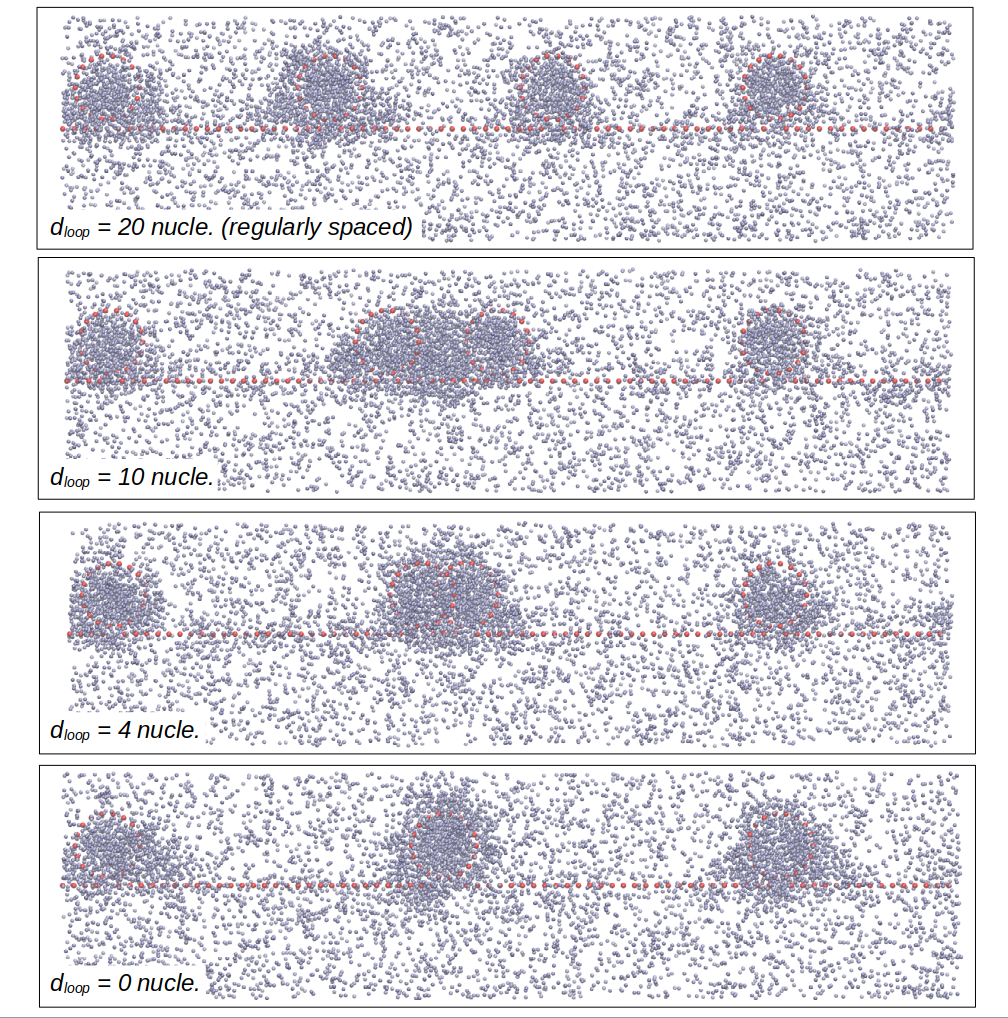}
        \label{fig:localisation_loops_SNAP}
    \end{subfigure}
    
    \begin{subfigure}{0.45\linewidth}
        \subcaption{}
        \includegraphics[width=\linewidth]{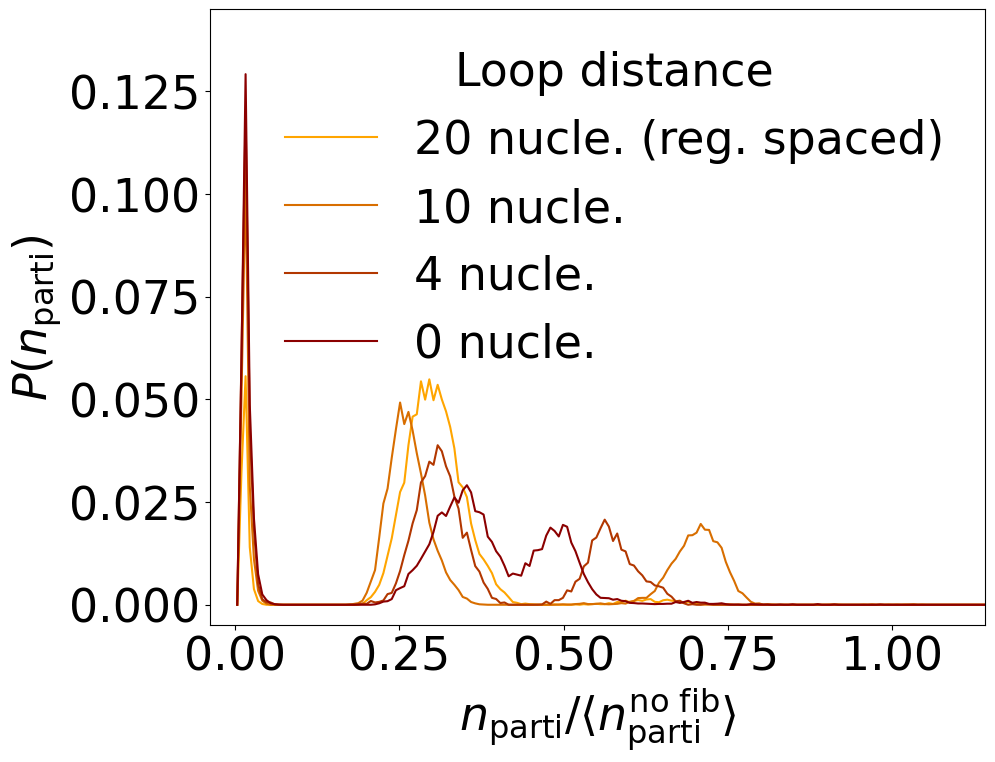}
        \label{fig:localisation_loops_DISTRIB}
    \end{subfigure}
    \hspace{0.0\linewidth}
    \begin{subfigure}{0.45\linewidth}
        \subcaption{}
        \includegraphics[width=\linewidth]{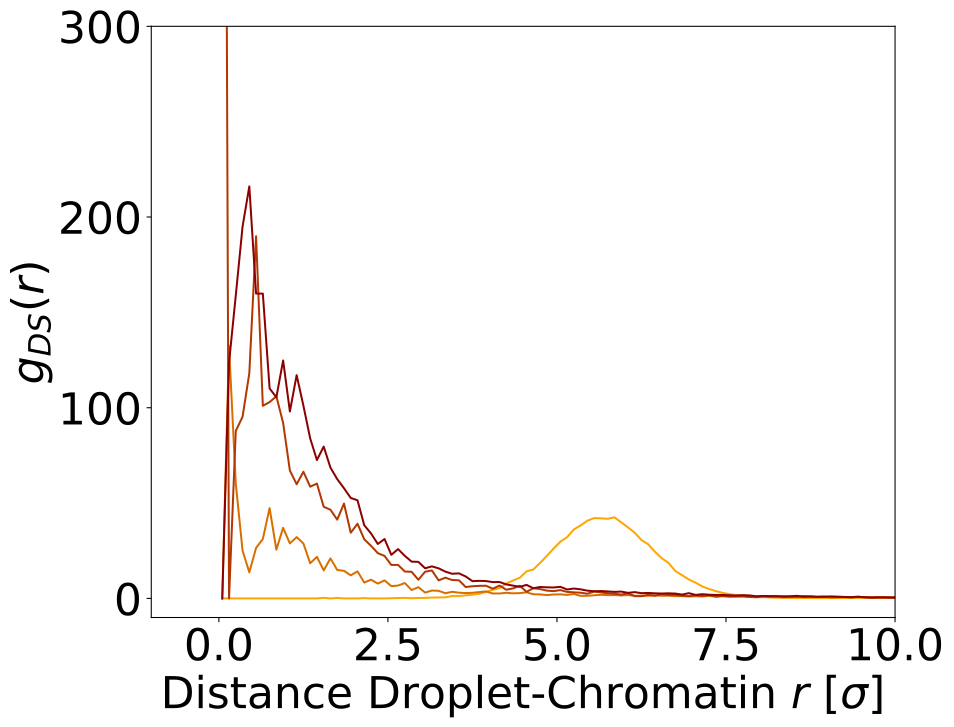}
        \label{fig:localisation_loops_GR_uneven}
    \end{subfigure}

    \caption{\subref{fig:localisation_loops_SNAP}. Representative snapshots of the system configurations, for systems with 4 loops of size 21 nucleosomes with irregular spacing. Selected distance between the two central loops $d_{\text{loop}}=[20$ (regular case) $10;4; 0]$ nucleosomes.  
    \subref{fig:localisation_loops_DISTRIB}. Droplet size distribution:  number of particles per droplet normalised by the number of particles in a single droplet in the absence of fibers. 
    \subref{fig:localisation_loops_GR_uneven}. Pair correlation function $g_{DS}(r)$ between the center of mass of the droplets and the nucleosomes from the specific site.  This site is fixed and located at the center of mass of the densest chromatin region. 
    }
    \label{fig:localisation_loops}
\end{figure}

We observe that for irregularly spaced loops, most of the matter gathers around the two central loops, while small droplets form around the outer loops. This is shown for a system with loops of size 21 nucleosomes in Figure \ref{fig:localisation_loops_SNAP}. The presence of asymmetry leads to droplet size polydispersity (Figure \ref{fig:localisation_loops_DISTRIB}), and a localization of the droplets around the specific site (Figure \ref{fig:localisation_loops_GR_uneven}). Interestingly, when the central loops get closer to each other, there is a maximum in the size of the central droplet at intermediate distances between loops (this can be seen in the snapshots, and in the peak on the size distribution corresponding to the largest droplet). This simple example shows how the geometry of chromatin domains attracting condensate proteins may finely control the extent of the condensate. In next section, we shall get further insight into such control mechanism with multi-fiber models.

\subsection{Fiber networks}

\begin{figure}
    \centering
    \begin{subfigure}{0.45\linewidth}
        \subcaption{$\varepsilon_{PC}=3.0$}
        \includegraphics[width=\linewidth]{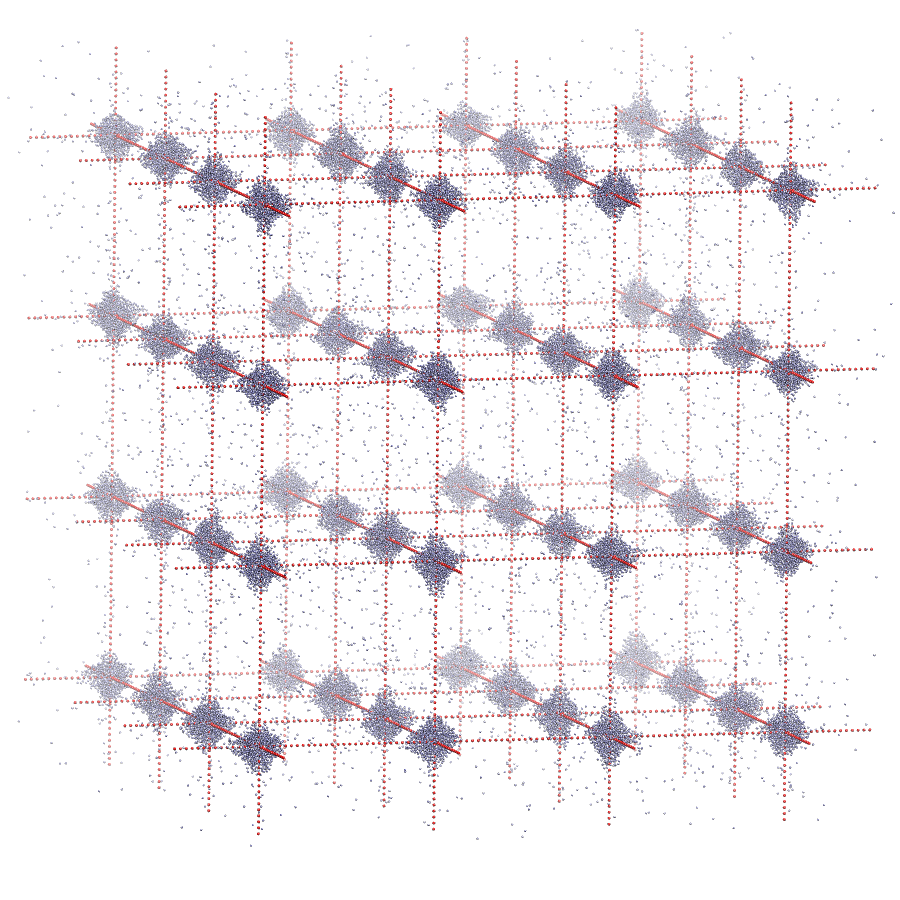}
        \label{fig:Network_Number_droplet_A}
    \end{subfigure}
    \hspace{0.0\linewidth}
    \begin{subfigure}{0.45\linewidth}
        \subcaption{$\varepsilon_{PC}=3.0$}
        \includegraphics[width=\linewidth]{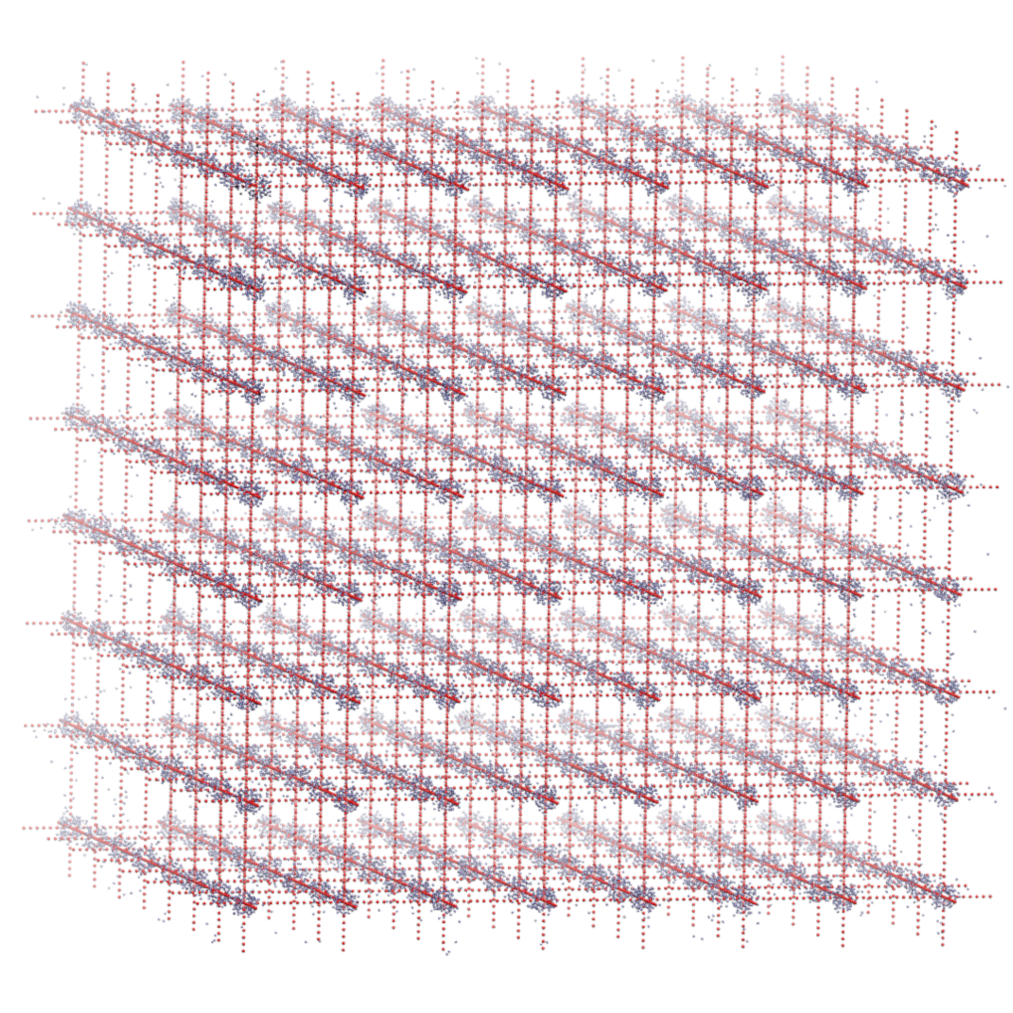}
        \label{fig:Network_Number_droplet_B}
    \end{subfigure}
    
    \begin{subfigure}{0.45\linewidth}
        \subcaption{$\varepsilon_{PC}=2.0$}
        \includegraphics[width=\linewidth]{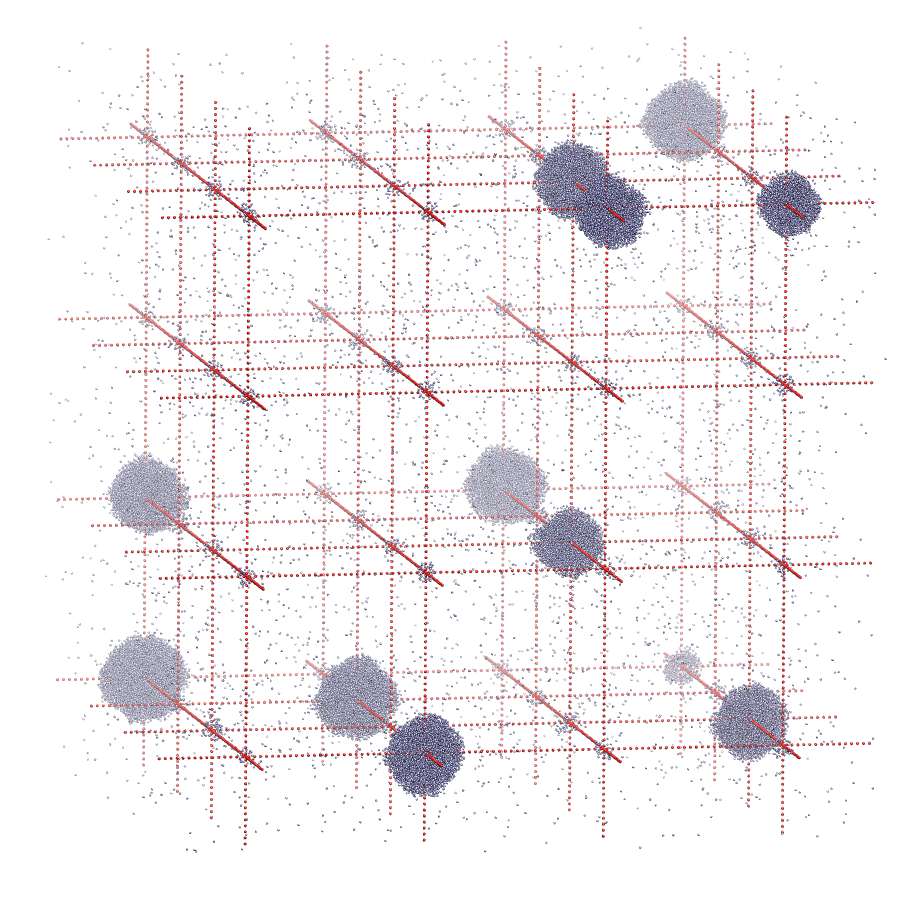} 
        \label{fig:Network_Number_droplet_C}
    \end{subfigure}
    \hspace{0.0\linewidth}
    \begin{subfigure}{0.45\linewidth}
        \subcaption{$\varepsilon_{PC}=2.0$}
        \includegraphics[width=\linewidth]{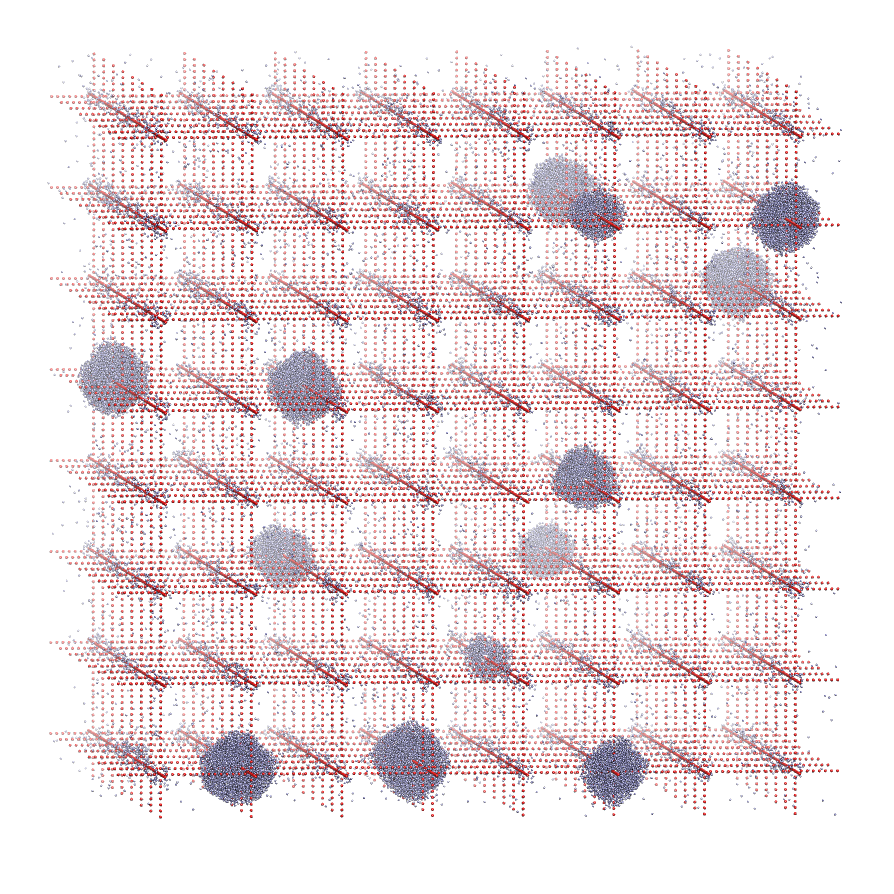} 
        \label{fig:Network_Number_droplet_D}
    \end{subfigure}
    
    \begin{subfigure}{0.6\linewidth}
        \subcaption{}
        \includegraphics[width=\linewidth]{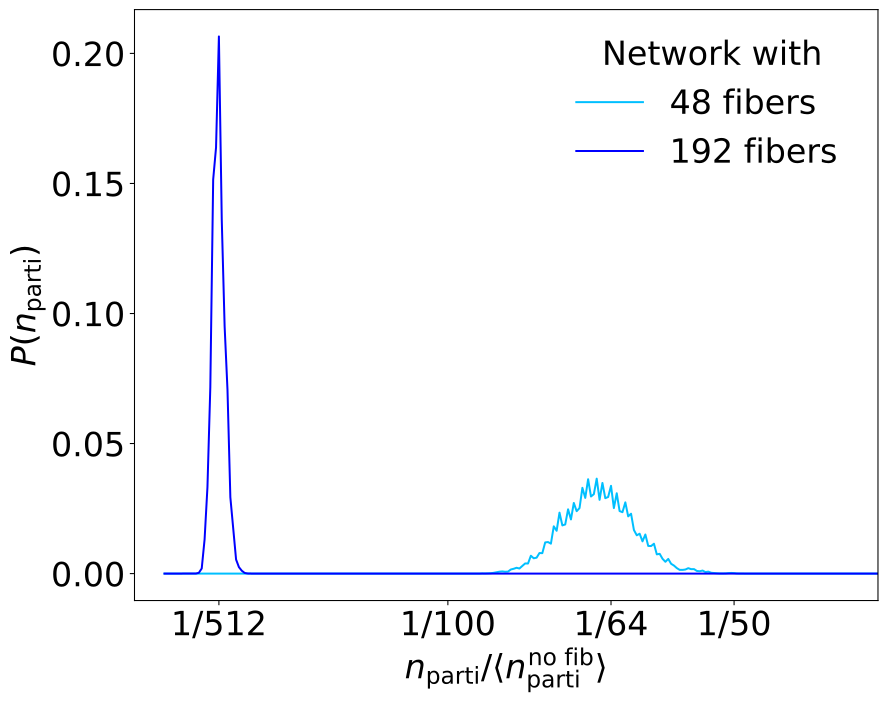}
        \label{fig:Network_Number_droplet_DISTRIB}
    \end{subfigure}
    
    
    \caption{ \subref{fig:Network_Number_droplet_A}-\subref{fig:Network_Number_droplet_D}. Representative snapshots of simulations after \num{15e7} time steps, for a regular network of 48 fibers with 64 intersections (\subref{fig:Network_Number_droplet_A}, \subref{fig:Network_Number_droplet_C}), or a regular network of 192 fibers with 512 intersections (\subref{fig:Network_Number_droplet_B}, \subref{fig:Network_Number_droplet_D}). In all cases, the protein-protein interaction parameter is $\varepsilon_{PP}=2.0$.  \subref{fig:Network_Number_droplet_DISTRIB}. Droplet size distribution for regular networks with $\varepsilon_{PP}=2.0$ and $\varepsilon_{PC}=3.0$. The droplet size is  the number of proteins per droplet. It is normalised by the total number of proteins in the dense phase in the absence of fibers $n^{\text{no fib}}_{\text{parti}}$. 
    }
    \label{fig:Network_Number_droplet}
\end{figure}

We then focus on networks of fibers, where intersections mimick regions where chromatin elements come close together in trans configurations. 
In regular networks, analogous to fibers with regularly spaced loops, droplets preferentially form near intersections (Figs. \ref{fig:Network_Number_droplet_A}-\ref{fig:Network_Number_droplet_D}), and show smaller size fluctuations for stronger protein-chromatin attraction (Fig. \ref{fig:Network_Number_droplet_DISTRIB}).
As with loops, the systems evolve towards two types of geometries for the dense phase: (i) one droplet per intersection (numerous small droplets, e.g., Fig.~\ref{fig:Network_Number_droplet_A} and \ref{fig:Network_Number_droplet_B}) and (ii) fewer, larger droplets (e.g., Fig.~\ref{fig:Network_Number_droplet_C} and \ref{fig:Network_Number_droplet_D}). This crossover is controlled by the balance between the energy gain from favourable protein-chromatin contacts at intersections and the energy cost of creating additional interfaces.

The first case (i), where droplets are stabilized at each intersection, is observed for sufficiently large values of $\varepsilon_{PC}$. For instance, at $\varepsilon_{PP}=2.0$ and $\varepsilon_{PC}=3.0$, all 64 intersections of a sparse network can host droplets, shown in Figure \ref{fig:Network_Number_droplet_A}, where the number of proteins per droplet fluctuates around $1/64$ times the number of proteins in the dense phase in the absence of fibers (Figure \ref{fig:Network_Number_droplet_DISTRIB}). In contrast, for a weaker protein-chromatin attraction $\varepsilon_{PC}\leq2.0$, only a fraction of the intersections are coated by proteins (case ii), resulting in a larger number of proteins per droplet. 
Please note that in system with networks, the time to reach steady state increases significantly, since some droplet may be locally trapped in some metastable positions on the network. In some of the system described below, there may still be some of these metastable droplets in the analysis. 
Nevertheless, in any case the simulation time is much larger than that of all previous systems, which are clearly at steady-state. 

Both the volume and the protein content of the dense-phase are poorly impacted by the number of fibers in the system and $\varepsilon_{PC}$ (Figure \ref{fig:Network_Number_droplet_DISTRIB}). Therefore, in the case where droplets are stabilized at all intersections, the number of preferred binding sites provided by the network determines the droplet size.

\begin{figure}
    \centering
    \begin{subfigure}{0.45\linewidth}
        \subcaption{}
        \includegraphics[width=\linewidth]{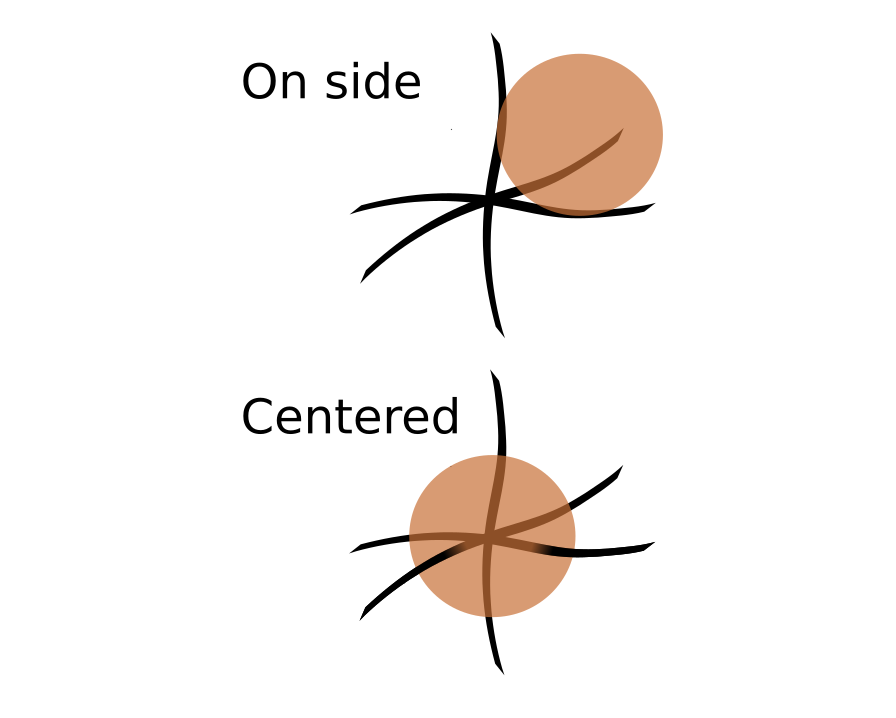}
        \label{fig:Number_size_droplet_network_SKETCH}
    \end{subfigure}
    \hspace{0.0\linewidth}
    \begin{subfigure}{0.45\linewidth}
        \subcaption{}
        \includegraphics[width=\linewidth]{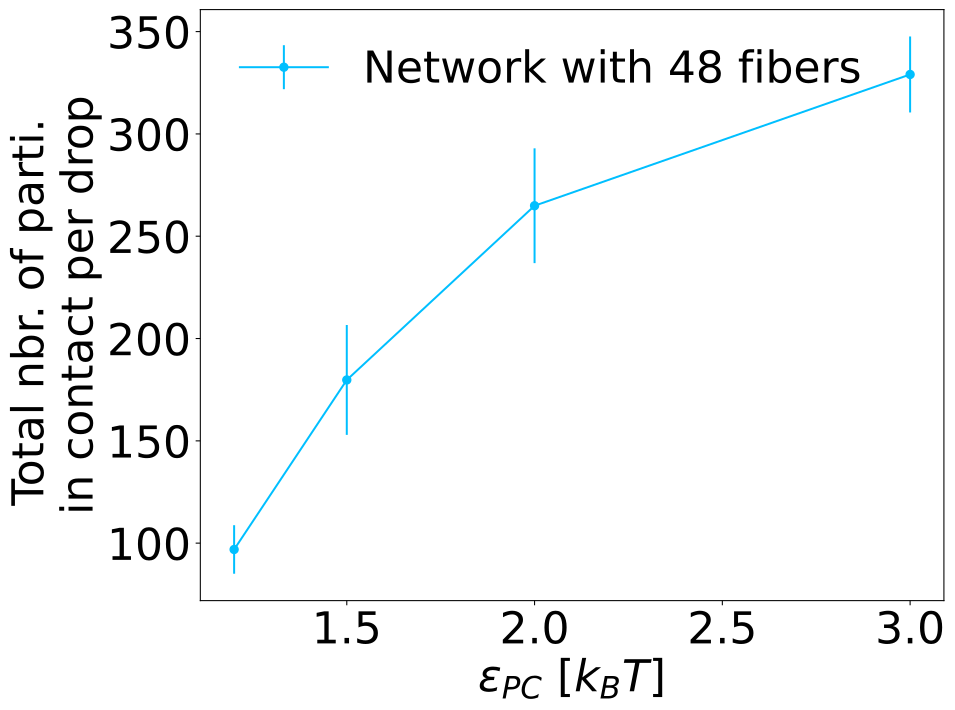}
        \label{fig:Number_size_droplet_network_NCONT}
    \end{subfigure}
    
    \begin{subfigure}{0.6\linewidth}
        \subcaption{}
        \includegraphics[width=\linewidth]{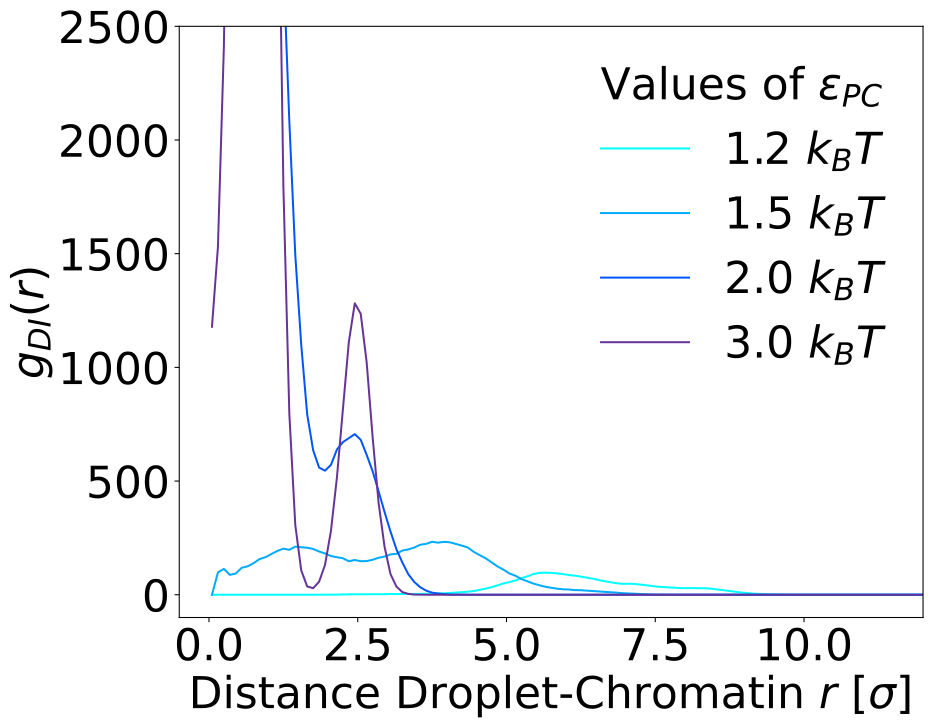}
        \label{fig:Number_size_droplet_network_GR4by4}
    \end{subfigure}

    \caption{\subref{fig:Number_size_droplet_network_SKETCH}. Illustrative sketch of the transition between different wetting states of the droplets. \subref{fig:Number_size_droplet_network_NCONT}. Number of particles in contact with fibers as a function of $\varepsilon_{PC}$. \subref{fig:Number_size_droplet_network_GR4by4}. $g_{DI}(r)$ pair correlation function between the droplet center of mass ($D$) and the network intersections (I). The results are given for systems with a regular network of 48 fibers, in which one droplet per intersection is observed in the simulations.}
    \label{fig:Number_size_droplet_network}
\end{figure}

Similarly to systems containing a single fiber, we observe distinct regimes analogous to a macroscopic wetting transition, as illustrated in Figure~\ref{fig:Number_size_droplet_network_SKETCH}. Upon varying the interaction parameters $\varepsilon_{PC}$ or $\varepsilon_{PP}$, droplets either form at the periphery of the intersections or become centered on them. This transition is characterized by an increase in the total number of proteins in contact with the fiber (Figure~\ref{fig:Number_size_droplet_network_NCONT}). It is further quantified using the pair correlation function $g_{DI}(r)$ between the droplet centers of mass and the intersections. For $\varepsilon_{PC} \geq 2.0,k_BT$, $g_{DI}(r)$ is maximized at $r=0$ (Figure~\ref{fig:Number_size_droplet_network_GR4by4}), corresponding to droplets centered on the fibers. In contrast, for weaker protein–chromatin interactions, $\varepsilon_{PC} < 2.0,k_BT$, the first peak of $g_{DI}(r)$ occurs at $r>0$. 

For regular networks, as all the intersections are equivalent, the dense phase is not expected to be localized at a specific location in the system. On the other hand, in the case of disordered networks, and more specifically when there is a region of higher chromatin density, the dense phase can be spatially localized.
In order to study the localization of the dense phase, we investigate phase separation in the presence of irregular networks, with and without a specific site where a bias is introduced to favor a high fiber concentration in a specific area of the simulation box (see details in Section~\ref{sec:model}).

\begin{figure}[!t]
    \centering
    \begin{subfigure}{0.45\linewidth}
        \subcaption{Low density contrast}
        \includegraphics[width=\linewidth]{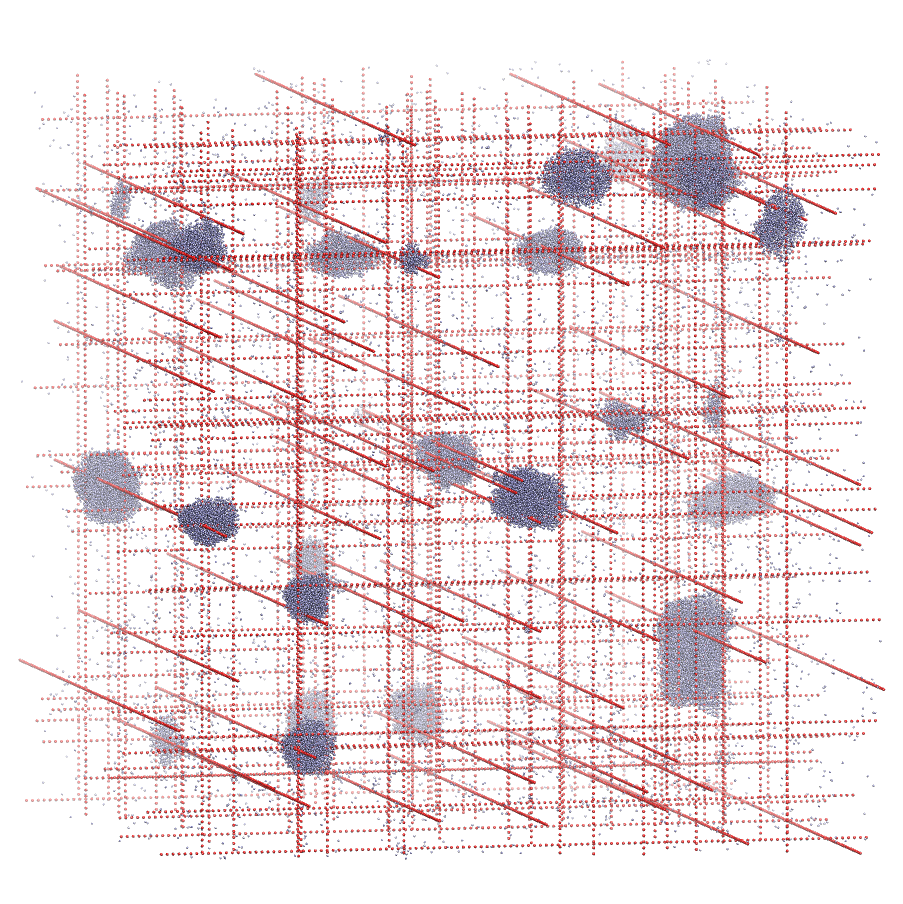}
        \label{fig:disordered_netwroks_SNAP_RAND}
    \end{subfigure}
    \hspace{0.0\linewidth}
    \begin{subfigure}{0.45\linewidth}
        \subcaption{High density contrast}
        \includegraphics[width=\linewidth]{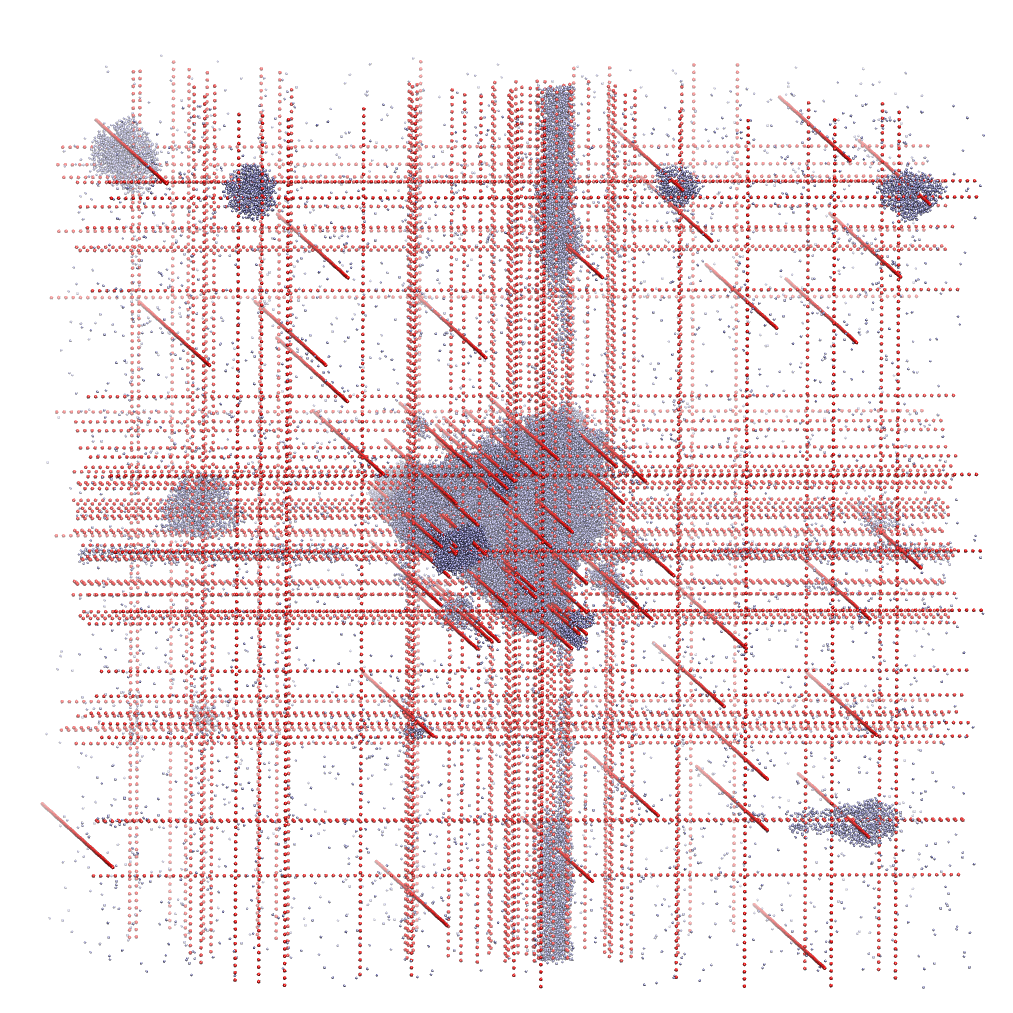}
        \label{fig:disordered_netwroks_SNAP_LOC}
    \end{subfigure}

    \begin{subfigure}{0.44\linewidth}
        \subcaption{}
        \includegraphics[width=\linewidth]{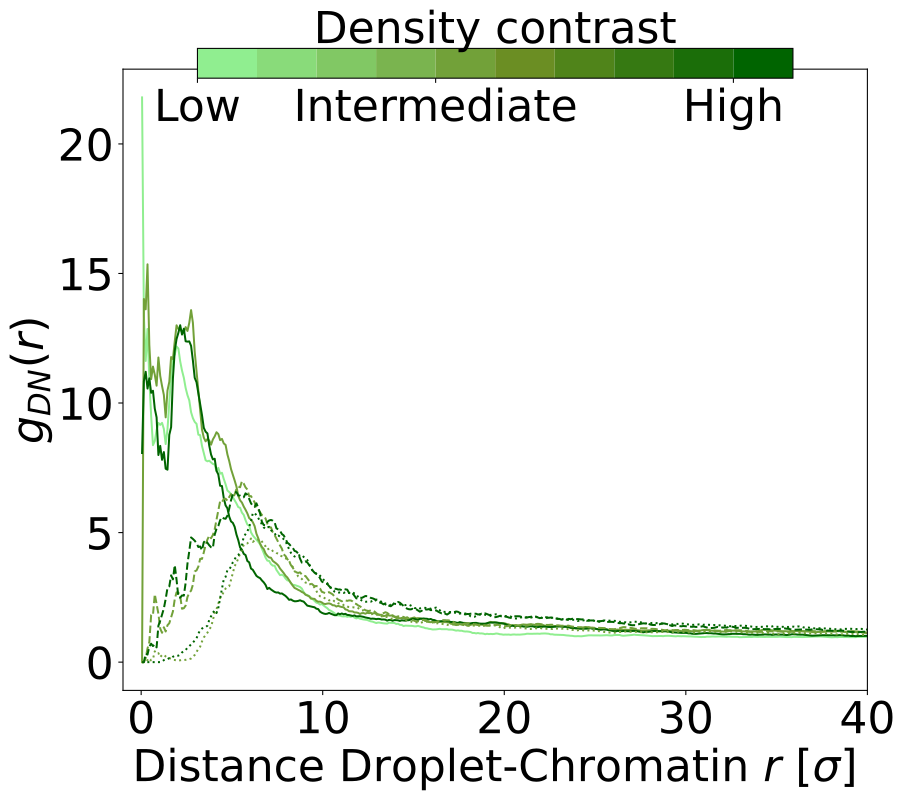}
        \label{fig:disordered_netwroks_GR_DN_all}
    \end{subfigure}
    \hspace{0.0\linewidth}
    
    \begin{subfigure}{0.50\linewidth}
        \subcaption{}
        \includegraphics[width=\linewidth]{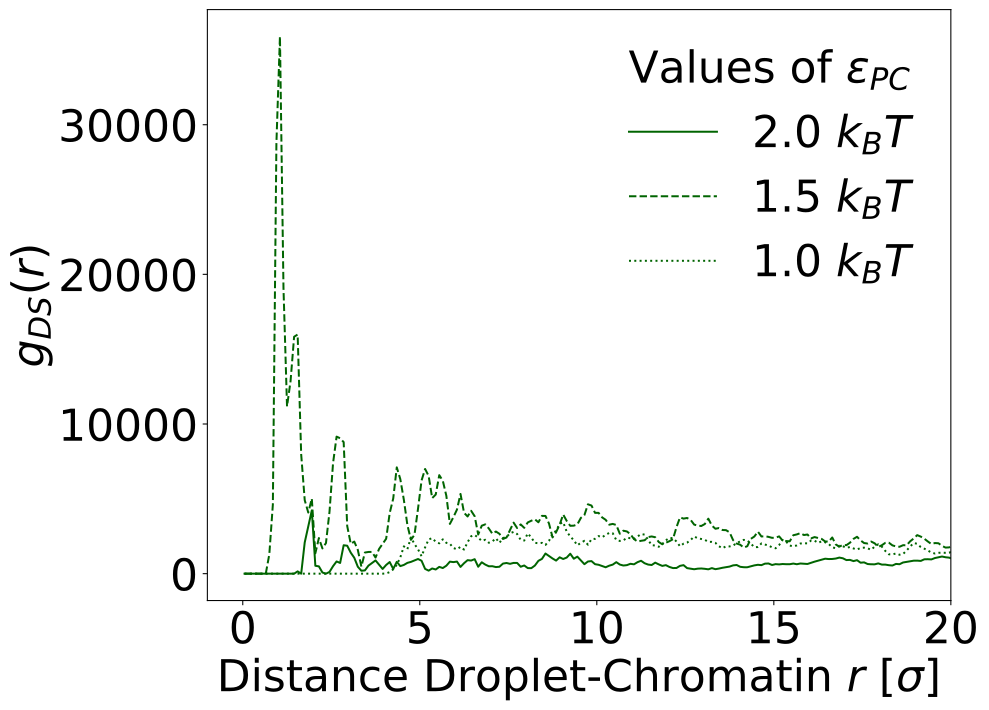}
        \label{fig:disordered_netwroks_GR_denser}
    \end{subfigure}
    
    \caption{\subref{fig:disordered_netwroks_SNAP_RAND}-\subref{fig:disordered_netwroks_SNAP_LOC}. Representative snapshots of disordered networks for $\varepsilon_{PC}=2.0$. In the case of a low density contrast, the fiber positions are randomly chosen in two directions from a  uniform probability distribution (\subref{fig:disordered_netwroks_SNAP_RAND}). In the case of a high density contrast,  the probability distribution is biased so that $50\%$ of the fibers are located in the central sub-volume (\subref{fig:disordered_netwroks_SNAP_LOC}). 
    \subref{fig:disordered_netwroks_GR_DN_all}. $g_{DN}(r)$ for irregular networks of 192 fibers. Three chromatin density contrasts are considered. In the case of low contrast, all the fibers are uniformly placed in the box. In the intermediate case, 30\% are constrained in the sub volume. In the High contrast case, 50\% of the fibers are concentrated in the sub volume. The $3$ cases correspond to light green, medium green and dark green, respectively. \subref{fig:disordered_netwroks_GR_denser}. $g_{DS}(r)$ for irregular networks of 192 fibers with $50\%$ of the fibers concentrated in the sub volume. 
    }
    \label{fig:disordered_netwroks}
\end{figure}

The asymmetry of the network strongly influences the size distribution and number of droplets in the system. 
In the studied parameter range, the number of droplets at steady-state is considerably reduced in comparison to regular networks with identical parameters (Figure \ref{fig:disordered_netwroks_SNAP_RAND}, \ref{fig:disordered_netwroks_SNAP_LOC}). 
Please note that, in these networks, the time required to reach steady state increases significantly, as some droplets may remain temporarily trapped in metastable positions within the network. Consequently, a fraction of the systems analyzed below may still contain such metastable droplets. Nevertheless, the simulation times considered here are substantially longer than those used for all previously studied systems, which were unambiguously found to be at steady state. 
The localisation of droplets is studied through the pair correlation function $g_{DN}(r)$ between droplet centers of mass and nucleosomes, shown in Figure \ref{fig:disordered_netwroks_GR_DN_all}. 
Droplets are more localized with respect to chromatin (shorter typical distance between droplet center of mass and chromatin) for stronger protein-chromatin attraction $\varepsilon_{PC}$, while keeping $\varepsilon_{PP}=2.0$ constant. 
Visualizing the system shows configurations with a large droplet coating the fibers around the specific site (chromatin dense region). A typical configuration is shown in Figure \ref{fig:disordered_netwroks_SNAP_LOC} for $\varepsilon_{PP}=\varepsilon_{PC}=2.0$. Note that in the range of parameters studied, density contrast does not seem to strongly affect the typical droplet-nucleosome distance (Figure \ref{fig:disordered_netwroks_GR_DN_all}). 

To confirm that the positions of droplets are correlated with the location of the specific site, and not only with respect to generic nucleosomes, we studied the pair correlation function $g_{DS}(r)$ between the center of mass of droplets and the nucleosomes within the specific site. This confirms the localization of droplets around the specific site for large values of $\varepsilon_{PC}$  (Figure \ref{fig:disordered_netwroks_GR_denser}), with the first peaks of $g_{DS}(r)$ located at $r<4$ for $\varepsilon_{PC}\geq 1.5 k_BT$.  Figure \ref{fig:disordered_netwroks_GR_denser} also shows a very low probability of finding a droplet at short distances around the specific site for $\varepsilon_{PC}=1.0$. This points towards the fact that the protein-chromatin interactions can play an important role in droplet localization in a fibrous environment. 

In the context of chromatin and protein condensates, there are only a few experimental studies characterizing at the same time the density of chromatin (e.g., of histone proteins), and the density of condensate-forming proteins. 
In a recent study, Mazzocca and coworkers~\cite{Mazzocca_2023} have shown that depending on protein-protein interactions within condensates and on protein-chromatin interactions, some condensates localize far from chromatin dense regions, while the opposite may also be found. In particular, when protein-protein interactions lead to solid-like properties of the condensates, proteins are more concentrated in less compact chromatin areas, which is consistent with the regime we observe when $\varepsilon_{PP}$ is larger than $\varepsilon_{PC}$.

\section{Conclusion}

The mechanisms governing condensate size selection and spatial localization in cells are numerous and strongly interdependent. In this work, we focused on the role of the surrounding environment in these processes, and in particular on how fibrous structures such as chromatin influence the organization of phase-separated condensates. These results corroborate and extend experimental observations on various condensates forming on fibers. 

Condensates and fibers can mutually affect one another through a coupling between wetting and phase separation. To isolate the contribution of fiber geometry and concentration to condensate properties, we employed minimal coarse-grained models combining phase-separating Lennard-Jones particles with fixed fibrous substrates. By keeping the fibers immobile, we decoupled geometric effects from scaffold rearrangements and focused on the equilibrium organization of the dense phase. This is partly motivated by the fact that chromatin dynamics is typically slower than that of diffusing proteins on the timescales relevant for condensate assembly. Nevertheless, our claim here is not to assume chromatin mobility is unimportant, but our works starts from the necessity of understanding the coupling of wetting and condensate formation for fixed fibers, before tackling the more complex case of a deformable substrate.  

We found that when protein-chromatin interactions are sufficiently strong for condensates to wet fibers, both local geometric features and large-scale network organization strongly influence droplet size, morphology, and number. In most of the parameter space explored, the presence of fibers only moderately affects the total volume and protein content of the dense phase, with more pronounced shifts observed at higher protein-chromatin affinity. These changes are associated with a wetting-like transition, in which droplets evolve from detached configurations at low protein-chromatin interaction strength to partially wetted states, and eventually to complete coating of the fibers. Similar wetting transitions of chromatin-associated condensates have been reported experimentally for transcription factors Klf4 and BRD4~\cite{Morin_2022, Strom_2024}. 

Local variations in fiber geometry, such as loops along a single fiber or intersections within a network, create regions of enhanced nucleosome density and increased available contact area. These sites act as preferential binding and nucleation points for the dense phase, as is was suggested from experimental observations of loop extrusions~\cite{Arnould_2020, Arnould_2021}. As a result, condensate organization reflects a competition between interfacial tension, which favors coarsening into a single droplet via Ostwald ripening, and the energetic gain associated with maximizing protein-chromatin contacts at these geometrically favorable locations. This competition enables the coexistence of multiple droplets when preferred sites are equivalent, providing an equilibrium mechanism for droplet size regulation that complements previously proposed non-equilibrium models based on biochemical turnover~\cite{Weber_2019,Zwicker_2022}. 

Beyond local effects, we showed that asymmetries in fiber organization at the system scale, such as uneven loop spacing or inhomogeneous fiber networks, can drive robust spatial localization of condensates. 
This behavior is consistent with recent experimental observations of DNA repair condensates that nucleate reproducibly at specific loci following local damage and consequent chromatin reorganization~\cite{GarciaFernandez_2025,Mine-hattab_2021,Heltberg_2022} and with observations of nuclear condensates whose locations are chromatin-concentration dependent~\cite{Mazzocca_2023}. Within such model, wetting on chromatin suggests the multiple droplets of water dispersed on a spider web, where droplets of various sizes coexist on several favorable locations of the web, or at some point may all fuse to the same region, such as the denser center of the spider web. 

In vivo, many nuclear condensates perform specialized functions at well-defined locations, including transcriptional regulation~\cite{Sabari_2018, Cho_2018,Mann_2023} and DNA repair~\cite{Lindahl_2000, Mine-hattab_2022}. Our results indicate that the interplay between protein-chromatin affinity and chromatin network organization can contribute significantly to the spatial control of such condensates. In particular, our findings are consistent with recent studies showing that condensate localization can correlate either positively or negatively with chromatin density depending on the relative strengths of protein-protein and protein-chromatin interactions~\cite{Mazzocca_2023, Shin_2018}. While additional layers of regulation—such as chromatin dynamics, active processes\cite{Zwicker_2022, Fries_active_2025, Berthin_2025}, and sequence-specific interactions—are undoubtedly important in living cells, our study demonstrates that geometric and energetic constraints imposed by chromatin architecture alone can already shape condensate size, coexistence, and localization.

\section{Data availability} Data are stored in a repository. The link will be made available on reasonable request to \href{mailto:vincent.dahirel@sorbonne-universite.fr}{vincent.dahirel@sorbonne-universite.fr}.

\appendix
\section{Methods}
\label{sec:methods}

\subsection{Droplet identification}
\label{sec:methods_track}

Two proteins are considered as part of the same droplet if they are closer than a threshold distance $r_{\text{cut}}$ of each other. The value of $r_{\text{cut}}$ is chosen as the distance at the first minimum in the protein-protein pair correlation function ($r_{\text{cut}}=1.65\sigma$ for the small systems with a single fiber and $r_{\text{cut}}=1.55 \sigma$ for the large systems with several fibers). In the presented data, a cluster of proteins is considered as a droplet if it is composed of more than 30 proteins.

\subsection{Estimating the droplet volume}
\label{sec:methods_vol}

Quantifying the volume of the distinct phases within a microscopic system presents significant challenges due to the absence of clearly defined phase boundaries. It is estimated from the positions $\rr_i=(r_{ix},r_{iy},r_{iz} )$ of the $n_{\text{parti}}$ proteins in each droplet. The center of mass of their positions along the $m$ coordinate is calculated as $\mathcal{R}_m=\frac{1}{n_{\text{parti}}}\sum_{i=1}^{n_{\text{parti}}}  r_{im} $, with $m, n \in \lbrace x,y,z \rbrace$. We then define the gyration tensor $S_{mn}$ as:
\begin{equation}
    S_{mn} = \frac{1}{n_{\text{parti}}} \sum_{i=1}^{n_{\text{parti}}} (r_{im}-\mathcal{R}_m)(r_{in}-\mathcal{R}_n).
\end{equation}

For spherical droplets, $S_{mn}$ is diagonal with eigenvalues $\lambda_1=\lambda_2=\lambda_3$, and the gyration radius is given by: $R^2_g= \lambda_1+\lambda_2+\lambda_3$. To include the case of ellipsoidal droplets, the droplet volume $V$ is estimated from the eigenvalues $\lambda_1,\lambda_2$ and $\lambda_3$ of $S_{mn}$, as:
\begin{equation}
    V\approx 4\pi \sqrt{3\lambda_1 \lambda_2 \lambda_3}
    \label{equ:volume_drop}
\end{equation}

If the droplet coats the fiber, the volume is slightly overestimated. In order to estimate the contribution from the fiber to the droplet volume, the contact length between the fiber and the droplet ($l_{\text{fib}}^{ \text{drop}}$) is measured. The volume occupied by the fiber is then estimated as that of a cylinder of radius $r_{\text{fib}} = (\sigma_{PC}-\sigma_{PP})2^{1/6} = 2^{1/6}$:
\begin{equation}
    V_{\text{fib}}^{\text{drop}} \approx \pi (r_{\text{fib}})^2 l_{\text{fib}}^{ \text{drop}}.
    \label{equ:volume_fiber_in_drop}
\end{equation}

\subsection{Droplet localization}
\label{sec:methods_gr}
To quantify the localization of the droplets with respect to the fibers, we calculate the pair correlation function $g_{DN}(r)$ between the center of mass of a droplet and the positions of all nucleosomes as:

\begin{equation}
    g_{DN}(r) = \left\langle \frac{1}{N_{\text{drop}} N_{\text{nucle}} } \sum_{i}^{N_{\text{drop}}} \sum_{j}^{N_{\text{nucle}}}\delta(d_{ij}-r)\right\rangle_t,
\end{equation}
$d_{ij}$ being the distance between the $i_{\text{th}}$ droplet center of mass and the $j_{\text{th}}$ nucleosome and $N_{\text{drop}}$ and $N_{\text{nucle}}$ being the number of droplets identified at time $t$ and the total number of nucleosomes in the system, respectively. 

To quantify whether proteins localize around preferred sites, as in particular in regions where chromatin is more packed, the same correlation function is calculated, but this time considering only a subset of nucleosomes. Two types of sites are considered: 
\begin{itemize}
    \item The intersections, with the associated pair correlation function denoted as $g_{DI}(r)$. This is the set of nucleosomes that are in close proximity (less than two typical nucleosome-nucleosome distances) from an intersecting fiber.
    \item  The specific site, with the corresponding pair correlation function $g_{DS}(r)$. This corresponds to the nucleosomes within the densest chromatin regions, which can be unambiguously defined in the case of irregular networks with a biased fiber distribution as explained hereafter. 
\end{itemize}

The local nucleosome density is used as a criterion to identify nucleosomes belonging to these preferred sites. For each nucleosome, the local nucleosome density is calculated as the number of neighbouring nucleosomes within a given radius divided by the volume considered. With this criterion, intersections correspond to a local increase in the nucleosome density compared to an isolated straight fiber (within a radius of $2.25d_{CC}\approx 3.76\sigma$). In practice, a regular intersection of three fibers corresponds to 12 selected nucleosomes. The specific site corresponds to nucleosomes with the 5\% highest large-scale nucleosome density, within a radius of $20.25 d_{CC}\approx 33.82 \sigma$. As a reference, the size of the cubic subvolume used to bias the fiber distribution in the irregular networks is $0.25\times200\sigma=50\sigma$. The radius is then chosen to be around the specific site length scale in order to select the nucleosomes at the center of the densest area. 

For disordered networks the density contrast is quantified as the variance of the distribution of the local nucleosome density (considering a radius of $20.25 d_{CC}\approx 33.82 \sigma$).

\section{Analytical model}
\label{Apx:Analytical_model}

From simple energy estimation one can show that the energy of a two-phase Lennard-Jones fluid is minimized by the formation of a single spherical droplet. The energy is estimated from the number of particles on the droplet surface and the number of inter-particle interactions. We consider a fixed total number of particles in the dense phase $N_{\text{parti}}$, and the number of interactions within the dilute phase as negligible. If all droplets have the same size ($N_{\text{drop}}$ droplets of radius $R$) the number of particles in a droplet is estimated as $n_{\text{parti}} = N_{\text{parti}}/N_{\text{drop}}$. Only the number of droplets $N_{\text{drop}}$ and the interaction potentials $\varepsilon_{PP}$ and $\varepsilon_{PC}$ vary.

The energy $E_{\text{no fib}}$ of the $N_{\text{drop}}$ droplets in the absence of fibers can be estimated as:
\begin{equation}
    E_{\text{no fib}} =  N_{\text{drop}} \left[  -n_{\text{parti}} n_{\text{neib}} \varepsilon_{PP} +\frac{n_{\text{neib}}\varepsilon_{PP}}{2} \frac{S_{\text{drop}}}{s_{\text{prot}}} \right]
\end{equation}
with $S_{\text{drop}}=4\pi R^2$ being the surface of the droplet, $ n_{\text{neib}}$ the average number neighbours around a protein when it is inside the droplet and $s_{\text{prot}}$ the effective surface occupied by a protein. The surface tension created by the energy cost of forming a surface leads to the minimisation of $E_{\text{no fib}}$ for $N_{\text{drop}}=1$. In this model, the dependency between the droplet radius $R$ and the number of particles composing the droplets $n_{\text{parti}}$ is inferred from the simulations data, by fitting the following formula: 
\begin{equation}
   R^3(n_{\text{parti}})  = n_{\text{parti}}A  +B
\end{equation}
We here extend the calculations to the case where the droplets form at a certain distance $d$ from an intersection of $n_{\text{fib}}$ fibers of radius $r_{\text{fib}}$, where $d$ is the distance between the center of the spherical droplet and the fibers. Two cases are studied:  

\begin{itemize}
    \item Case in which a single droplet forms on a single fiber. ($n_{\text{fib}}=1$ )
    \item Case in which $N_{\text{drop}}$ droplets form around or near an intersection of a regular network ($n_{\text{fib}}=3$). An illustrative sketch of an intersection is shown in Figure \ref{fig:Number_size_droplet_network_SKETCH}.
\end{itemize}

The presence of fibers is included as an additional surface $S_{\text{fib}}$, which is the surface of contact between the droplet and the fiber: 
\begin{equation}
    S_{\text{fib}}(R, d) =  n_{\text{fib}}\left[ 2\pi r_{\text{fib}} l_{\text{fib}}^{\text{drop}}(R, d) \right] \alpha(R,d) ;
\end{equation}
with $l_{\text{fib}}^{\text{drop}}$ being the total fiber length in contact with the droplet, and $\alpha(R,d)$ being the proportion of the fiber being coated by the droplet. To account for the portion of the droplet surface taken up by the fiber, $S_{\text{drop}}$ becomes:
\begin{equation}
    S_{\text{drop}}(R, d)\approx  4\pi R^2 -\alpha(R,d)2n_{\text{fib}}\times \pi r_{\text{fib}}^2.
\end{equation}

 The energy $E_{\text{wetting}}$ of the $N_{\text{drop}}$ droplets is calculated as:
\begin{equation}
\begin{split}
     E_{\text{wetting}} = N_{\text{drop}} & \left[- n_{\text{parti}} n_{\text{neib}} \varepsilon_{PP} +  n_{\text{neib}}\frac{\varepsilon_{PP}}{2} \frac{S_{\text{drop}}}{s_{\text{prot}}} \right. \\ 
     &+ \left.  \left( n_{\text{neib}}\frac{\varepsilon_{PP}}{2}  -n_{\text{neib}}^{PC}\varepsilon_{PC}\right)\frac{S_{\text{fib}}}{s_{\text{prot}}}\right]
     \end{split}
     \label{equ: E_wetting}
\end{equation}
with $n_{\text{neib}}^{PC}$ being the average number of nucleosomes a protein in contact with the fiber interacts with.

For comparison with the simulations, one can extract from this analytical description the number of proteins in contact with the fiber at the energy minima. This leads to a sharp transition similar to the one shown in Figure \ref{fig:linear_localize_NCONT}. 
From there, the transition from detached to on the side of the fiber is located where the number of particles in contact becomes greater than 1. The transition from on the side to the centered state corresponds to the values of $\varepsilon_{PC}$ where the number of particles in contact reaches the plateau. Both transitions are shown in Figure \ref{fig:linear_localize_C}.

\bibliography{Ma_bibliothèque_06mai26_Clean}

\begin{thebibliography}{75}%
\makeatletter
\providecommand \@ifxundefined [1]{%
 \@ifx{#1\undefined}
}%
\providecommand \@ifnum [1]{%
 \ifnum #1\expandafter \@firstoftwo
 \else \expandafter \@secondoftwo
 \fi
}%
\providecommand \@ifx [1]{%
 \ifx #1\expandafter \@firstoftwo
 \else \expandafter \@secondoftwo
 \fi
}%
\providecommand \natexlab [1]{#1}%
\providecommand \enquote  [1]{``#1''}%
\providecommand \bibnamefont  [1]{#1}%
\providecommand \bibfnamefont [1]{#1}%
\providecommand \citenamefont [1]{#1}%
\providecommand \href@noop [0]{\@secondoftwo}%
\providecommand \href [0]{\begingroup \@sanitize@url \@href}%
\providecommand \@href[1]{\@@startlink{#1}\@@href}%
\providecommand \@@href[1]{\endgroup#1\@@endlink}%
\providecommand \@sanitize@url [0]{\catcode `\\12\catcode `\$12\catcode
  `\&12\catcode `\#12\catcode `\^12\catcode `\_12\catcode `\%12\relax}%
\providecommand \@@startlink[1]{}%
\providecommand \@@endlink[0]{}%
\providecommand \url  [0]{\begingroup\@sanitize@url \@url }%
\providecommand \@url [1]{\endgroup\@href {#1}{\urlprefix }}%
\providecommand \urlprefix  [0]{URL }%
\providecommand \Eprint [0]{\href }%
\providecommand \doibase [0]{https://doi.org/}%
\providecommand \selectlanguage [0]{\@gobble}%
\providecommand \bibinfo  [0]{\@secondoftwo}%
\providecommand \bibfield  [0]{\@secondoftwo}%
\providecommand \translation [1]{[#1]}%
\providecommand \BibitemOpen [0]{}%
\providecommand \bibitemStop [0]{}%
\providecommand \bibitemNoStop [0]{.\EOS\space}%
\providecommand \EOS [0]{\spacefactor3000\relax}%
\providecommand \BibitemShut  [1]{\csname bibitem#1\endcsname}%
\let\auto@bib@innerbib\@empty
\bibitem [{\citenamefont {Brangwynne}\ \emph {et~al.}(2009)\citenamefont
  {Brangwynne}, \citenamefont {Eckmann}, \citenamefont {Courson}, \citenamefont
  {Rybarska}, \citenamefont {Hoege}, \citenamefont {Gharakhani}, \citenamefont
  {Jülicher},\ and\ \citenamefont {Hyman}}]{Brangwynne_2009}%
  \BibitemOpen
  \bibfield  {author} {\bibinfo {author} {\bibfnamefont {C.~P.}\ \bibnamefont
  {Brangwynne}}, \bibinfo {author} {\bibfnamefont {C.~R.}\ \bibnamefont
  {Eckmann}}, \bibinfo {author} {\bibfnamefont {D.~S.}\ \bibnamefont
  {Courson}}, \bibinfo {author} {\bibfnamefont {A.}~\bibnamefont {Rybarska}},
  \bibinfo {author} {\bibfnamefont {C.}~\bibnamefont {Hoege}}, \bibinfo
  {author} {\bibfnamefont {J.}~\bibnamefont {Gharakhani}}, \bibinfo {author}
  {\bibfnamefont {F.}~\bibnamefont {Jülicher}},\ and\ \bibinfo {author}
  {\bibfnamefont {A.~A.}\ \bibnamefont {Hyman}},\ }\bibfield  {title} {\bibinfo
  {title} {Germline {P} {Granules} {Are} {Liquid} {Droplets} {That} {Localize}
  by {Controlled} {Dissolution}/{Condensation}},\ }\href
  {https://doi.org/10.1126/science.1172046} {\bibfield  {journal} {\bibinfo
  {journal} {Science}\ }\textbf {\bibinfo {volume} {324}},\ \bibinfo {pages}
  {1729} (\bibinfo {year} {2009})}\BibitemShut {NoStop}%
\bibitem [{\citenamefont {Shin}\ and\ \citenamefont
  {Brangwynne}(2017)}]{Shin_2017}%
  \BibitemOpen
  \bibfield  {author} {\bibinfo {author} {\bibfnamefont {Y.}~\bibnamefont
  {Shin}}\ and\ \bibinfo {author} {\bibfnamefont {C.~P.}\ \bibnamefont
  {Brangwynne}},\ }\bibfield  {title} {\bibinfo {title} {Liquid phase
  condensation in cell physiology and disease},\ }\bibfield  {journal}
  {\bibinfo  {journal} {Science}\ }\textbf {\bibinfo {volume} {357}},\ \href
  {https://doi.org/10.1126/science.aaf4382} {10.1126/science.aaf4382} (\bibinfo
  {year} {2017})\BibitemShut {NoStop}%
\bibitem [{\citenamefont {Shin}\ \emph {et~al.}(2018)\citenamefont {Shin},
  \citenamefont {Chang}, \citenamefont {Lee}, \citenamefont {Berry},
  \citenamefont {Sanders}, \citenamefont {Ronceray}, \citenamefont {Wingreen},
  \citenamefont {Haataja},\ and\ \citenamefont {Brangwynne}}]{Shin_2018}%
  \BibitemOpen
  \bibfield  {author} {\bibinfo {author} {\bibfnamefont {Y.}~\bibnamefont
  {Shin}}, \bibinfo {author} {\bibfnamefont {Y.-C.}\ \bibnamefont {Chang}},
  \bibinfo {author} {\bibfnamefont {D.~S.}\ \bibnamefont {Lee}}, \bibinfo
  {author} {\bibfnamefont {J.}~\bibnamefont {Berry}}, \bibinfo {author}
  {\bibfnamefont {D.~W.}\ \bibnamefont {Sanders}}, \bibinfo {author}
  {\bibfnamefont {P.}~\bibnamefont {Ronceray}}, \bibinfo {author}
  {\bibfnamefont {N.~S.}\ \bibnamefont {Wingreen}}, \bibinfo {author}
  {\bibfnamefont {M.}~\bibnamefont {Haataja}},\ and\ \bibinfo {author}
  {\bibfnamefont {C.~P.}\ \bibnamefont {Brangwynne}},\ }\bibfield  {title}
  {\bibinfo {title} {Liquid {Nuclear} {Condensates} {Mechanically} {Sense} and
  {Restructure} the {Genome}},\ }\href
  {https://doi.org/10.1016/j.cell.2018.10.057} {\bibfield  {journal} {\bibinfo
  {journal} {Cell}\ }\textbf {\bibinfo {volume} {175}},\ \bibinfo {pages}
  {1481} (\bibinfo {year} {2018})}\BibitemShut {NoStop}%
\bibitem [{\citenamefont {Gibson}\ \emph {et~al.}(2019)\citenamefont {Gibson},
  \citenamefont {Doolittle}, \citenamefont {Schneider}, \citenamefont {Jensen},
  \citenamefont {Gamarra}, \citenamefont {Henry}, \citenamefont {Gerlich},
  \citenamefont {Redding},\ and\ \citenamefont {Rosen}}]{Gibson_2019}%
  \BibitemOpen
  \bibfield  {author} {\bibinfo {author} {\bibfnamefont {B.~A.}\ \bibnamefont
  {Gibson}}, \bibinfo {author} {\bibfnamefont {L.~K.}\ \bibnamefont
  {Doolittle}}, \bibinfo {author} {\bibfnamefont {M.~W.}\ \bibnamefont
  {Schneider}}, \bibinfo {author} {\bibfnamefont {L.~E.}\ \bibnamefont
  {Jensen}}, \bibinfo {author} {\bibfnamefont {N.}~\bibnamefont {Gamarra}},
  \bibinfo {author} {\bibfnamefont {L.}~\bibnamefont {Henry}}, \bibinfo
  {author} {\bibfnamefont {D.~W.}\ \bibnamefont {Gerlich}}, \bibinfo {author}
  {\bibfnamefont {S.}~\bibnamefont {Redding}},\ and\ \bibinfo {author}
  {\bibfnamefont {M.~K.}\ \bibnamefont {Rosen}},\ }\bibfield  {title} {\bibinfo
  {title} {Organization of {Chromatin} by {Intrinsic} and {Regulated} {Phase}
  {Separation}},\ }\href {https://doi.org/10.1016/j.cell.2019.08.037}
  {\bibfield  {journal} {\bibinfo  {journal} {Cell}\ }\textbf {\bibinfo
  {volume} {179}},\ \bibinfo {pages} {470} (\bibinfo {year}
  {2019})}\BibitemShut {NoStop}%
\bibitem [{\citenamefont {Weber}\ \emph {et~al.}(2019)\citenamefont {Weber},
  \citenamefont {Zwicker}, \citenamefont {Jülicher},\ and\ \citenamefont
  {Lee}}]{Weber_2019}%
  \BibitemOpen
  \bibfield  {author} {\bibinfo {author} {\bibfnamefont {C.~A.}\ \bibnamefont
  {Weber}}, \bibinfo {author} {\bibfnamefont {D.}~\bibnamefont {Zwicker}},
  \bibinfo {author} {\bibfnamefont {F.}~\bibnamefont {Jülicher}},\ and\
  \bibinfo {author} {\bibfnamefont {C.~F.}\ \bibnamefont {Lee}},\ }\bibfield
  {title} {\bibinfo {title} {Physics of active emulsions},\ }\href
  {https://doi.org/10.1088/1361-6633/ab052b} {\bibfield  {journal} {\bibinfo
  {journal} {Reports on Progress in Physics}\ }\textbf {\bibinfo {volume}
  {82}},\ \bibinfo {pages} {064601} (\bibinfo {year} {2019})}\BibitemShut
  {NoStop}%
\bibitem [{\citenamefont {Sabari}\ \emph {et~al.}(2020)\citenamefont {Sabari},
  \citenamefont {Dall’Agnese},\ and\ \citenamefont {Young}}]{Sabari_2020}%
  \BibitemOpen
  \bibfield  {author} {\bibinfo {author} {\bibfnamefont {B.~R.}\ \bibnamefont
  {Sabari}}, \bibinfo {author} {\bibfnamefont {A.}~\bibnamefont
  {Dall’Agnese}},\ and\ \bibinfo {author} {\bibfnamefont {R.~A.}\
  \bibnamefont {Young}},\ }\bibfield  {title} {\bibinfo {title} {Biomolecular
  {Condensates} in the {Nucleus}},\ }\href
  {https://doi.org/10.1016/j.tibs.2020.06.007} {\bibfield  {journal} {\bibinfo
  {journal} {Trends in Biochemical Sciences}\ }\textbf {\bibinfo {volume}
  {45}},\ \bibinfo {pages} {961} (\bibinfo {year} {2020})}\BibitemShut
  {NoStop}%
\bibitem [{\citenamefont {Gouveia}\ \emph {et~al.}(2022)\citenamefont
  {Gouveia}, \citenamefont {Kim}, \citenamefont {Shaevitz}, \citenamefont
  {Petry}, \citenamefont {Stone},\ and\ \citenamefont
  {Brangwynne}}]{Gouveia_2022}%
  \BibitemOpen
  \bibfield  {author} {\bibinfo {author} {\bibfnamefont {B.}~\bibnamefont
  {Gouveia}}, \bibinfo {author} {\bibfnamefont {Y.}~\bibnamefont {Kim}},
  \bibinfo {author} {\bibfnamefont {J.~W.}\ \bibnamefont {Shaevitz}}, \bibinfo
  {author} {\bibfnamefont {S.}~\bibnamefont {Petry}}, \bibinfo {author}
  {\bibfnamefont {H.~A.}\ \bibnamefont {Stone}},\ and\ \bibinfo {author}
  {\bibfnamefont {C.~P.}\ \bibnamefont {Brangwynne}},\ }\bibfield  {title}
  {\bibinfo {title} {Capillary forces generated by biomolecular condensates},\
  }\href {https://doi.org/10.1038/s41586-022-05138-6} {\bibfield  {journal}
  {\bibinfo  {journal} {Nature}\ }\textbf {\bibinfo {volume} {609}},\ \bibinfo
  {pages} {255} (\bibinfo {year} {2022})}\BibitemShut {NoStop}%
\bibitem [{\citenamefont {Rouches}\ and\ \citenamefont
  {Machta}(2024)}]{Rouches_2024}%
  \BibitemOpen
  \bibfield  {author} {\bibinfo {author} {\bibfnamefont {M.~N.}\ \bibnamefont
  {Rouches}}\ and\ \bibinfo {author} {\bibfnamefont {B.~B.}\ \bibnamefont
  {Machta}},\ }\href {https://doi.org/10.1101/2024.04.29.591767} {\bibinfo
  {title} {Polymer {Collapse} \& {Liquid}-{Liquid} {Phase}-{Separation} are
  {Coupled} in a {Generalized} {Prewetting} {Transition}}} (\bibinfo {year}
  {2024})\BibitemShut {NoStop}%
\bibitem [{\citenamefont {Hnisz}\ \emph {et~al.}(2017)\citenamefont {Hnisz},
  \citenamefont {Shrinivas}, \citenamefont {Young}, \citenamefont
  {Chakraborty},\ and\ \citenamefont {Sharp}}]{Hnisz_2017}%
  \BibitemOpen
  \bibfield  {author} {\bibinfo {author} {\bibfnamefont {D.}~\bibnamefont
  {Hnisz}}, \bibinfo {author} {\bibfnamefont {K.}~\bibnamefont {Shrinivas}},
  \bibinfo {author} {\bibfnamefont {R.~A.}\ \bibnamefont {Young}}, \bibinfo
  {author} {\bibfnamefont {A.~K.}\ \bibnamefont {Chakraborty}},\ and\ \bibinfo
  {author} {\bibfnamefont {P.~A.}\ \bibnamefont {Sharp}},\ }\bibfield  {title}
  {\bibinfo {title} {A {Phase} {Separation} {Model} for {Transcriptional}
  {Control}},\ }\href {https://doi.org/10.1016/j.cell.2017.02.007} {\bibfield
  {journal} {\bibinfo  {journal} {Cell}\ }\textbf {\bibinfo {volume} {169}},\
  \bibinfo {pages} {13} (\bibinfo {year} {2017})}\BibitemShut {NoStop}%
\bibitem [{\citenamefont {Rippe}(2022)}]{Rippe_2022}%
  \BibitemOpen
  \bibfield  {author} {\bibinfo {author} {\bibfnamefont {K.}~\bibnamefont
  {Rippe}},\ }\bibfield  {title} {\bibinfo {title} {Liquid–{Liquid} {Phase}
  {Separation} in {Chromatin}},\ }\href
  {https://doi.org/10.1101/cshperspect.a040683} {\bibfield  {journal} {\bibinfo
   {journal} {Cold Spring Harbor Perspectives in Biology}\ }\textbf {\bibinfo
  {volume} {14}},\ \bibinfo {pages} {a040683} (\bibinfo {year}
  {2022})}\BibitemShut {NoStop}%
\bibitem [{\citenamefont {Wu}\ \emph {et~al.}(2022)\citenamefont {Wu},
  \citenamefont {Chen}, \citenamefont {Liu}, \citenamefont {Ma}, \citenamefont
  {Huang},\ and\ \citenamefont {Lin}}]{Wu_2022}%
  \BibitemOpen
  \bibfield  {author} {\bibinfo {author} {\bibfnamefont {J.}~\bibnamefont
  {Wu}}, \bibinfo {author} {\bibfnamefont {B.}~\bibnamefont {Chen}}, \bibinfo
  {author} {\bibfnamefont {Y.}~\bibnamefont {Liu}}, \bibinfo {author}
  {\bibfnamefont {L.}~\bibnamefont {Ma}}, \bibinfo {author} {\bibfnamefont
  {W.}~\bibnamefont {Huang}},\ and\ \bibinfo {author} {\bibfnamefont
  {Y.}~\bibnamefont {Lin}},\ }\bibfield  {title} {\bibinfo {title} {Modulating
  gene regulation function by chemically controlled transcription factor
  clustering},\ }\href {https://doi.org/10.1038/s41467-022-30397-2} {\bibfield
  {journal} {\bibinfo  {journal} {Nature Communications}\ }\textbf {\bibinfo
  {volume} {13}},\ \bibinfo {pages} {2663} (\bibinfo {year}
  {2022})}\BibitemShut {NoStop}%
\bibitem [{\citenamefont {Mann}\ and\ \citenamefont
  {Notani}(2023)}]{Mann_2023}%
  \BibitemOpen
  \bibfield  {author} {\bibinfo {author} {\bibfnamefont {R.}~\bibnamefont
  {Mann}}\ and\ \bibinfo {author} {\bibfnamefont {D.}~\bibnamefont {Notani}},\
  }\bibfield  {title} {\bibinfo {title} {Transcription factor condensates and
  signaling driven transcription},\ }\href
  {https://doi.org/10.1080/19491034.2023.2205758} {\bibfield  {journal}
  {\bibinfo  {journal} {Nucleus}\ }\textbf {\bibinfo {volume} {14}},\ \bibinfo
  {pages} {2205758} (\bibinfo {year} {2023})}\BibitemShut {NoStop}%
\bibitem [{\citenamefont {Brangwynne}\ \emph {et~al.}(2015)\citenamefont
  {Brangwynne}, \citenamefont {Tompa},\ and\ \citenamefont
  {Pappu}}]{Brangwynne_2015}%
  \BibitemOpen
  \bibfield  {author} {\bibinfo {author} {\bibfnamefont {C.}~\bibnamefont
  {Brangwynne}}, \bibinfo {author} {\bibfnamefont {P.}~\bibnamefont {Tompa}},\
  and\ \bibinfo {author} {\bibfnamefont {R.}~\bibnamefont {Pappu}},\ }\bibfield
   {title} {\bibinfo {title} {Polymer physics of intracellular phase
  transitions},\ }\href {https://doi.org/10.1038/nphys3532} {\bibfield
  {journal} {\bibinfo  {journal} {Nature Physics}\ }\textbf {\bibinfo {volume}
  {11}},\ \bibinfo {pages} {899} (\bibinfo {year} {2015})}\BibitemShut
  {NoStop}%
\bibitem [{\citenamefont {Banani}\ \emph {et~al.}(2017)\citenamefont {Banani},
  \citenamefont {Lee}, \citenamefont {Hyman},\ and\ \citenamefont
  {Rosen}}]{Banani_2017}%
  \BibitemOpen
  \bibfield  {author} {\bibinfo {author} {\bibfnamefont {S.~F.}\ \bibnamefont
  {Banani}}, \bibinfo {author} {\bibfnamefont {H.~O.}\ \bibnamefont {Lee}},
  \bibinfo {author} {\bibfnamefont {A.~A.}\ \bibnamefont {Hyman}},\ and\
  \bibinfo {author} {\bibfnamefont {M.~K.}\ \bibnamefont {Rosen}},\ }\bibfield
  {title} {\bibinfo {title} {Biomolecular condensates: organizers of cellular
  biochemistry},\ }\href {https://doi.org/10.1038/nrm.2017.7} {\bibfield
  {journal} {\bibinfo  {journal} {Nature Reviews Molecular Cell Biology}\
  }\textbf {\bibinfo {volume} {18}},\ \bibinfo {pages} {285} (\bibinfo {year}
  {2017})}\BibitemShut {NoStop}%
\bibitem [{\citenamefont {Kilic}\ \emph {et~al.}(2019)\citenamefont {Kilic},
  \citenamefont {Lezaja}, \citenamefont {Gatti}, \citenamefont {Bianco},
  \citenamefont {Michelena}, \citenamefont {Imhof},\ and\ \citenamefont
  {Altmeyer}}]{Kilic_2019}%
  \BibitemOpen
  \bibfield  {author} {\bibinfo {author} {\bibfnamefont {S.}~\bibnamefont
  {Kilic}}, \bibinfo {author} {\bibfnamefont {A.}~\bibnamefont {Lezaja}},
  \bibinfo {author} {\bibfnamefont {M.}~\bibnamefont {Gatti}}, \bibinfo
  {author} {\bibfnamefont {E.}~\bibnamefont {Bianco}}, \bibinfo {author}
  {\bibfnamefont {J.}~\bibnamefont {Michelena}}, \bibinfo {author}
  {\bibfnamefont {R.}~\bibnamefont {Imhof}},\ and\ \bibinfo {author}
  {\bibfnamefont {M.}~\bibnamefont {Altmeyer}},\ }\bibfield  {title} {\bibinfo
  {title} {Phase separation of {53BP1} determines liquid‐like behavior of
  {DNA} repair compartments},\ }\href
  {https://doi.org/10.15252/embj.2018101379} {\bibfield  {journal} {\bibinfo
  {journal} {The EMBO Journal}\ }\textbf {\bibinfo {volume} {38}},\ \bibinfo
  {pages} {e101379} (\bibinfo {year} {2019})}\BibitemShut {NoStop}%
\bibitem [{\citenamefont {Hyman}\ \emph {et~al.}(2014)\citenamefont {Hyman},
  \citenamefont {Weber},\ and\ \citenamefont {Jülicher}}]{Hyman_2014}%
  \BibitemOpen
  \bibfield  {author} {\bibinfo {author} {\bibfnamefont {A.~A.}\ \bibnamefont
  {Hyman}}, \bibinfo {author} {\bibfnamefont {C.~A.}\ \bibnamefont {Weber}},\
  and\ \bibinfo {author} {\bibfnamefont {F.}~\bibnamefont {Jülicher}},\
  }\bibfield  {title} {\bibinfo {title} {Liquid-{Liquid} {Phase} {Separation}
  in {Biology}},\ }\href
  {https://doi.org/10.1146/annurev-cellbio-100913-013325} {\bibfield  {journal}
  {\bibinfo  {journal} {Annual Review of Cell and Developmental Biology}\
  }\textbf {\bibinfo {volume} {30}},\ \bibinfo {pages} {39} (\bibinfo {year}
  {2014})}\BibitemShut {NoStop}%
\bibitem [{\citenamefont {Alberti}\ \emph {et~al.}(2019)\citenamefont
  {Alberti}, \citenamefont {Gladfelter},\ and\ \citenamefont
  {Mittag}}]{Alberti_2019}%
  \BibitemOpen
  \bibfield  {author} {\bibinfo {author} {\bibfnamefont {S.}~\bibnamefont
  {Alberti}}, \bibinfo {author} {\bibfnamefont {A.}~\bibnamefont
  {Gladfelter}},\ and\ \bibinfo {author} {\bibfnamefont {T.}~\bibnamefont
  {Mittag}},\ }\bibfield  {title} {\bibinfo {title} {Considerations and
  {Challenges} in {Studying} {Liquid}-{Liquid} {Phase} {Separation} and
  {Biomolecular} {Condensates}},\ }\href
  {https://doi.org/10.1016/j.cell.2018.12.035} {\bibfield  {journal} {\bibinfo
  {journal} {Cell}\ }\textbf {\bibinfo {volume} {176}},\ \bibinfo {pages} {419}
  (\bibinfo {year} {2019})}\BibitemShut {NoStop}%
\bibitem [{\citenamefont {Heltberg}\ \emph {et~al.}(2021)\citenamefont
  {Heltberg}, \citenamefont {Miné-Hattab}, \citenamefont {Taddei},
  \citenamefont {Walczak},\ and\ \citenamefont {Mora}}]{Heltberg_2021}%
  \BibitemOpen
  \bibfield  {author} {\bibinfo {author} {\bibfnamefont {M.~L.}\ \bibnamefont
  {Heltberg}}, \bibinfo {author} {\bibfnamefont {J.}~\bibnamefont
  {Miné-Hattab}}, \bibinfo {author} {\bibfnamefont {A.}~\bibnamefont
  {Taddei}}, \bibinfo {author} {\bibfnamefont {A.~M.}\ \bibnamefont
  {Walczak}},\ and\ \bibinfo {author} {\bibfnamefont {T.}~\bibnamefont
  {Mora}},\ }\bibfield  {title} {\bibinfo {title} {Physical observables to
  determine the nature of membrane-less cellular sub-compartments},\ }\href
  {https://doi.org/10.7554/eLife.69181} {\bibfield  {journal} {\bibinfo
  {journal} {eLife}\ }\textbf {\bibinfo {volume} {10}},\ \bibinfo {pages}
  {e69181} (\bibinfo {year} {2021})}\BibitemShut {NoStop}%
\bibitem [{\citenamefont {Miné-Hattab}\ \emph {et~al.}(2021)\citenamefont
  {Miné-Hattab}, \citenamefont {Heltberg}, \citenamefont {Villemeur},
  \citenamefont {Guedj}, \citenamefont {Mora}, \citenamefont {Walczak},
  \citenamefont {Dahan},\ and\ \citenamefont {Taddei}}]{Mine-hattab_2021}%
  \BibitemOpen
  \bibfield  {author} {\bibinfo {author} {\bibfnamefont {J.}~\bibnamefont
  {Miné-Hattab}}, \bibinfo {author} {\bibfnamefont {M.}~\bibnamefont
  {Heltberg}}, \bibinfo {author} {\bibfnamefont {M.}~\bibnamefont {Villemeur}},
  \bibinfo {author} {\bibfnamefont {C.}~\bibnamefont {Guedj}}, \bibinfo
  {author} {\bibfnamefont {T.}~\bibnamefont {Mora}}, \bibinfo {author}
  {\bibfnamefont {A.~M.}\ \bibnamefont {Walczak}}, \bibinfo {author}
  {\bibfnamefont {M.}~\bibnamefont {Dahan}},\ and\ \bibinfo {author}
  {\bibfnamefont {A.}~\bibnamefont {Taddei}},\ }\bibfield  {title} {\bibinfo
  {title} {Single molecule microscopy reveals key physical features of repair
  foci in living cells},\ }\href {https://doi.org/10.7554/eLife.60577}
  {\bibfield  {journal} {\bibinfo  {journal} {eLife}\ }\textbf {\bibinfo
  {volume} {10}},\ \bibinfo {pages} {e60577} (\bibinfo {year}
  {2021})}\BibitemShut {NoStop}%
\bibitem [{\citenamefont {Heltberg}\ \emph {et~al.}(2022)\citenamefont
  {Heltberg}, \citenamefont {Lucchetti}, \citenamefont {Hsieh}, \citenamefont
  {Minh~Nguyen}, \citenamefont {Chen},\ and\ \citenamefont
  {Jensen}}]{Heltberg_2022}%
  \BibitemOpen
  \bibfield  {author} {\bibinfo {author} {\bibfnamefont {M.~S.}\ \bibnamefont
  {Heltberg}}, \bibinfo {author} {\bibfnamefont {A.}~\bibnamefont {Lucchetti}},
  \bibinfo {author} {\bibfnamefont {F.-S.}\ \bibnamefont {Hsieh}}, \bibinfo
  {author} {\bibfnamefont {D.~P.}\ \bibnamefont {Minh~Nguyen}}, \bibinfo
  {author} {\bibfnamefont {S.-h.}\ \bibnamefont {Chen}},\ and\ \bibinfo
  {author} {\bibfnamefont {M.~H.}\ \bibnamefont {Jensen}},\ }\bibfield  {title}
  {\bibinfo {title} {Enhanced {DNA} repair through droplet formation and p53
  oscillations},\ }\href {https://doi.org/10.1016/j.cell.2022.10.004}
  {\bibfield  {journal} {\bibinfo  {journal} {Cell}\ }\textbf {\bibinfo
  {volume} {185}},\ \bibinfo {pages} {4394} (\bibinfo {year}
  {2022})}\BibitemShut {NoStop}%
\bibitem [{\citenamefont {García~Fernández}\ \emph
  {et~al.}(2023)\citenamefont {García~Fernández}, \citenamefont {Huet},\ and\
  \citenamefont {Miné-Hattab}}]{GarciaFernandez_2023}%
  \BibitemOpen
  \bibfield  {author} {\bibinfo {author} {\bibfnamefont {F.}~\bibnamefont
  {García~Fernández}}, \bibinfo {author} {\bibfnamefont {S.}~\bibnamefont
  {Huet}},\ and\ \bibinfo {author} {\bibfnamefont {J.}~\bibnamefont
  {Miné-Hattab}},\ }\bibfield  {title} {\bibinfo {title} {Multi-{Scale}
  {Imaging} of the {Dynamic} {Organization} of {Chromatin}},\ }\href
  {https://doi.org/10.3390/ijms242115975} {\bibfield  {journal} {\bibinfo
  {journal} {International Journal of Molecular Sciences}\ }\textbf {\bibinfo
  {volume} {24}},\ \bibinfo {pages} {15975} (\bibinfo {year}
  {2023})}\BibitemShut {NoStop}%
\bibitem [{\citenamefont {Filion}\ \emph {et~al.}(2010)\citenamefont {Filion},
  \citenamefont {Van~Bemmel}, \citenamefont {Braunschweig}, \citenamefont
  {Talhout}, \citenamefont {Kind}, \citenamefont {Ward}, \citenamefont
  {Brugman}, \citenamefont {De~Castro}, \citenamefont {Kerkhoven},
  \citenamefont {Bussemaker},\ and\ \citenamefont
  {van Steensel}}]{Filion_2010}%
  \BibitemOpen
  \bibfield  {author} {\bibinfo {author} {\bibfnamefont {G.~J.}\ \bibnamefont
  {Filion}}, \bibinfo {author} {\bibfnamefont {J.~G.}\ \bibnamefont
  {Van~Bemmel}}, \bibinfo {author} {\bibfnamefont {U.}~\bibnamefont
  {Braunschweig}}, \bibinfo {author} {\bibfnamefont {W.}~\bibnamefont
  {Talhout}}, \bibinfo {author} {\bibfnamefont {J.}~\bibnamefont {Kind}},
  \bibinfo {author} {\bibfnamefont {L.~D.}\ \bibnamefont {Ward}}, \bibinfo
  {author} {\bibfnamefont {W.}~\bibnamefont {Brugman}}, \bibinfo {author}
  {\bibfnamefont {I.~J.}\ \bibnamefont {De~Castro}}, \bibinfo {author}
  {\bibfnamefont {R.~M.}\ \bibnamefont {Kerkhoven}}, \bibinfo {author}
  {\bibfnamefont {H.~J.}\ \bibnamefont {Bussemaker}},\ and\ \bibinfo {author}
  {\bibfnamefont {B.}~\bibnamefont {van Steensel}},\ }\bibfield  {title}
  {\bibinfo {title} {Systematic {Protein} {Location} {Mapping} {Reveals} {Five}
  {Principal} {Chromatin} {Types} in {Drosophila} {Cells}},\ }\href
  {https://doi.org/10.1016/j.cell.2010.09.009} {\bibfield  {journal} {\bibinfo
  {journal} {Cell}\ }\textbf {\bibinfo {volume} {143}},\ \bibinfo {pages} {212}
  (\bibinfo {year} {2010})}\BibitemShut {NoStop}%
\bibitem [{\citenamefont {Boettiger}\ \emph {et~al.}(2016)\citenamefont
  {Boettiger}, \citenamefont {Bintu}, \citenamefont {Moffitt}, \citenamefont
  {Wang}, \citenamefont {Beliveau}, \citenamefont {Fudenberg}, \citenamefont
  {Imakaev}, \citenamefont {Mirny}, \citenamefont {Wu},\ and\ \citenamefont
  {Zhuang}}]{Boettiger_2016}%
  \BibitemOpen
  \bibfield  {author} {\bibinfo {author} {\bibfnamefont {A.~N.}\ \bibnamefont
  {Boettiger}}, \bibinfo {author} {\bibfnamefont {B.}~\bibnamefont {Bintu}},
  \bibinfo {author} {\bibfnamefont {J.~R.}\ \bibnamefont {Moffitt}}, \bibinfo
  {author} {\bibfnamefont {S.}~\bibnamefont {Wang}}, \bibinfo {author}
  {\bibfnamefont {B.~J.}\ \bibnamefont {Beliveau}}, \bibinfo {author}
  {\bibfnamefont {G.}~\bibnamefont {Fudenberg}}, \bibinfo {author}
  {\bibfnamefont {M.}~\bibnamefont {Imakaev}}, \bibinfo {author} {\bibfnamefont
  {L.~A.}\ \bibnamefont {Mirny}}, \bibinfo {author} {\bibfnamefont {C.-t.}\
  \bibnamefont {Wu}},\ and\ \bibinfo {author} {\bibfnamefont {X.}~\bibnamefont
  {Zhuang}},\ }\bibfield  {title} {\bibinfo {title} {Super-resolution imaging
  reveals distinct chromatin folding for different epigenetic states},\ }\href
  {https://doi.org/10.1038/nature16496} {\bibfield  {journal} {\bibinfo
  {journal} {Nature}\ }\textbf {\bibinfo {volume} {529}},\ \bibinfo {pages}
  {418} (\bibinfo {year} {2016})}\BibitemShut {NoStop}%
\bibitem [{\citenamefont {Cattoni}\ \emph {et~al.}(2017)\citenamefont
  {Cattoni}, \citenamefont {Cardozo~Gizzi}, \citenamefont {Georgieva},
  \citenamefont {Di~Stefano}, \citenamefont {Valeri}, \citenamefont
  {Chamousset}, \citenamefont {Houbron}, \citenamefont {Déjardin},
  \citenamefont {Fiche}, \citenamefont {González}, \citenamefont {Chang},
  \citenamefont {Sexton}, \citenamefont {Marti-Renom}, \citenamefont
  {Bantignies}, \citenamefont {Cavalli},\ and\ \citenamefont
  {Nollmann}}]{Cattoni_2017}%
  \BibitemOpen
  \bibfield  {author} {\bibinfo {author} {\bibfnamefont {D.~I.}\ \bibnamefont
  {Cattoni}}, \bibinfo {author} {\bibfnamefont {A.~M.}\ \bibnamefont
  {Cardozo~Gizzi}}, \bibinfo {author} {\bibfnamefont {M.}~\bibnamefont
  {Georgieva}}, \bibinfo {author} {\bibfnamefont {M.}~\bibnamefont
  {Di~Stefano}}, \bibinfo {author} {\bibfnamefont {A.}~\bibnamefont {Valeri}},
  \bibinfo {author} {\bibfnamefont {D.}~\bibnamefont {Chamousset}}, \bibinfo
  {author} {\bibfnamefont {C.}~\bibnamefont {Houbron}}, \bibinfo {author}
  {\bibfnamefont {S.}~\bibnamefont {Déjardin}}, \bibinfo {author}
  {\bibfnamefont {J.-B.}\ \bibnamefont {Fiche}}, \bibinfo {author}
  {\bibfnamefont {I.}~\bibnamefont {González}}, \bibinfo {author}
  {\bibfnamefont {J.-M.}\ \bibnamefont {Chang}}, \bibinfo {author}
  {\bibfnamefont {T.}~\bibnamefont {Sexton}}, \bibinfo {author} {\bibfnamefont
  {M.~A.}\ \bibnamefont {Marti-Renom}}, \bibinfo {author} {\bibfnamefont
  {F.}~\bibnamefont {Bantignies}}, \bibinfo {author} {\bibfnamefont
  {G.}~\bibnamefont {Cavalli}},\ and\ \bibinfo {author} {\bibfnamefont
  {M.}~\bibnamefont {Nollmann}},\ }\bibfield  {title} {\bibinfo {title}
  {Single-cell absolute contact probability detection reveals chromosomes are
  organized by multiple low-frequency yet specific interactions},\ }\href
  {https://doi.org/10.1038/s41467-017-01962-x} {\bibfield  {journal} {\bibinfo
  {journal} {Nature Communications}\ }\textbf {\bibinfo {volume} {8}},\
  \bibinfo {pages} {1753} (\bibinfo {year} {2017})}\BibitemShut {NoStop}%
\bibitem [{\citenamefont {Szabo}\ \emph {et~al.}(2018)\citenamefont {Szabo},
  \citenamefont {Jost}, \citenamefont {Chang}, \citenamefont {Cattoni},
  \citenamefont {Papadopoulos}, \citenamefont {Bonev}, \citenamefont {Sexton},
  \citenamefont {Gurgo}, \citenamefont {Jacquier}, \citenamefont {Nollmann},
  \citenamefont {Bantignies},\ and\ \citenamefont {Cavalli}}]{Szabo_2018}%
  \BibitemOpen
  \bibfield  {author} {\bibinfo {author} {\bibfnamefont {Q.}~\bibnamefont
  {Szabo}}, \bibinfo {author} {\bibfnamefont {D.}~\bibnamefont {Jost}},
  \bibinfo {author} {\bibfnamefont {J.-M.}\ \bibnamefont {Chang}}, \bibinfo
  {author} {\bibfnamefont {D.~I.}\ \bibnamefont {Cattoni}}, \bibinfo {author}
  {\bibfnamefont {G.~L.}\ \bibnamefont {Papadopoulos}}, \bibinfo {author}
  {\bibfnamefont {B.}~\bibnamefont {Bonev}}, \bibinfo {author} {\bibfnamefont
  {T.}~\bibnamefont {Sexton}}, \bibinfo {author} {\bibfnamefont
  {J.}~\bibnamefont {Gurgo}}, \bibinfo {author} {\bibfnamefont
  {C.}~\bibnamefont {Jacquier}}, \bibinfo {author} {\bibfnamefont
  {M.}~\bibnamefont {Nollmann}}, \bibinfo {author} {\bibfnamefont
  {F.}~\bibnamefont {Bantignies}},\ and\ \bibinfo {author} {\bibfnamefont
  {G.}~\bibnamefont {Cavalli}},\ }\bibfield  {title} {\bibinfo {title} {{TADs}
  are {3D} structural units of higher-order chromosome organization in
  \textit{{Drosophila}}},\ }\href {https://doi.org/10.1126/sciadv.aar8082}
  {\bibfield  {journal} {\bibinfo  {journal} {Science Advances}\ }\textbf
  {\bibinfo {volume} {4}},\ \bibinfo {pages} {eaar8082} (\bibinfo {year}
  {2018})}\BibitemShut {NoStop}%
\bibitem [{\citenamefont {Sexton}\ \emph {et~al.}(2012)\citenamefont {Sexton},
  \citenamefont {Yaffe}, \citenamefont {Kenigsberg}, \citenamefont
  {Bantignies}, \citenamefont {Leblanc}, \citenamefont {Hoichman},
  \citenamefont {Parrinello}, \citenamefont {Tanay},\ and\ \citenamefont
  {Cavalli}}]{Sexton_2012}%
  \BibitemOpen
  \bibfield  {author} {\bibinfo {author} {\bibfnamefont {T.}~\bibnamefont
  {Sexton}}, \bibinfo {author} {\bibfnamefont {E.}~\bibnamefont {Yaffe}},
  \bibinfo {author} {\bibfnamefont {E.}~\bibnamefont {Kenigsberg}}, \bibinfo
  {author} {\bibfnamefont {F.}~\bibnamefont {Bantignies}}, \bibinfo {author}
  {\bibfnamefont {B.}~\bibnamefont {Leblanc}}, \bibinfo {author} {\bibfnamefont
  {M.}~\bibnamefont {Hoichman}}, \bibinfo {author} {\bibfnamefont
  {H.}~\bibnamefont {Parrinello}}, \bibinfo {author} {\bibfnamefont
  {A.}~\bibnamefont {Tanay}},\ and\ \bibinfo {author} {\bibfnamefont
  {G.}~\bibnamefont {Cavalli}},\ }\bibfield  {title} {\bibinfo {title}
  {Three-{Dimensional} {Folding} and {Functional} {Organization} {Principles}
  of the {Drosophila} {Genome}},\ }\href
  {https://doi.org/10.1016/j.cell.2012.01.010} {\bibfield  {journal} {\bibinfo
  {journal} {Cell}\ }\textbf {\bibinfo {volume} {148}},\ \bibinfo {pages} {458}
  (\bibinfo {year} {2012})}\BibitemShut {NoStop}%
\bibitem [{\citenamefont {Arnould}\ and\ \citenamefont
  {Legube}(2020)}]{Arnould_2020}%
  \BibitemOpen
  \bibfield  {author} {\bibinfo {author} {\bibfnamefont {C.}~\bibnamefont
  {Arnould}}\ and\ \bibinfo {author} {\bibfnamefont {G.}~\bibnamefont
  {Legube}},\ }\bibfield  {title} {\bibinfo {title} {The {Secret} {Life} of
  {Chromosome} {Loops} upon {DNA} {Double}-{Strand} {Break}},\ }\href
  {https://doi.org/10.1016/j.jmb.2019.07.036} {\bibfield  {journal} {\bibinfo
  {journal} {Journal of Molecular Biology}\ }\textbf {\bibinfo {volume}
  {432}},\ \bibinfo {pages} {724} (\bibinfo {year} {2020})}\BibitemShut
  {NoStop}%
\bibitem [{\citenamefont {Arnould}\ \emph {et~al.}(2021)\citenamefont
  {Arnould}, \citenamefont {Rocher}, \citenamefont {Finoux}, \citenamefont
  {Clouaire}, \citenamefont {Li}, \citenamefont {Zhou}, \citenamefont {Caron},
  \citenamefont {Mangeot}, \citenamefont {Ricci}, \citenamefont {Mourad},
  \citenamefont {Haber}, \citenamefont {Noordermeer},\ and\ \citenamefont
  {Legube}}]{Arnould_2021}%
  \BibitemOpen
  \bibfield  {author} {\bibinfo {author} {\bibfnamefont {C.}~\bibnamefont
  {Arnould}}, \bibinfo {author} {\bibfnamefont {V.}~\bibnamefont {Rocher}},
  \bibinfo {author} {\bibfnamefont {A.-L.}\ \bibnamefont {Finoux}}, \bibinfo
  {author} {\bibfnamefont {T.}~\bibnamefont {Clouaire}}, \bibinfo {author}
  {\bibfnamefont {K.}~\bibnamefont {Li}}, \bibinfo {author} {\bibfnamefont
  {F.}~\bibnamefont {Zhou}}, \bibinfo {author} {\bibfnamefont {P.}~\bibnamefont
  {Caron}}, \bibinfo {author} {\bibfnamefont {P.~E.}\ \bibnamefont {Mangeot}},
  \bibinfo {author} {\bibfnamefont {E.~P.}\ \bibnamefont {Ricci}}, \bibinfo
  {author} {\bibfnamefont {R.}~\bibnamefont {Mourad}}, \bibinfo {author}
  {\bibfnamefont {J.~E.}\ \bibnamefont {Haber}}, \bibinfo {author}
  {\bibfnamefont {D.}~\bibnamefont {Noordermeer}},\ and\ \bibinfo {author}
  {\bibfnamefont {G.}~\bibnamefont {Legube}},\ }\bibfield  {title} {\bibinfo
  {title} {Loop extrusion as a mechanism for formation of {DNA} damage repair
  foci},\ }\href {https://doi.org/10.1038/s41586-021-03193-z} {\bibfield
  {journal} {\bibinfo  {journal} {Nature}\ }\textbf {\bibinfo {volume} {590}},\
  \bibinfo {pages} {660} (\bibinfo {year} {2021})}\BibitemShut {NoStop}%
\bibitem [{\citenamefont {Arnould}\ \emph {et~al.}(2023)\citenamefont
  {Arnould}, \citenamefont {Rocher}, \citenamefont {Saur}, \citenamefont
  {Bader}, \citenamefont {Muzzopappa}, \citenamefont {Collins}, \citenamefont
  {Lesage}, \citenamefont {Le~Bozec}, \citenamefont {Puget}, \citenamefont
  {Clouaire}, \citenamefont {Mangeat}, \citenamefont {Mourad}, \citenamefont
  {Ahituv}, \citenamefont {Noordermeer}, \citenamefont {Erdel}, \citenamefont
  {Bushell}, \citenamefont {Marnef},\ and\ \citenamefont
  {Legube}}]{Arnould_2023}%
  \BibitemOpen
  \bibfield  {author} {\bibinfo {author} {\bibfnamefont {C.}~\bibnamefont
  {Arnould}}, \bibinfo {author} {\bibfnamefont {V.}~\bibnamefont {Rocher}},
  \bibinfo {author} {\bibfnamefont {F.}~\bibnamefont {Saur}}, \bibinfo {author}
  {\bibfnamefont {A.~S.}\ \bibnamefont {Bader}}, \bibinfo {author}
  {\bibfnamefont {F.}~\bibnamefont {Muzzopappa}}, \bibinfo {author}
  {\bibfnamefont {S.}~\bibnamefont {Collins}}, \bibinfo {author} {\bibfnamefont
  {E.}~\bibnamefont {Lesage}}, \bibinfo {author} {\bibfnamefont
  {B.}~\bibnamefont {Le~Bozec}}, \bibinfo {author} {\bibfnamefont
  {N.}~\bibnamefont {Puget}}, \bibinfo {author} {\bibfnamefont
  {T.}~\bibnamefont {Clouaire}}, \bibinfo {author} {\bibfnamefont
  {T.}~\bibnamefont {Mangeat}}, \bibinfo {author} {\bibfnamefont
  {R.}~\bibnamefont {Mourad}}, \bibinfo {author} {\bibfnamefont
  {N.}~\bibnamefont {Ahituv}}, \bibinfo {author} {\bibfnamefont
  {D.}~\bibnamefont {Noordermeer}}, \bibinfo {author} {\bibfnamefont
  {F.}~\bibnamefont {Erdel}}, \bibinfo {author} {\bibfnamefont
  {M.}~\bibnamefont {Bushell}}, \bibinfo {author} {\bibfnamefont
  {A.}~\bibnamefont {Marnef}},\ and\ \bibinfo {author} {\bibfnamefont
  {G.}~\bibnamefont {Legube}},\ }\bibfield  {title} {\bibinfo {title}
  {Chromatin compartmentalization regulates the response to {DNA} damage},\
  }\href {https://doi.org/10.1038/s41586-023-06635-y} {\bibfield  {journal}
  {\bibinfo  {journal} {Nature}\ }\textbf {\bibinfo {volume} {623}},\ \bibinfo
  {pages} {183} (\bibinfo {year} {2023})}\BibitemShut {NoStop}%
\bibitem [{\citenamefont {Shin}\ and\ \citenamefont
  {Kolomeisky}(2019)}]{Shin_2019}%
  \BibitemOpen
  \bibfield  {author} {\bibinfo {author} {\bibfnamefont {J.}~\bibnamefont
  {Shin}}\ and\ \bibinfo {author} {\bibfnamefont {A.~B.}\ \bibnamefont
  {Kolomeisky}},\ }\bibfield  {title} {\bibinfo {title} {Facilitation of {DNA}
  loop formation by protein-{DNA} non-specific interactions},\ }\href
  {https://doi.org/10.1039/C9SM00671K} {\bibfield  {journal} {\bibinfo
  {journal} {Soft Matter}\ }\textbf {\bibinfo {volume} {15}},\ \bibinfo {pages}
  {5255} (\bibinfo {year} {2019})}\BibitemShut {NoStop}%
\bibitem [{\citenamefont {Keber}\ \emph {et~al.}(2024)\citenamefont {Keber},
  \citenamefont {Nguyen}, \citenamefont {Mariossi}, \citenamefont
  {Brangwynne},\ and\ \citenamefont {Wühr}}]{Keber_2024}%
  \BibitemOpen
  \bibfield  {author} {\bibinfo {author} {\bibfnamefont {F.~C.}\ \bibnamefont
  {Keber}}, \bibinfo {author} {\bibfnamefont {T.}~\bibnamefont {Nguyen}},
  \bibinfo {author} {\bibfnamefont {A.}~\bibnamefont {Mariossi}}, \bibinfo
  {author} {\bibfnamefont {C.~P.}\ \bibnamefont {Brangwynne}},\ and\ \bibinfo
  {author} {\bibfnamefont {M.}~\bibnamefont {Wühr}},\ }\bibfield  {title}
  {\bibinfo {title} {Evidence for widespread cytoplasmic structuring into
  mesoscale condensates},\ }\href {https://doi.org/10.1038/s41556-024-01363-5}
  {\bibfield  {journal} {\bibinfo  {journal} {Nature Cell Biology}\ }\textbf
  {\bibinfo {volume} {26}},\ \bibinfo {pages} {346} (\bibinfo {year}
  {2024})}\BibitemShut {NoStop}%
\bibitem [{\citenamefont {Broedersz}\ \emph {et~al.}(2014)\citenamefont
  {Broedersz}, \citenamefont {Wang}, \citenamefont {Meir}, \citenamefont
  {Loparo}, \citenamefont {Rudner},\ and\ \citenamefont
  {Wingreen}}]{Broedersz_2014}%
  \BibitemOpen
  \bibfield  {author} {\bibinfo {author} {\bibfnamefont {C.~P.}\ \bibnamefont
  {Broedersz}}, \bibinfo {author} {\bibfnamefont {X.}~\bibnamefont {Wang}},
  \bibinfo {author} {\bibfnamefont {Y.}~\bibnamefont {Meir}}, \bibinfo {author}
  {\bibfnamefont {J.~J.}\ \bibnamefont {Loparo}}, \bibinfo {author}
  {\bibfnamefont {D.~Z.}\ \bibnamefont {Rudner}},\ and\ \bibinfo {author}
  {\bibfnamefont {N.~S.}\ \bibnamefont {Wingreen}},\ }\bibfield  {title}
  {\bibinfo {title} {Condensation and localization of the partitioning protein
  {ParB} on the bacterial chromosome},\ }\href
  {https://doi.org/10.1073/pnas.1402529111} {\bibfield  {journal} {\bibinfo
  {journal} {Proceedings of the National Academy of Sciences}\ }\textbf
  {\bibinfo {volume} {111}},\ \bibinfo {pages} {8809} (\bibinfo {year}
  {2014})}\BibitemShut {NoStop}%
\bibitem [{\citenamefont {Zwicker}(2022)}]{Zwicker_2022}%
  \BibitemOpen
  \bibfield  {author} {\bibinfo {author} {\bibfnamefont {D.}~\bibnamefont
  {Zwicker}},\ }\bibfield  {title} {\bibinfo {title} {The intertwined physics
  of active chemical reactions and phase separation},\ }\href
  {https://doi.org/10.1016/j.cocis.2022.101606} {\bibfield  {journal} {\bibinfo
   {journal} {Current Opinion in Colloid \& Interface Science}\ }\textbf
  {\bibinfo {volume} {61}},\ \bibinfo {pages} {101606} (\bibinfo {year}
  {2022})}\BibitemShut {NoStop}%
\bibitem [{\citenamefont {Strom}\ \emph {et~al.}(2024)\citenamefont {Strom},
  \citenamefont {Eeftens}, \citenamefont {Polyachenko}, \citenamefont {Weaver},
  \citenamefont {Watanabe}, \citenamefont {Bracha}, \citenamefont {Orlovsky},
  \citenamefont {Jumper}, \citenamefont {Jacobs},\ and\ \citenamefont
  {Brangwynne}}]{Strom_2024}%
  \BibitemOpen
  \bibfield  {author} {\bibinfo {author} {\bibfnamefont {A.~R.}\ \bibnamefont
  {Strom}}, \bibinfo {author} {\bibfnamefont {J.~M.}\ \bibnamefont {Eeftens}},
  \bibinfo {author} {\bibfnamefont {Y.}~\bibnamefont {Polyachenko}}, \bibinfo
  {author} {\bibfnamefont {C.~J.}\ \bibnamefont {Weaver}}, \bibinfo {author}
  {\bibfnamefont {H.-F.}\ \bibnamefont {Watanabe}}, \bibinfo {author}
  {\bibfnamefont {D.}~\bibnamefont {Bracha}}, \bibinfo {author} {\bibfnamefont
  {N.~D.}\ \bibnamefont {Orlovsky}}, \bibinfo {author} {\bibfnamefont {C.~C.}\
  \bibnamefont {Jumper}}, \bibinfo {author} {\bibfnamefont {W.~M.}\
  \bibnamefont {Jacobs}},\ and\ \bibinfo {author} {\bibfnamefont {C.~P.}\
  \bibnamefont {Brangwynne}},\ }\bibfield  {title} {\bibinfo {title} {Interplay
  of condensation and chromatin binding underlies {BRD4} targeting},\ }\href
  {https://doi.org/10.1091/mbc.E24-01-0046} {\bibfield  {journal} {\bibinfo
  {journal} {Molecular Biology of the Cell}\ }\textbf {\bibinfo {volume}
  {35}},\ \bibinfo {pages} {ar88} (\bibinfo {year} {2024})}\BibitemShut
  {NoStop}%
\bibitem [{\citenamefont {Du}\ \emph {et~al.}(2024)\citenamefont {Du},
  \citenamefont {Stitzinger}, \citenamefont {Spille}, \citenamefont {Cho},
  \citenamefont {Lee}, \citenamefont {Hijaz}, \citenamefont {Quintana},\ and\
  \citenamefont {Cissé}}]{Du_2024}%
  \BibitemOpen
  \bibfield  {author} {\bibinfo {author} {\bibfnamefont {M.}~\bibnamefont
  {Du}}, \bibinfo {author} {\bibfnamefont {S.~H.}\ \bibnamefont {Stitzinger}},
  \bibinfo {author} {\bibfnamefont {J.-H.}\ \bibnamefont {Spille}}, \bibinfo
  {author} {\bibfnamefont {W.-K.}\ \bibnamefont {Cho}}, \bibinfo {author}
  {\bibfnamefont {C.}~\bibnamefont {Lee}}, \bibinfo {author} {\bibfnamefont
  {M.}~\bibnamefont {Hijaz}}, \bibinfo {author} {\bibfnamefont
  {A.}~\bibnamefont {Quintana}},\ and\ \bibinfo {author} {\bibfnamefont
  {I.~I.}\ \bibnamefont {Cissé}},\ }\bibfield  {title} {\bibinfo {title}
  {Direct observation of a condensate effect on super-enhancer controlled gene
  bursting},\ }\href {https://doi.org/10.1016/j.cell.2023.12.005} {\bibfield
  {journal} {\bibinfo  {journal} {Cell}\ }\textbf {\bibinfo {volume} {187}},\
  \bibinfo {pages} {331} (\bibinfo {year} {2024})}\BibitemShut {NoStop}%
\bibitem [{\citenamefont {Rayleigh}(1878)}]{Rayleigh_1878}%
  \BibitemOpen
  \bibfield  {author} {\bibinfo {author} {\bibfnamefont {L.}~\bibnamefont
  {Rayleigh}},\ }\bibfield  {title} {\bibinfo {title} {On {The} {Instability}
  {Of} {Jets}},\ }\href {https://doi.org/10.1112/plms/s1-10.1.4} {\bibfield
  {journal} {\bibinfo  {journal} {Proceedings of the London Mathematical
  Society}\ }\textbf {\bibinfo {volume} {s1-10}},\ \bibinfo {pages} {4}
  (\bibinfo {year} {1878})}\BibitemShut {NoStop}%
\bibitem [{\citenamefont {De~Gennes}\ \emph {et~al.}(2004)\citenamefont
  {De~Gennes}, \citenamefont {Brochard-Wyart},\ and\ \citenamefont
  {Quéré}}]{DeGennes_2004}%
  \BibitemOpen
  \bibfield  {author} {\bibinfo {author} {\bibfnamefont {P.-G.}\ \bibnamefont
  {De~Gennes}}, \bibinfo {author} {\bibfnamefont {F.}~\bibnamefont
  {Brochard-Wyart}},\ and\ \bibinfo {author} {\bibfnamefont {D.}~\bibnamefont
  {Quéré}},\ }\href {https://doi.org/10.1007/978-0-387-21656-0} {\emph
  {\bibinfo {title} {Capillarity and {Wetting} {Phenomena}}}}\ (\bibinfo
  {publisher} {Springer New York},\ \bibinfo {address} {New York, NY},\
  \bibinfo {year} {2004})\BibitemShut {NoStop}%
\bibitem [{\citenamefont {Eggers}\ and\ \citenamefont
  {Villermaux}(2008)}]{Eggers_2008}%
  \BibitemOpen
  \bibfield  {author} {\bibinfo {author} {\bibfnamefont {J.}~\bibnamefont
  {Eggers}}\ and\ \bibinfo {author} {\bibfnamefont {E.}~\bibnamefont
  {Villermaux}},\ }\bibfield  {title} {\bibinfo {title} {Physics of liquid
  jets},\ }\href {https://doi.org/10.1088/0034-4885/71/3/036601} {\bibfield
  {journal} {\bibinfo  {journal} {Reports on Progress in Physics}\ }\textbf
  {\bibinfo {volume} {71}},\ \bibinfo {pages} {036601} (\bibinfo {year}
  {2008})}\BibitemShut {NoStop}%
\bibitem [{\citenamefont {Zhang}\ \emph
  {et~al.}(2021{\natexlab{a}})\citenamefont {Zhang}, \citenamefont
  {Sprittles},\ and\ \citenamefont {Lockerby}}]{Zhang_2021}%
  \BibitemOpen
  \bibfield  {author} {\bibinfo {author} {\bibfnamefont {Y.}~\bibnamefont
  {Zhang}}, \bibinfo {author} {\bibfnamefont {J.}~\bibnamefont {Sprittles}},\
  and\ \bibinfo {author} {\bibfnamefont {D.}~\bibnamefont {Lockerby}},\
  }\bibfield  {title} {\bibinfo {title} {Thermal capillary wave growth and
  surface roughening of nanoscale liquid films},\ }\href
  {https://doi.org/10.1017/jfm.2021.164} {\bibfield  {journal} {\bibinfo
  {journal} {Journal of Fluid Mechanics}\ }\textbf {\bibinfo {volume} {915}},\
  \bibinfo {pages} {A135} (\bibinfo {year} {2021}{\natexlab{a}})}\BibitemShut
  {NoStop}%
\bibitem [{\citenamefont {Gopan}\ and\ \citenamefont
  {Sathian}(2014)}]{Gopan_2014}%
  \BibitemOpen
  \bibfield  {author} {\bibinfo {author} {\bibfnamefont {N.}~\bibnamefont
  {Gopan}}\ and\ \bibinfo {author} {\bibfnamefont {S.~P.}\ \bibnamefont
  {Sathian}},\ }\bibfield  {title} {\bibinfo {title} {Rayleigh instability at
  small length scales},\ }\href {https://doi.org/10.1103/PhysRevE.90.033001}
  {\bibfield  {journal} {\bibinfo  {journal} {Physical Review E}\ }\textbf
  {\bibinfo {volume} {90}},\ \bibinfo {pages} {033001} (\bibinfo {year}
  {2014})}\BibitemShut {NoStop}%
\bibitem [{\citenamefont {Zhang}\ \emph {et~al.}(2020)\citenamefont {Zhang},
  \citenamefont {Sprittles},\ and\ \citenamefont {Lockerby}}]{Zhang_2020}%
  \BibitemOpen
  \bibfield  {author} {\bibinfo {author} {\bibfnamefont {Y.}~\bibnamefont
  {Zhang}}, \bibinfo {author} {\bibfnamefont {J.~E.}\ \bibnamefont
  {Sprittles}},\ and\ \bibinfo {author} {\bibfnamefont {D.~A.}\ \bibnamefont
  {Lockerby}},\ }\bibfield  {title} {\bibinfo {title} {Nanoscale thin-film
  flows with thermal fluctuations and slip},\ }\href
  {https://doi.org/10.1103/PhysRevE.102.053105} {\bibfield  {journal} {\bibinfo
   {journal} {Physical Review E}\ }\textbf {\bibinfo {volume} {102}},\ \bibinfo
  {pages} {053105} (\bibinfo {year} {2020})}\BibitemShut {NoStop}%
\bibitem [{\citenamefont {Flory}(1942)}]{Flory_1942}%
  \BibitemOpen
  \bibfield  {author} {\bibinfo {author} {\bibfnamefont {P.~J.}\ \bibnamefont
  {Flory}},\ }\bibfield  {title} {\bibinfo {title} {Thermodynamics of {High}
  {Polymer} {Solutions}},\ }\href {https://doi.org/10.1063/1.1723621}
  {\bibfield  {journal} {\bibinfo  {journal} {The Journal of Chemical Physics}\
  }\textbf {\bibinfo {volume} {10}},\ \bibinfo {pages} {51} (\bibinfo {year}
  {1942})}\BibitemShut {NoStop}%
\bibitem [{\citenamefont {Huggins}(1942)}]{Huggins_1942}%
  \BibitemOpen
  \bibfield  {author} {\bibinfo {author} {\bibfnamefont {M.~L.}\ \bibnamefont
  {Huggins}},\ }\bibfield  {title} {\bibinfo {title} {Some {Properties} of
  {Solutions} of {Long}-chain {Compounds}.},\ }\href
  {https://doi.org/10.1021/j150415a018} {\bibfield  {journal} {\bibinfo
  {journal} {The Journal of Physical Chemistry}\ }\textbf {\bibinfo {volume}
  {46}},\ \bibinfo {pages} {151} (\bibinfo {year} {1942})}\BibitemShut
  {NoStop}%
\bibitem [{\citenamefont {Tiani}\ \emph {et~al.}(2025)\citenamefont {Tiani},
  \citenamefont {Jardat},\ and\ \citenamefont {Dahirel}}]{Tiani_2025}%
  \BibitemOpen
  \bibfield  {author} {\bibinfo {author} {\bibfnamefont {R.}~\bibnamefont
  {Tiani}}, \bibinfo {author} {\bibfnamefont {M.}~\bibnamefont {Jardat}},\ and\
  \bibinfo {author} {\bibfnamefont {V.}~\bibnamefont {Dahirel}},\ }\bibfield
  {title} {\bibinfo {title} {Phase transitions in chromatin: {Mesoscopic} and
  mean-field approaches},\ }\href@noop {} {\bibfield  {journal} {\bibinfo
  {journal} {The Journal}\ } (\bibinfo {year} {2025})}\BibitemShut {NoStop}%
\bibitem [{\citenamefont {Ronceray}\ \emph {et~al.}(2022)\citenamefont
  {Ronceray}, \citenamefont {Mao}, \citenamefont {Košmrlj},\ and\
  \citenamefont {Haataja}}]{Ronceray_liquid_2022}%
  \BibitemOpen
  \bibfield  {author} {\bibinfo {author} {\bibfnamefont {P.}~\bibnamefont
  {Ronceray}}, \bibinfo {author} {\bibfnamefont {S.}~\bibnamefont {Mao}},
  \bibinfo {author} {\bibfnamefont {A.}~\bibnamefont {Košmrlj}},\ and\
  \bibinfo {author} {\bibfnamefont {M.~P.}\ \bibnamefont {Haataja}},\
  }\bibfield  {title} {\bibinfo {title} {Liquid demixing in elastic networks:
  {Cavitation}, permeation, or size selection?},\ }\href
  {https://doi.org/10.1209/0295-5075/ac56ac} {\bibfield  {journal} {\bibinfo
  {journal} {Europhysics Letters}\ }\textbf {\bibinfo {volume} {137}},\
  \bibinfo {pages} {67001} (\bibinfo {year} {2022})}\BibitemShut {NoStop}%
\bibitem [{\citenamefont {Rosowski}\ \emph {et~al.}(2020)\citenamefont
  {Rosowski}, \citenamefont {Sai}, \citenamefont {Vidal-Henriquez},
  \citenamefont {Zwicker}, \citenamefont {Style},\ and\ \citenamefont
  {Dufresne}}]{Rosowski_2020}%
  \BibitemOpen
  \bibfield  {author} {\bibinfo {author} {\bibfnamefont {K.~A.}\ \bibnamefont
  {Rosowski}}, \bibinfo {author} {\bibfnamefont {T.}~\bibnamefont {Sai}},
  \bibinfo {author} {\bibfnamefont {E.}~\bibnamefont {Vidal-Henriquez}},
  \bibinfo {author} {\bibfnamefont {D.}~\bibnamefont {Zwicker}}, \bibinfo
  {author} {\bibfnamefont {R.~W.}\ \bibnamefont {Style}},\ and\ \bibinfo
  {author} {\bibfnamefont {E.~R.}\ \bibnamefont {Dufresne}},\ }\bibfield
  {title} {\bibinfo {title} {Elastic ripening and inhibition of liquid–liquid
  phase separation},\ }\href {https://doi.org/10.1038/s41567-019-0767-2}
  {\bibfield  {journal} {\bibinfo  {journal} {Nature Physics}\ }\textbf
  {\bibinfo {volume} {16}},\ \bibinfo {pages} {422} (\bibinfo {year}
  {2020})}\BibitemShut {NoStop}%
\bibitem [{\citenamefont {Style}\ \emph {et~al.}(2018)\citenamefont {Style},
  \citenamefont {Sai}, \citenamefont {Fanelli}, \citenamefont {Ijavi},
  \citenamefont {Smith-Mannschott}, \citenamefont {Xu}, \citenamefont {Wilen},\
  and\ \citenamefont {Dufresne}}]{Style_2018}%
  \BibitemOpen
  \bibfield  {author} {\bibinfo {author} {\bibfnamefont {R.~W.}\ \bibnamefont
  {Style}}, \bibinfo {author} {\bibfnamefont {T.}~\bibnamefont {Sai}}, \bibinfo
  {author} {\bibfnamefont {N.}~\bibnamefont {Fanelli}}, \bibinfo {author}
  {\bibfnamefont {M.}~\bibnamefont {Ijavi}}, \bibinfo {author} {\bibfnamefont
  {K.}~\bibnamefont {Smith-Mannschott}}, \bibinfo {author} {\bibfnamefont
  {Q.}~\bibnamefont {Xu}}, \bibinfo {author} {\bibfnamefont {L.~A.}\
  \bibnamefont {Wilen}},\ and\ \bibinfo {author} {\bibfnamefont {E.~R.}\
  \bibnamefont {Dufresne}},\ }\bibfield  {title} {\bibinfo {title}
  {Liquid-{Liquid} {Phase} {Separation} in an {Elastic} {Network}},\ }\href
  {https://doi.org/10.1103/PhysRevX.8.011028} {\bibfield  {journal} {\bibinfo
  {journal} {Physical Review X}\ }\textbf {\bibinfo {volume} {8}},\ \bibinfo
  {pages} {011028} (\bibinfo {year} {2018})}\BibitemShut {NoStop}%
\bibitem [{\citenamefont {Lee}\ \emph {et~al.}(2021)\citenamefont {Lee},
  \citenamefont {Wingreen},\ and\ \citenamefont {Brangwynne}}]{Lee_2021}%
  \BibitemOpen
  \bibfield  {author} {\bibinfo {author} {\bibfnamefont {D.~S.~W.}\
  \bibnamefont {Lee}}, \bibinfo {author} {\bibfnamefont {N.~S.}\ \bibnamefont
  {Wingreen}},\ and\ \bibinfo {author} {\bibfnamefont {C.~P.}\ \bibnamefont
  {Brangwynne}},\ }\bibfield  {title} {\bibinfo {title} {Chromatin mechanics
  dictates subdiffusion and coarsening dynamics of embedded condensates},\
  }\href {https://doi.org/10.1038/s41567-020-01125-8} {\bibfield  {journal}
  {\bibinfo  {journal} {Nature Physics}\ }\textbf {\bibinfo {volume} {17}},\
  \bibinfo {pages} {531} (\bibinfo {year} {2021})}\BibitemShut {NoStop}%
\bibitem [{\citenamefont {Qi}\ and\ \citenamefont {Zhang}(2021)}]{Qi_2021}%
  \BibitemOpen
  \bibfield  {author} {\bibinfo {author} {\bibfnamefont {Y.}~\bibnamefont
  {Qi}}\ and\ \bibinfo {author} {\bibfnamefont {B.}~\bibnamefont {Zhang}},\
  }\bibfield  {title} {\bibinfo {title} {Chromatin network retards nucleoli
  coalescence},\ }\href {https://doi.org/10.1038/s41467-021-27123-9} {\bibfield
   {journal} {\bibinfo  {journal} {Nature Communications}\ }\textbf {\bibinfo
  {volume} {12}},\ \bibinfo {pages} {6824} (\bibinfo {year}
  {2021})}\BibitemShut {NoStop}%
\bibitem [{\citenamefont {Allen}\ and\ \citenamefont
  {Tildesley}(2017)}]{Allen_2017}%
  \BibitemOpen
  \bibfield  {author} {\bibinfo {author} {\bibfnamefont {M.~P.}\ \bibnamefont
  {Allen}}\ and\ \bibinfo {author} {\bibfnamefont {D.~J.}\ \bibnamefont
  {Tildesley}},\ }\href@noop {} {\emph {\bibinfo {title} {Computer simulation
  of liquids}}},\ \bibinfo {edition} {second edition}\ ed.\ (\bibinfo
  {publisher} {Oxford University Press},\ \bibinfo {address} {Oxford},\
  \bibinfo {year} {2017})\BibitemShut {NoStop}%
\bibitem [{\citenamefont {Thompson}\ \emph {et~al.}(2022)\citenamefont
  {Thompson}, \citenamefont {Aktulga}, \citenamefont {Berger}, \citenamefont
  {Bolintineanu}, \citenamefont {Brown}, \citenamefont {Crozier}, \citenamefont
  {In~'T~Veld}, \citenamefont {Kohlmeyer}, \citenamefont {Moore}, \citenamefont
  {Nguyen}, \citenamefont {Shan}, \citenamefont {Stevens}, \citenamefont
  {Tranchida}, \citenamefont {Trott},\ and\ \citenamefont
  {Plimpton}}]{Thompson_2022}%
  \BibitemOpen
  \bibfield  {author} {\bibinfo {author} {\bibfnamefont {A.~P.}\ \bibnamefont
  {Thompson}}, \bibinfo {author} {\bibfnamefont {H.~M.}\ \bibnamefont
  {Aktulga}}, \bibinfo {author} {\bibfnamefont {R.}~\bibnamefont {Berger}},
  \bibinfo {author} {\bibfnamefont {D.~S.}\ \bibnamefont {Bolintineanu}},
  \bibinfo {author} {\bibfnamefont {W.~M.}\ \bibnamefont {Brown}}, \bibinfo
  {author} {\bibfnamefont {P.~S.}\ \bibnamefont {Crozier}}, \bibinfo {author}
  {\bibfnamefont {P.~J.}\ \bibnamefont {In~'T~Veld}}, \bibinfo {author}
  {\bibfnamefont {A.}~\bibnamefont {Kohlmeyer}}, \bibinfo {author}
  {\bibfnamefont {S.~G.}\ \bibnamefont {Moore}}, \bibinfo {author}
  {\bibfnamefont {T.~D.}\ \bibnamefont {Nguyen}}, \bibinfo {author}
  {\bibfnamefont {R.}~\bibnamefont {Shan}}, \bibinfo {author} {\bibfnamefont
  {M.~J.}\ \bibnamefont {Stevens}}, \bibinfo {author} {\bibfnamefont
  {J.}~\bibnamefont {Tranchida}}, \bibinfo {author} {\bibfnamefont
  {C.}~\bibnamefont {Trott}},\ and\ \bibinfo {author} {\bibfnamefont {S.~J.}\
  \bibnamefont {Plimpton}},\ }\bibfield  {title} {\bibinfo {title} {{LAMMPS} -
  a flexible simulation tool for particle-based materials modeling at the
  atomic, meso, and continuum scales},\ }\href
  {https://doi.org/10.1016/j.cpc.2021.108171} {\bibfield  {journal} {\bibinfo
  {journal} {Computer Physics Communications}\ }\textbf {\bibinfo {volume}
  {271}},\ \bibinfo {pages} {108171} (\bibinfo {year} {2022})}\BibitemShut
  {NoStop}%
\bibitem [{\citenamefont {Brackley}\ \emph {et~al.}(2016)\citenamefont
  {Brackley}, \citenamefont {Michieletto}, \citenamefont {Mouvet},
  \citenamefont {Johnson}, \citenamefont {Kelly}, \citenamefont {Cook},\ and\
  \citenamefont {Marenduzzo}}]{Brackley_2016}%
  \BibitemOpen
  \bibfield  {author} {\bibinfo {author} {\bibfnamefont {C.~A.}\ \bibnamefont
  {Brackley}}, \bibinfo {author} {\bibfnamefont {D.}~\bibnamefont
  {Michieletto}}, \bibinfo {author} {\bibfnamefont {F.}~\bibnamefont {Mouvet}},
  \bibinfo {author} {\bibfnamefont {J.}~\bibnamefont {Johnson}}, \bibinfo
  {author} {\bibfnamefont {S.}~\bibnamefont {Kelly}}, \bibinfo {author}
  {\bibfnamefont {P.~R.}\ \bibnamefont {Cook}},\ and\ \bibinfo {author}
  {\bibfnamefont {D.}~\bibnamefont {Marenduzzo}},\ }\bibfield  {title}
  {\bibinfo {title} {Simulating topological domains in human chromosomes with a
  fitting-free model},\ }\href {https://doi.org/10.1080/19491034.2016.1239684}
  {\bibfield  {journal} {\bibinfo  {journal} {Nucleus}\ }\textbf {\bibinfo
  {volume} {7}},\ \bibinfo {pages} {453} (\bibinfo {year} {2016})}\BibitemShut
  {NoStop}%
\bibitem [{\citenamefont {Zhang}\ \emph
  {et~al.}(2021{\natexlab{b}})\citenamefont {Zhang}, \citenamefont {Lee},
  \citenamefont {Meir}, \citenamefont {Brangwynne},\ and\ \citenamefont
  {Wingreen}}]{Zhang_mechanical_2021}%
  \BibitemOpen
  \bibfield  {author} {\bibinfo {author} {\bibfnamefont {Y.}~\bibnamefont
  {Zhang}}, \bibinfo {author} {\bibfnamefont {D.~S.}\ \bibnamefont {Lee}},
  \bibinfo {author} {\bibfnamefont {Y.}~\bibnamefont {Meir}}, \bibinfo {author}
  {\bibfnamefont {C.~P.}\ \bibnamefont {Brangwynne}},\ and\ \bibinfo {author}
  {\bibfnamefont {N.~S.}\ \bibnamefont {Wingreen}},\ }\bibfield  {title}
  {\bibinfo {title} {Mechanical {Frustration} of {Phase} {Separation} in the
  {Cell} {Nucleus} by {Chromatin}},\ }\href
  {https://doi.org/10.1103/PhysRevLett.126.258102} {\bibfield  {journal}
  {\bibinfo  {journal} {Physical Review Letters}\ }\textbf {\bibinfo {volume}
  {126}},\ \bibinfo {pages} {258102} (\bibinfo {year}
  {2021}{\natexlab{b}})}\BibitemShut {NoStop}%
\bibitem [{\citenamefont {Tortora}\ \emph {et~al.}(2023)\citenamefont
  {Tortora}, \citenamefont {Brennan}, \citenamefont {Karpen},\ and\
  \citenamefont {Jost}}]{Tortora_2023}%
  \BibitemOpen
  \bibfield  {author} {\bibinfo {author} {\bibfnamefont {M.~M.~C.}\
  \bibnamefont {Tortora}}, \bibinfo {author} {\bibfnamefont {L.~D.}\
  \bibnamefont {Brennan}}, \bibinfo {author} {\bibfnamefont {G.}~\bibnamefont
  {Karpen}},\ and\ \bibinfo {author} {\bibfnamefont {D.}~\bibnamefont {Jost}},\
  }\bibfield  {title} {\bibinfo {title} {{HP1}-driven phase separation
  recapitulates the thermodynamics and kinetics of heterochromatin condensate
  formation},\ }\href {https://doi.org/10.1073/pnas.2211855120} {\bibfield
  {journal} {\bibinfo  {journal} {Proceedings of the National Academy of
  Sciences}\ }\textbf {\bibinfo {volume} {120}},\ \bibinfo {pages}
  {e2211855120} (\bibinfo {year} {2023})}\BibitemShut {NoStop}%
\bibitem [{\citenamefont {McGhee}\ and\ \citenamefont
  {Felsenfeld}(1980)}]{Mcghee_1980}%
  \BibitemOpen
  \bibfield  {author} {\bibinfo {author} {\bibfnamefont {J.~D.}\ \bibnamefont
  {McGhee}}\ and\ \bibinfo {author} {\bibfnamefont {G.}~\bibnamefont
  {Felsenfeld}},\ }\bibfield  {title} {\bibinfo {title} {Nucleosome
  {Structure}},\ }\href {https://doi.org/10.1146/annurev.bi.49.070180.005343}
  {\bibfield  {journal} {\bibinfo  {journal} {Annual Review of Biochemistry}\
  }\textbf {\bibinfo {volume} {49}},\ \bibinfo {pages} {1115} (\bibinfo {year}
  {1980})}\BibitemShut {NoStop}%
\bibitem [{\citenamefont {Olins}\ and\ \citenamefont
  {Olins}(2003)}]{Olins_2003}%
  \BibitemOpen
  \bibfield  {author} {\bibinfo {author} {\bibfnamefont {D.~E.}\ \bibnamefont
  {Olins}}\ and\ \bibinfo {author} {\bibfnamefont {A.~L.}\ \bibnamefont
  {Olins}},\ }\bibfield  {title} {\bibinfo {title} {Chromatin history: our view
  from the bridge},\ }\href {https://doi.org/10.1038/nrm1225} {\bibfield
  {journal} {\bibinfo  {journal} {Nature Reviews Molecular Cell Biology}\
  }\textbf {\bibinfo {volume} {4}},\ \bibinfo {pages} {809} (\bibinfo {year}
  {2003})}\BibitemShut {NoStop}%
\bibitem [{\citenamefont {Widom}(1992)}]{Widom_1992}%
  \BibitemOpen
  \bibfield  {author} {\bibinfo {author} {\bibfnamefont {J.}~\bibnamefont
  {Widom}},\ }\bibfield  {title} {\bibinfo {title} {A relationship between the
  helical twist of {DNA} and the ordered positioning of nucleosomes in all
  eukaryotic cells.},\ }\href {https://doi.org/10.1073/pnas.89.3.1095}
  {\bibfield  {journal} {\bibinfo  {journal} {Proceedings of the National
  Academy of Sciences}\ }\textbf {\bibinfo {volume} {89}},\ \bibinfo {pages}
  {1095} (\bibinfo {year} {1992})}\BibitemShut {NoStop}%
\bibitem [{\citenamefont {Schalch}\ \emph {et~al.}(2005)\citenamefont
  {Schalch}, \citenamefont {Duda}, \citenamefont {Sargent},\ and\ \citenamefont
  {Richmond}}]{Schalch_2005}%
  \BibitemOpen
  \bibfield  {author} {\bibinfo {author} {\bibfnamefont {T.}~\bibnamefont
  {Schalch}}, \bibinfo {author} {\bibfnamefont {S.}~\bibnamefont {Duda}},
  \bibinfo {author} {\bibfnamefont {D.~F.}\ \bibnamefont {Sargent}},\ and\
  \bibinfo {author} {\bibfnamefont {T.~J.}\ \bibnamefont {Richmond}},\
  }\bibfield  {title} {\bibinfo {title} {X-ray structure of a tetranucleosome
  and its implications for the chromatin fibre},\ }\href
  {https://doi.org/10.1038/nature03686} {\bibfield  {journal} {\bibinfo
  {journal} {Nature}\ }\textbf {\bibinfo {volume} {436}},\ \bibinfo {pages}
  {138} (\bibinfo {year} {2005})}\BibitemShut {NoStop}%
\bibitem [{\citenamefont {Lieberman-Aiden}\ \emph {et~al.}(2009)\citenamefont
  {Lieberman-Aiden}, \citenamefont {Van~Berkum}, \citenamefont {Williams},
  \citenamefont {Imakaev}, \citenamefont {Ragoczy}, \citenamefont {Telling},
  \citenamefont {Amit}, \citenamefont {Lajoie}, \citenamefont {Sabo},
  \citenamefont {Dorschner}, \citenamefont {Sandstrom}, \citenamefont
  {Bernstein}, \citenamefont {Bender}, \citenamefont {Groudine}, \citenamefont
  {Gnirke}, \citenamefont {Stamatoyannopoulos}, \citenamefont {Mirny},
  \citenamefont {Lander},\ and\ \citenamefont {Dekker}}]{Lieberman-aiden_2009}%
  \BibitemOpen
  \bibfield  {author} {\bibinfo {author} {\bibfnamefont {E.}~\bibnamefont
  {Lieberman-Aiden}}, \bibinfo {author} {\bibfnamefont {N.~L.}\ \bibnamefont
  {Van~Berkum}}, \bibinfo {author} {\bibfnamefont {L.}~\bibnamefont
  {Williams}}, \bibinfo {author} {\bibfnamefont {M.}~\bibnamefont {Imakaev}},
  \bibinfo {author} {\bibfnamefont {T.}~\bibnamefont {Ragoczy}}, \bibinfo
  {author} {\bibfnamefont {A.}~\bibnamefont {Telling}}, \bibinfo {author}
  {\bibfnamefont {I.}~\bibnamefont {Amit}}, \bibinfo {author} {\bibfnamefont
  {B.~R.}\ \bibnamefont {Lajoie}}, \bibinfo {author} {\bibfnamefont {P.~J.}\
  \bibnamefont {Sabo}}, \bibinfo {author} {\bibfnamefont {M.~O.}\ \bibnamefont
  {Dorschner}}, \bibinfo {author} {\bibfnamefont {R.}~\bibnamefont
  {Sandstrom}}, \bibinfo {author} {\bibfnamefont {B.}~\bibnamefont
  {Bernstein}}, \bibinfo {author} {\bibfnamefont {M.~A.}\ \bibnamefont
  {Bender}}, \bibinfo {author} {\bibfnamefont {M.}~\bibnamefont {Groudine}},
  \bibinfo {author} {\bibfnamefont {A.}~\bibnamefont {Gnirke}}, \bibinfo
  {author} {\bibfnamefont {J.}~\bibnamefont {Stamatoyannopoulos}}, \bibinfo
  {author} {\bibfnamefont {L.~A.}\ \bibnamefont {Mirny}}, \bibinfo {author}
  {\bibfnamefont {E.~S.}\ \bibnamefont {Lander}},\ and\ \bibinfo {author}
  {\bibfnamefont {J.}~\bibnamefont {Dekker}},\ }\bibfield  {title} {\bibinfo
  {title} {Comprehensive {Mapping} of {Long}-{Range} {Interactions} {Reveals}
  {Folding} {Principles} of the {Human} {Genome}},\ }\href
  {https://doi.org/10.1126/science.1181369} {\bibfield  {journal} {\bibinfo
  {journal} {Science}\ }\textbf {\bibinfo {volume} {326}},\ \bibinfo {pages}
  {289} (\bibinfo {year} {2009})}\BibitemShut {NoStop}%
\bibitem [{\citenamefont {Fudenberg}\ \emph {et~al.}(2016)\citenamefont
  {Fudenberg}, \citenamefont {Imakaev}, \citenamefont {Lu}, \citenamefont
  {Goloborodko}, \citenamefont {Abdennur},\ and\ \citenamefont
  {Mirny}}]{Fudenberg_2016}%
  \BibitemOpen
  \bibfield  {author} {\bibinfo {author} {\bibfnamefont {G.}~\bibnamefont
  {Fudenberg}}, \bibinfo {author} {\bibfnamefont {M.}~\bibnamefont {Imakaev}},
  \bibinfo {author} {\bibfnamefont {C.}~\bibnamefont {Lu}}, \bibinfo {author}
  {\bibfnamefont {A.}~\bibnamefont {Goloborodko}}, \bibinfo {author}
  {\bibfnamefont {N.}~\bibnamefont {Abdennur}},\ and\ \bibinfo {author}
  {\bibfnamefont {L.}~\bibnamefont {Mirny}},\ }\bibfield  {title} {\bibinfo
  {title} {Formation of {Chromosomal} {Domains} by {Loop} {Extrusion}},\ }\href
  {https://doi.org/10.1016/j.celrep.2016.04.085} {\bibfield  {journal}
  {\bibinfo  {journal} {Cell Reports}\ }\textbf {\bibinfo {volume} {15}},\
  \bibinfo {pages} {2038} (\bibinfo {year} {2016})}\BibitemShut {NoStop}%
\bibitem [{\citenamefont {Cahn}(1977)}]{Cahn_1977}%
  \BibitemOpen
  \bibfield  {author} {\bibinfo {author} {\bibfnamefont {J.~W.}\ \bibnamefont
  {Cahn}},\ }\bibfield  {title} {\bibinfo {title} {Critical point wetting},\
  }\href {https://doi.org/10.1063/1.434402} {\bibfield  {journal} {\bibinfo
  {journal} {The Journal of Chemical Physics}\ }\textbf {\bibinfo {volume}
  {66}},\ \bibinfo {pages} {3667} (\bibinfo {year} {1977})}\BibitemShut
  {NoStop}%
\bibitem [{\citenamefont {De~Gennes}(1985)}]{DeGennes_1985}%
  \BibitemOpen
  \bibfield  {author} {\bibinfo {author} {\bibfnamefont {P.~G.}\ \bibnamefont
  {De~Gennes}},\ }\bibfield  {title} {\bibinfo {title} {Wetting: statics and
  dynamics},\ }\href {https://doi.org/10.1103/RevModPhys.57.827} {\bibfield
  {journal} {\bibinfo  {journal} {Reviews of Modern Physics}\ }\textbf
  {\bibinfo {volume} {57}},\ \bibinfo {pages} {827} (\bibinfo {year}
  {1985})}\BibitemShut {NoStop}%
\bibitem [{\citenamefont {Binder}\ \emph {et~al.}(2003)\citenamefont {Binder},
  \citenamefont {Landau},\ and\ \citenamefont {Müller}}]{Binder_2003}%
  \BibitemOpen
  \bibfield  {author} {\bibinfo {author} {\bibfnamefont {K.}~\bibnamefont
  {Binder}}, \bibinfo {author} {\bibfnamefont {D.}~\bibnamefont {Landau}},\
  and\ \bibinfo {author} {\bibfnamefont {M.}~\bibnamefont {Müller}},\
  }\bibfield  {title} {\bibinfo {title} {Monte {Carlo} {Studies} of {Wetting},
  {Interface} {Localization} and {Capillary} {Condensation}},\ }\href
  {https://doi.org/10.1023/A:1022173600263} {\bibfield  {journal} {\bibinfo
  {journal} {Journal of Statistical Physics}\ }\textbf {\bibinfo {volume}
  {110}},\ \bibinfo {pages} {1411} (\bibinfo {year} {2003})}\BibitemShut
  {NoStop}%
\bibitem [{\citenamefont {Morin}\ \emph {et~al.}(2020)\citenamefont {Morin},
  \citenamefont {Wittmann}, \citenamefont {Choubey}, \citenamefont {Klosin},
  \citenamefont {Golfier}, \citenamefont {Hyman}, \citenamefont {Jülicher},\
  and\ \citenamefont {Grill}}]{Morin_2020}%
  \BibitemOpen
  \bibfield  {author} {\bibinfo {author} {\bibfnamefont {J.~A.}\ \bibnamefont
  {Morin}}, \bibinfo {author} {\bibfnamefont {S.}~\bibnamefont {Wittmann}},
  \bibinfo {author} {\bibfnamefont {S.}~\bibnamefont {Choubey}}, \bibinfo
  {author} {\bibfnamefont {A.}~\bibnamefont {Klosin}}, \bibinfo {author}
  {\bibfnamefont {S.}~\bibnamefont {Golfier}}, \bibinfo {author} {\bibfnamefont
  {A.~A.}\ \bibnamefont {Hyman}}, \bibinfo {author} {\bibfnamefont
  {F.}~\bibnamefont {Jülicher}},\ and\ \bibinfo {author} {\bibfnamefont
  {S.~W.}\ \bibnamefont {Grill}},\ }\href
  {https://doi.org/10.1101/2020.09.24.311712} {\bibinfo {title} {Surface
  condensation of a pioneer transcription factor on {DNA}}} (\bibinfo {year}
  {2020})\BibitemShut {NoStop}%
\bibitem [{\citenamefont {Quail}\ \emph {et~al.}(2020)\citenamefont {Quail},
  \citenamefont {Golfier}, \citenamefont {Elsner}, \citenamefont {Ishihara},
  \citenamefont {Murugesan}, \citenamefont {Renger}, \citenamefont
  {Jülicher},\ and\ \citenamefont {Brugués}}]{Quail_2020}%
  \BibitemOpen
  \bibfield  {author} {\bibinfo {author} {\bibfnamefont {T.}~\bibnamefont
  {Quail}}, \bibinfo {author} {\bibfnamefont {S.}~\bibnamefont {Golfier}},
  \bibinfo {author} {\bibfnamefont {M.}~\bibnamefont {Elsner}}, \bibinfo
  {author} {\bibfnamefont {K.}~\bibnamefont {Ishihara}}, \bibinfo {author}
  {\bibfnamefont {V.}~\bibnamefont {Murugesan}}, \bibinfo {author}
  {\bibfnamefont {R.}~\bibnamefont {Renger}}, \bibinfo {author} {\bibfnamefont
  {F.}~\bibnamefont {Jülicher}},\ and\ \bibinfo {author} {\bibfnamefont
  {J.}~\bibnamefont {Brugués}},\ }\href
  {https://doi.org/10.1101/2020.09.17.302299} {\bibinfo {title} {Force
  generation by protein-{DNA} co-condensation}} (\bibinfo {year}
  {2020})\BibitemShut {NoStop}%
\bibitem [{\citenamefont {Alemasova}\ and\ \citenamefont
  {Lavrik}(2022)}]{Alemasova_polyadp-ribose_2022}%
  \BibitemOpen
  \bibfield  {author} {\bibinfo {author} {\bibfnamefont {E.~E.}\ \bibnamefont
  {Alemasova}}\ and\ \bibinfo {author} {\bibfnamefont {O.~I.}\ \bibnamefont
  {Lavrik}},\ }\bibfield  {title} {\bibinfo {title} {Poly({ADP}-ribose) in
  {Condensates}: {The} {PARtnership} of {Phase} {Separation} and
  {Site}-{Specific} {Interactions}},\ }\href
  {https://doi.org/10.3390/ijms232214075} {\bibfield  {journal} {\bibinfo
  {journal} {International Journal of Molecular Sciences}\ }\textbf {\bibinfo
  {volume} {23}},\ \bibinfo {pages} {14075} (\bibinfo {year}
  {2022})}\BibitemShut {NoStop}%
\bibitem [{\citenamefont {Mazzocca}\ \emph {et~al.}(2023)\citenamefont
  {Mazzocca}, \citenamefont {Loffreda}, \citenamefont {Colombo}, \citenamefont
  {Fillot}, \citenamefont {Gnani}, \citenamefont {Falletta}, \citenamefont
  {Monteleone}, \citenamefont {Capozi}, \citenamefont {Bertrand}, \citenamefont
  {Legube}, \citenamefont {Lavagnino}, \citenamefont {Tacchetti},\ and\
  \citenamefont {Mazza}}]{Mazzocca_2023}%
  \BibitemOpen
  \bibfield  {author} {\bibinfo {author} {\bibfnamefont {M.}~\bibnamefont
  {Mazzocca}}, \bibinfo {author} {\bibfnamefont {A.}~\bibnamefont {Loffreda}},
  \bibinfo {author} {\bibfnamefont {E.}~\bibnamefont {Colombo}}, \bibinfo
  {author} {\bibfnamefont {T.}~\bibnamefont {Fillot}}, \bibinfo {author}
  {\bibfnamefont {D.}~\bibnamefont {Gnani}}, \bibinfo {author} {\bibfnamefont
  {P.}~\bibnamefont {Falletta}}, \bibinfo {author} {\bibfnamefont
  {E.}~\bibnamefont {Monteleone}}, \bibinfo {author} {\bibfnamefont
  {S.}~\bibnamefont {Capozi}}, \bibinfo {author} {\bibfnamefont
  {E.}~\bibnamefont {Bertrand}}, \bibinfo {author} {\bibfnamefont
  {G.}~\bibnamefont {Legube}}, \bibinfo {author} {\bibfnamefont
  {Z.}~\bibnamefont {Lavagnino}}, \bibinfo {author} {\bibfnamefont
  {C.}~\bibnamefont {Tacchetti}},\ and\ \bibinfo {author} {\bibfnamefont
  {D.}~\bibnamefont {Mazza}},\ }\bibfield  {title} {\bibinfo {title} {Chromatin
  organization drives the search mechanism of nuclear factors},\ }\href
  {https://doi.org/10.1038/s41467-023-42133-5} {\bibfield  {journal} {\bibinfo
  {journal} {Nature Communications}\ }\textbf {\bibinfo {volume} {14}},\
  \bibinfo {pages} {6433} (\bibinfo {year} {2023})}\BibitemShut {NoStop}%
\bibitem [{\citenamefont {Morin}\ \emph {et~al.}(2022)\citenamefont {Morin},
  \citenamefont {Wittmann}, \citenamefont {Choubey}, \citenamefont {Klosin},
  \citenamefont {Golfier}, \citenamefont {Hyman}, \citenamefont {Jülicher},\
  and\ \citenamefont {Grill}}]{Morin_2022}%
  \BibitemOpen
  \bibfield  {author} {\bibinfo {author} {\bibfnamefont {J.~A.}\ \bibnamefont
  {Morin}}, \bibinfo {author} {\bibfnamefont {S.}~\bibnamefont {Wittmann}},
  \bibinfo {author} {\bibfnamefont {S.}~\bibnamefont {Choubey}}, \bibinfo
  {author} {\bibfnamefont {A.}~\bibnamefont {Klosin}}, \bibinfo {author}
  {\bibfnamefont {S.}~\bibnamefont {Golfier}}, \bibinfo {author} {\bibfnamefont
  {A.~A.}\ \bibnamefont {Hyman}}, \bibinfo {author} {\bibfnamefont
  {F.}~\bibnamefont {Jülicher}},\ and\ \bibinfo {author} {\bibfnamefont
  {S.~W.}\ \bibnamefont {Grill}},\ }\bibfield  {title} {\bibinfo {title}
  {Sequence-dependent surface condensation of a pioneer transcription factor on
  {DNA}},\ }\href {https://doi.org/10.1038/s41567-021-01462-2} {\bibfield
  {journal} {\bibinfo  {journal} {Nature Physics}\ }\textbf {\bibinfo {volume}
  {18}},\ \bibinfo {pages} {271} (\bibinfo {year} {2022})}\BibitemShut
  {NoStop}%
\bibitem [{\citenamefont {García~Fernández}\ \emph
  {et~al.}(2025)\citenamefont {García~Fernández}, \citenamefont {Park},
  \citenamefont {Chapuis}, \citenamefont {Pinto~Jurado}, \citenamefont
  {Imburchia}, \citenamefont {Smith}, \citenamefont {José~Longarini},
  \citenamefont {Taddei}, \citenamefont {Hubert}, \citenamefont {Sokolovska},
  \citenamefont {Matić}, \citenamefont {Huet},\ and\ \citenamefont
  {Miné-Hattab}}]{GarciaFernandez_2025}%
  \BibitemOpen
  \bibfield  {author} {\bibinfo {author} {\bibfnamefont {F.}~\bibnamefont
  {García~Fernández}}, \bibinfo {author} {\bibfnamefont {J.}~\bibnamefont
  {Park}}, \bibinfo {author} {\bibfnamefont {C.}~\bibnamefont {Chapuis}},
  \bibinfo {author} {\bibfnamefont {E.}~\bibnamefont {Pinto~Jurado}}, \bibinfo
  {author} {\bibfnamefont {V.}~\bibnamefont {Imburchia}}, \bibinfo {author}
  {\bibfnamefont {R.}~\bibnamefont {Smith}}, \bibinfo {author} {\bibfnamefont
  {E.}~\bibnamefont {José~Longarini}}, \bibinfo {author} {\bibfnamefont
  {A.}~\bibnamefont {Taddei}}, \bibinfo {author} {\bibfnamefont
  {C.}~\bibnamefont {Hubert}}, \bibinfo {author} {\bibfnamefont
  {N.}~\bibnamefont {Sokolovska}}, \bibinfo {author} {\bibfnamefont
  {I.}~\bibnamefont {Matić}}, \bibinfo {author} {\bibfnamefont
  {S.}~\bibnamefont {Huet}},\ and\ \bibinfo {author} {\bibfnamefont
  {J.}~\bibnamefont {Miné-Hattab}},\ }\bibfield  {title} {\bibinfo {title}
  {Single nucleosome imaging reveals principles of transient multiscale
  chromatin reorganization triggered by histone {ADP}-ribosylation at {DNA}
  lesions},\ }\href {https://doi.org/10.1038/s41467-025-61834-7} {\bibfield
  {journal} {\bibinfo  {journal} {Nature Communications}\ }\textbf {\bibinfo
  {volume} {16}},\ \bibinfo {pages} {6652} (\bibinfo {year}
  {2025})}\BibitemShut {NoStop}%
\bibitem [{\citenamefont {Sabari}\ \emph {et~al.}(2018)\citenamefont {Sabari},
  \citenamefont {Dall’Agnese}, \citenamefont {Boija}, \citenamefont {Klein},
  \citenamefont {Coffey}, \citenamefont {Shrinivas}, \citenamefont {Abraham},
  \citenamefont {Hannett}, \citenamefont {Zamudio}, \citenamefont {Manteiga},
  \citenamefont {Li}, \citenamefont {Guo}, \citenamefont {Day}, \citenamefont
  {Schuijers}, \citenamefont {Vasile}, \citenamefont {Malik}, \citenamefont
  {Hnisz}, \citenamefont {Lee}, \citenamefont {Cisse}, \citenamefont {Roeder},
  \citenamefont {Sharp}, \citenamefont {Chakraborty},\ and\ \citenamefont
  {Young}}]{Sabari_2018}%
  \BibitemOpen
  \bibfield  {author} {\bibinfo {author} {\bibfnamefont {B.~R.}\ \bibnamefont
  {Sabari}}, \bibinfo {author} {\bibfnamefont {A.}~\bibnamefont
  {Dall’Agnese}}, \bibinfo {author} {\bibfnamefont {A.}~\bibnamefont
  {Boija}}, \bibinfo {author} {\bibfnamefont {I.~A.}\ \bibnamefont {Klein}},
  \bibinfo {author} {\bibfnamefont {E.~L.}\ \bibnamefont {Coffey}}, \bibinfo
  {author} {\bibfnamefont {K.}~\bibnamefont {Shrinivas}}, \bibinfo {author}
  {\bibfnamefont {B.~J.}\ \bibnamefont {Abraham}}, \bibinfo {author}
  {\bibfnamefont {N.~M.}\ \bibnamefont {Hannett}}, \bibinfo {author}
  {\bibfnamefont {A.~V.}\ \bibnamefont {Zamudio}}, \bibinfo {author}
  {\bibfnamefont {J.~C.}\ \bibnamefont {Manteiga}}, \bibinfo {author}
  {\bibfnamefont {C.~H.}\ \bibnamefont {Li}}, \bibinfo {author} {\bibfnamefont
  {Y.~E.}\ \bibnamefont {Guo}}, \bibinfo {author} {\bibfnamefont {D.~S.}\
  \bibnamefont {Day}}, \bibinfo {author} {\bibfnamefont {J.}~\bibnamefont
  {Schuijers}}, \bibinfo {author} {\bibfnamefont {E.}~\bibnamefont {Vasile}},
  \bibinfo {author} {\bibfnamefont {S.}~\bibnamefont {Malik}}, \bibinfo
  {author} {\bibfnamefont {D.}~\bibnamefont {Hnisz}}, \bibinfo {author}
  {\bibfnamefont {T.~I.}\ \bibnamefont {Lee}}, \bibinfo {author} {\bibfnamefont
  {I.~I.}\ \bibnamefont {Cisse}}, \bibinfo {author} {\bibfnamefont {R.~G.}\
  \bibnamefont {Roeder}}, \bibinfo {author} {\bibfnamefont {P.~A.}\
  \bibnamefont {Sharp}}, \bibinfo {author} {\bibfnamefont {A.~K.}\ \bibnamefont
  {Chakraborty}},\ and\ \bibinfo {author} {\bibfnamefont {R.~A.}\ \bibnamefont
  {Young}},\ }\bibfield  {title} {\bibinfo {title} {Coactivator condensation at
  super-enhancers links phase separation and gene control},\ }\href
  {https://doi.org/10.1126/science.aar3958} {\bibfield  {journal} {\bibinfo
  {journal} {Science}\ }\textbf {\bibinfo {volume} {361}},\ \bibinfo {pages}
  {eaar3958} (\bibinfo {year} {2018})}\BibitemShut {NoStop}%
\bibitem [{\citenamefont {Cho}\ \emph {et~al.}(2018)\citenamefont {Cho},
  \citenamefont {Spille}, \citenamefont {Hecht}, \citenamefont {Lee},
  \citenamefont {Li}, \citenamefont {Grube},\ and\ \citenamefont
  {Cisse}}]{Cho_2018}%
  \BibitemOpen
  \bibfield  {author} {\bibinfo {author} {\bibfnamefont {W.-K.}\ \bibnamefont
  {Cho}}, \bibinfo {author} {\bibfnamefont {J.-H.}\ \bibnamefont {Spille}},
  \bibinfo {author} {\bibfnamefont {M.}~\bibnamefont {Hecht}}, \bibinfo
  {author} {\bibfnamefont {C.}~\bibnamefont {Lee}}, \bibinfo {author}
  {\bibfnamefont {C.}~\bibnamefont {Li}}, \bibinfo {author} {\bibfnamefont
  {V.}~\bibnamefont {Grube}},\ and\ \bibinfo {author} {\bibfnamefont {I.~I.}\
  \bibnamefont {Cisse}},\ }\bibfield  {title} {\bibinfo {title} {Mediator and
  {RNA} polymerase {II} clusters associate in transcription-dependent
  condensates},\ }\href {https://doi.org/10.1126/science.aar4199} {\bibfield
  {journal} {\bibinfo  {journal} {Science}\ }\textbf {\bibinfo {volume}
  {361}},\ \bibinfo {pages} {412} (\bibinfo {year} {2018})}\BibitemShut
  {NoStop}%
\bibitem [{\citenamefont {Lindahl}\ and\ \citenamefont
  {Barnes}(2000)}]{Lindahl_2000}%
  \BibitemOpen
  \bibfield  {author} {\bibinfo {author} {\bibfnamefont {T.}~\bibnamefont
  {Lindahl}}\ and\ \bibinfo {author} {\bibfnamefont {D.}~\bibnamefont
  {Barnes}},\ }\bibfield  {title} {\bibinfo {title} {Repair of {Endogenous}
  {DNA} {Damage}},\ }\href {https://doi.org/10.1101/sqb.2000.65.127} {\bibfield
   {journal} {\bibinfo  {journal} {Cold Spring Harbor Symposia on Quantitative
  Biology}\ }\textbf {\bibinfo {volume} {65}},\ \bibinfo {pages} {127}
  (\bibinfo {year} {2000})}\BibitemShut {NoStop}%
\bibitem [{\citenamefont {Miné-Hattab}\ \emph {et~al.}(2022)\citenamefont
  {Miné-Hattab}, \citenamefont {Liu},\ and\ \citenamefont
  {Taddei}}]{Mine-hattab_2022}%
  \BibitemOpen
  \bibfield  {author} {\bibinfo {author} {\bibfnamefont {J.}~\bibnamefont
  {Miné-Hattab}}, \bibinfo {author} {\bibfnamefont {S.}~\bibnamefont {Liu}},\
  and\ \bibinfo {author} {\bibfnamefont {A.}~\bibnamefont {Taddei}},\
  }\bibfield  {title} {\bibinfo {title} {Repair {Foci} as {Liquid} {Phase}
  {Separation}: {Evidence} and {Limitations}},\ }\href
  {https://doi.org/10.3390/genes13101846} {\bibfield  {journal} {\bibinfo
  {journal} {Genes}\ }\textbf {\bibinfo {volume} {13}},\ \bibinfo {pages}
  {1846} (\bibinfo {year} {2022})}\BibitemShut {NoStop}%
\bibitem [{\citenamefont {Fries}\ \emph {et~al.}(2025)\citenamefont {Fries},
  \citenamefont {Diaz}, \citenamefont {Jardat}, \citenamefont {Pagonabarraga},
  \citenamefont {Illien},\ and\ \citenamefont {Dahirel}}]{Fries_active_2025}%
  \BibitemOpen
  \bibfield  {author} {\bibinfo {author} {\bibfnamefont {J.}~\bibnamefont
  {Fries}}, \bibinfo {author} {\bibfnamefont {J.}~\bibnamefont {Diaz}},
  \bibinfo {author} {\bibfnamefont {M.}~\bibnamefont {Jardat}}, \bibinfo
  {author} {\bibfnamefont {I.}~\bibnamefont {Pagonabarraga}}, \bibinfo {author}
  {\bibfnamefont {P.}~\bibnamefont {Illien}},\ and\ \bibinfo {author}
  {\bibfnamefont {V.}~\bibnamefont {Dahirel}},\ }\bibfield  {title} {\bibinfo
  {title} {Active droplets controlled by enzymatic reactions},\ }\href
  {https://doi.org/10.1098/rsif.2024.0803} {\bibfield  {journal} {\bibinfo
  {journal} {Journal of The Royal Society Interface}\ }\textbf {\bibinfo
  {volume} {22}},\ \bibinfo {pages} {20240803} (\bibinfo {year}
  {2025})}\BibitemShut {NoStop}%
\bibitem [{\citenamefont {Berthin}\ \emph {et~al.}(2025)\citenamefont
  {Berthin}, \citenamefont {Fries}, \citenamefont {Jardat}, \citenamefont
  {Dahirel},\ and\ \citenamefont {Illien}}]{Berthin_2025}%
  \BibitemOpen
  \bibfield  {author} {\bibinfo {author} {\bibfnamefont {R.}~\bibnamefont
  {Berthin}}, \bibinfo {author} {\bibfnamefont {J.~D.}\ \bibnamefont {Fries}},
  \bibinfo {author} {\bibfnamefont {M.}~\bibnamefont {Jardat}}, \bibinfo
  {author} {\bibfnamefont {V.}~\bibnamefont {Dahirel}},\ and\ \bibinfo {author}
  {\bibfnamefont {P.}~\bibnamefont {Illien}},\ }\bibfield  {title} {\bibinfo
  {title} {Microscopic and stochastic simulations of chemically active
  droplets},\ }\href {https://doi.org/10.1103/PhysRevE.111.L023403} {\bibfield
  {journal} {\bibinfo  {journal} {Physical Review E}\ }\textbf {\bibinfo
  {volume} {111}},\ \bibinfo {pages} {L023403} (\bibinfo {year}
  {2025})}\BibitemShut {NoStop}%
\end{thebibliography}%

\end{document}